\documentclass[twocolumn]{svjour3}
\usepackage{epsfig,graphics}
\usepackage{xcolor}
\usepackage{epstopdf}
\definecolor{mydarkblue}{rgb}{0,0.08,0.45}
\usepackage{url}
\usepackage{times}
\usepackage{amsmath,amssymb}
\usepackage{mathrsfs}
\usepackage{mhchem}
\usepackage{caption}
\usepackage{authblk}
\usepackage[ruled]{algorithm2e}
\usepackage{subfigure}
\usepackage[top=1.5in,bottom=1.5in,left=1.1in,right=1.1in]{geometry}
\usepackage{newclude}
\usepackage{afterpage}

\newcommand{\ind}{\mathbf{1}}
\newcommand{\rd}{\textrm{d}}
\newcommand{\eps}{\epsilon}
\newcommand{\vz}{\mathbf{z}}

\newcommand{\vk}{\mathbf{k}}

\newcommand{\mD}{\mathbf{D}}

\newcommand{\mM}{\mathbf{M}}
\newcommand{\mW}{\mathbf{W}(t)}

\newcommand{\mQ}{\mathbf{Q}}
\newcommand{\mG}{\mathbf{G}}
\newcommand{\mI}{\mathbf{I}}

\newcommand{\vf}{\mathbf{f}}
\newcommand{\mF}{\mathbf{F}}
\newcommand{\vx}{\mathbf{x}}
\newcommand{\vy}{\mathbf{y}}
\newcommand{\vzero}{\mathbf{0}}
\def\veta{\mbox{\boldmath$\eta$}}

\newcommand{\ptrans}[3]{p(#2|#1;#3)}

\newcommand{\postcap}{\vspace*{-0.25in}}

\def\keywords{\vspace{.5em}
{\textit{Keywords}:\,\relax%
}}

\begin{document}

\title{Irreversible Samplers from Jump and Continuous Markov Processes}

\author{Yi-An Ma \and Emily B. Fox \and Tianqi Chen \and Lei Wu}
\institute{Yi-An Ma \at Department of Electrical Engineering and Computer Sciences, University of California, Berkeley
\\\email{yianma@berkeley.edu}
\and
Emily B. Fox \at Department of Statistics, University of Washington
\\\email{ebfox@stat.washington.edu}
\and
Tianqi Chen \at Department of Computer Science and Engineering, University of Washington
\and
Lei Wu \at School of Mathematical Sciences, Peking University}

\maketitle

\begin{abstract}
In this paper, we propose irreversible versions of the Metropolis Hastings (MH) and Metropolis adjusted Langevin algorithm (MALA) with a main focus on the latter.
For the former, we show how one can simply switch between different proposal and acceptance distributions upon rejection to obtain an irreversible jump sampler (I-Jump).
The resulting algorithm has a simple implementation akin to MH, but with the demonstrated benefits of irreversibility.
We then show how the previously proposed MALA method can also be extended to exploit irreversible stochastic dynamics as proposal distributions in the I-Jump sampler.
Our experiments explore how irreversibility can increase the efficiency of the samplers in different situations.
\end{abstract}

\keywords{Bayesian inference; Hamiltonian Monte Carlo; Irreversible samplers; Jump processes; Markov chain Monte Carlo; Metropolis-Hastings}

\section{Introduction}
\label{sec:intro}
Markov chain Monte Carlo (MCMC) methods are the defacto tools for inference in Bayesian models \cite{JunLiu,XiAn}.
The Metropolis-Hastings (MH) algorithm is often used as the default approach because of its ease of implementation.
One designs a proposal distribution to generate samples and uses an accept-reject procedure to ensure that the target distribution is maintained.
A focus has been on developing clever proposals \cite{RAM,HeavyTail,JunLiu} to specify a Markov process with good mixing rates, but traditional methods are often strongly coupled to a specific challenge setting, like multimodal targets \cite{RAM} or heavy tailed distributions \cite{HeavyTail}.
In practice, one often does not know the structure of the target distribution, which might additionally exhibit a combination of these factors.

To address some of these limitations, the Metropolis adjusted Langevin algorithm (MALA) \cite{MALA,RMALA} and Hamiltonian Monte Carlo (HMC) \cite{Duane:1987HMC,NealHMC} have been proposed.
These methods are constructed to use the local gradient information of the target distribution in the proposal.
In these approaches, the task of constructing good samplers is translated to finding continuous Markov dynamics, in the form of differential equations (stochastic or deterministic), that can generate better MH proposals for general target distributions.
A lot of recent attention has been on extensions to the Langevin and Hamiltonian dynamics, such as through use of second order information of the target distribution resulting in Riemann manifold variants \cite{RMALA,RiemannianMALA}.

One restriction of all the above approaches is that regardless of the properties of the underlying continuous dynamics, direct use of the MH correction necessarily results in the
whole sampler being reversible (the forward path is statistically indistinguishable from the backward one).
This represents a serous limitation since irreversibility has been shown to increase the mixing rate of samplers in general \cite{NealNonReversible,Diaconis,Chen_Hwang,Chen_Lovasz_Pak}.

Our goal is to address this issue by defining easy-to-specify and computationally straightforward irreversible samplers.
In particular, we focus in on defining an irreversible MALA algorithm (I-MALA); the construction of this algorithm consists of two parts.
One part is to construct irreversible continuous Markov dynamics that leaves the target distribution invariant (see Sec.~\ref{sec:framework}).
Then, to correct for the discretization error involved in simulating the continuous Markov dynamics---while avoiding use of the reversible MH procedure---we develop an irreversible jump sampler (I-Jump).
The I-Jump algorithm has an implementation similar to the simplicity of the MH algorithm, but allows for irreversibility (see Sec.~\ref{sec:jump}).

We start in Sec.~\ref{sec:framework} by providing background on samplers using continuous Markov processes and present a general stochastic differential equation (SDE) based framework;
a preliminary version of this work appeared in \cite{completesample}.
Within the framework, general irreversible continuous dynamics are defined through specifying two matrices: a positive semidefinite matrix and a skew-symmetric matrix.
We prove that for any choice of these matrices, the continuous dynamics leave the target distribution invariant.
We likewise prove that any continuous Markov process with the correct stationary distribution has a representation in this framework.

We then turn our attention in Sec.~\ref{sec:jump} to introducing the I-Jump sampler alternative to the MH algorithm.
We first reparameterize the jump process and arrive at a straightforward set of constraints on the transition probabilities that ensures that the target distribution is the stationary distribution.
We then revise the MH algorithm to allow for irreversibility, while at the same time satisfying the derived constraints for correctness.
The resulting sampler implementation has the ease and efficiency of the standard MH method.
Importantly, our method outperforms existing approaches \cite{MH1,MH2} in terms of mixing rate versus runtime in a range of settings, from heavy-tailed to multimodal targets.  We further demonstrate these performance gains in a set of challenging real world applications.

In Sec.~\ref{sec:combined}, we then make use of the general irreversible stochastic dynamics framework to define better proposal distributions in the I-Jump sampler, leading to our sought-after I-MALA algorithm.
%
Similar to the MALA or HMC algorithm, we use a discretization of the continuous dynamics to propose a sample, but then use our I-Jump sampler rather than the reversible MH correction.
Importantly, local gradient information of the target distribution is taken into account in the SDE for better efficiency than the I-Jump algorithm with standard independent proposals.
Furthermore, the simplicity of the MH algorithm is still inherited while the overall sampling process remains irreversible.
We can view the benefits of this approach from two angles: (i) the SDE can provide an efficient proposal distribution for our I-Jump sampler or (ii) the accept-reject scheme allows us to correct for the bias introduced by sampling the continuous dynamics via a discretized SDE.
This also opens up the possibility to combine, for example, the fast mixing of Langevin diffusion within local modes with the fast traversing of Hamiltonian dynamics.
We examine the increased efficiency of our proposed I-MALA algorithm within the context of a Bayesian logistic regression model and stochastic volatility model.



\section{Samplers Using Continuous Markov Processes}
\label{sec:framework}
We start with the standard MCMC goal of drawing samples from an unnormalized target distribution $\pi(\theta)$.
It is often customary to include auxiliary variables $r$ (e.g. in the HMC algorithm) to facilitate the sampling process.
Hence we write $\vz = (\theta, r)$ to contain all variables being sampled according to the joint distribution $\pi(\vz)$.
In this section, we focus on continuous Markov processes and
discuss possible choices of continuous dynamics that leave $\pi(\vz)$ invariant.
If the stochastic process is further ergodic, then simulating the stationary stochastic dynamics equates with providing samples from $\pi(\vz)$.
These processes will be deployed in the I-MALA algorithm of Sec.~\ref{sec:combined}.


A realization of a continuous Markov process can be represented as the following stochastic differential equation (SDE):
\begin{align}
\rd \vz = \vf(\vz) \rd t + \sqrt{2 \mD(\vz)} \rd \mW, \label{Eq:ItoSDE}
\end{align}
where $\vz$: $[0,+\infty)\times \Omega \rightarrow \mathbb{R}^d$ is a real random vector defined on a probability space $(\Omega, \Sigma, P)$, parameterized by time $t$; $\vf: \mathbb{R}^d \rightarrow \mathbb{R}^d$ is a real vector valued function; $\mD(\vz)$ is a $d\times d$ positive semidefinite diffusion matrix; and $\mW$ is a $d$ dimensional Wiener process.
Following It\^o's convention, \eqref{Eq:ItoSDE} defines the following diffusion process (with an abuse of notation, we also use $\vz$ to denote value of the random vector $\vz$):
\begin{align}
\centering
\dfrac{\partial}{\partial t} p(\vz;t)
= &\sum_{i,j = 1}^d \dfrac{\partial^2}{\partial \vz_i \partial \vz_j} \big[\mD_{ij}(\vz) p(\vz;t)\big] \nonumber\\
&- \sum_{i=1}^d \dfrac{\partial}{\partial \vz_i} \big[\vf_i(\vz) p(\vz;t) \big], \label{Eq:ItoFPE}
\end{align}
where $p(\vz;t)$ is the probability density function (assuming it exists for all $t$) of random vector $\vz$ at time $t$.
The continuous Markov process, \eqref{Eq:ItoSDE}, can be used to generate samples from $\pi(\vz)$ if $\pi(\vz)$ is a stationary solution to \eqref{Eq:ItoFPE}.

Although \eqref{Eq:ItoSDE} provides a way to simulate the continuous dynamics and obtain samples from the Markov process, it is not clear which choices of $\vf$ and $\mD$ will result in a stationary distribution of \eqref{Eq:ItoFPE} equal to the target distribution $\pi(\vz)$.
For a given $\vf$ and $\mD$, \eqref{Eq:ItoFPE} allows us to analyze this stationary distribution, but it is very challenging to define the set of $\vf$ and $\mD$ that yield a specified stationary distribution.
Researchers have resorted to special cases such as overdamped Langevin \cite{MALA,SGLD}, underdamped Langevin \cite{Horowitz:1991,SGHMC} and Nos\'e-Hoover \cite{SGNHT,Shang} dynamics in the statistical physics literature for inspiration.

\subsection{Complete Recipe with Continuous Markov Dynamics}
\label{sec:cont}
We propose using an alternative form for \eqref{Eq:ItoSDE} specified via two matrices $\mD$ and $\mQ$, as well as the target distribution $\pi(\vz)$, as first considered in \cite{completesample,JShi}:
\begin{align}
\rd \vz =& \Big[ - \big(\mD(\vz)+\mQ(\vz)\big) \nabla H(\vz) + \Gamma(\vz)\Big] \rd t  \nonumber\\
&+ \sqrt{2 \mD(\vz)} \rd \mW,
\nonumber\\
\Gamma_i(\vz) &= \sum_{j=1}^d \dfrac{\partial}{\partial \vz_j} \big({\mD}_{ij}(\vz) + {\mQ}_{ij}(\vz)\big)
\label{Eq:SDE}
\end{align}
Here, $H(\vz) = -\log(\pi(\vz))$; $\mD(\vz)$ is a positive semidefinite diffusion matrix and $\mQ(\vz)$ a skew-symmetric matrix.
As discussed in Appendix~\ref{appendix:rev}, \eqref{Eq:SDE} decomposes into reversible and irreversible process.
Matrix $\mD(\vz)$ corresponds to the reversible part while matrix $\mQ(\vz)$ determines the irreversible part.
In the following, we use the Fokker-Planck equation associated with \eqref{Eq:SDE} to verify two properties of this representation: One is that \eqref{Eq:SDE} has $\pi(\vz)$ as its invariant distribution; the other is that any continuous Markov process with $\pi(\vz)$ as the invariant distribution can be written in the form of \eqref{Eq:SDE} (i.e., there exists a $\mD(\vz)$ and $\mQ(\vz)$ that place the process in this representation).
Together, these two properties (i) allow us to very straightforwardly explore a set of valid samplers by specifying pairs $(\mD(\vz),\mQ(\vz))$ of positive semidefinite and skew-symmetric matrices, respectively, and (ii) ensure that as we span the space of all possible $(\mD(\vz),\mQ(\vz))$, we know we have covered all possible valid samplers.
That is, our representation is \emph{complete}.
A preliminary version of this complete framework appeared in \cite{completesample}.

The Fokker-Planck equation (following It\^o's convention) associated with \eqref{Eq:SDE} is (see Appendix~\ref{appnd_CompleteProof}):
\begin{align}
\centering
&\dfrac{\partial}{\partial t} p(\vz;t) \label{Eq:FPE} \\ \nonumber
&= \nabla^T \cdot
	\bigg(  \big[\mD(\vz) + \mQ(\vz)\big] \left[
p(\vz;t) \nabla H(\vz) + \nabla p(\vz;t) \right]\bigg),
\end{align}
%
%
%
where $\nabla^T \cdot \vf(\vz) = \sum_{i=1}^d \dfrac{\partial f_i(\vz)}{\partial \vz_i}$.
It is straightforward to verify that $p^s(\vz)\propto\pi(\vz) = e^{-H(\vz)}$ is a stationary solution to \eqref{Eq:FPE}.  More significantly, Theorem~\ref{thm:complete} states that any process \eqref{Eq:ItoSDE} and \eqref{Eq:ItoFPE} with stationary distribution $\pi(\vz)$ has a representation in our framework.
\begin{theorem}
Suppose \eqref{Eq:ItoFPE}
has stationary probability density function $p^s(\vz)\propto\pi(\vz)$.  Further assume that \sloppy
$\left[\vf_i(\vz) \pi(\vz) - \sum_{j=1}^d \dfrac{\partial}{\partial \theta_j} \Big({\mD}_{ij}(\vz) \pi(\vz)\Big)\right]$ is Lebesgue integrable.
Then, there exists a skew-symmetric matrix $\mQ(\vz)$ such that the right hand side of \eqref{Eq:ItoFPE} is equivalent to the right hand side of \eqref{Eq:FPE}.
\label{thm:complete}
\end{theorem}
For the reader's convenience, we include the constructive proof of Theorem~\ref{thm:complete} in Appendix~\ref{appnd_CompleteProof}.
To the best of our knowledge, the exact form of \eqref{Eq:SDE} was first presented in the statistical mechanics literature \cite{PAo,JShi}; however, the completeness of the representation of continuous Markov processes was made only later in our preliminary paper \cite{completesample}.
The proof of Theorem~\ref{thm:complete} is comprised of two sets of ideas stemming from different fields:
In the study of continuous Markov processes, earlier work \cite{qian-jsp,Dembo,qian_jmp,large_deviations,Villani,PavBook} realized that diffusion processes with stationary probability density function $\pi(\vz)$ can be decomposed into reversible and irreversible parts preserving $\pi(\vz)$ as the invariant measure.
In stochastic models of fluid dynamics and homogenization, earlier work \cite{stochastic_fluid_dynamics} found that divergenceless vector fields can be represented as the divergence of an anti-symmetric matrix valued potential.
Combining both ideas leads to the discovery of \eqref{Eq:FPE} that underlies the proof of Theorem~\ref{thm:complete}.
Similar structures have also been examined when one or both of $\mD(\vz)$ and $\mQ(\vz)$ are constant matrices \cite{hwang1993,Thouless}.

\subsection{Continuous Markov Process Sampling Algorithm}
\label{sec:discretization}
We can simulate from \eqref{Eq:SDE} by using the following $\epsilon$-discretization of the SDE:
\begin{align}
\vz_{t+1} \leftarrow & \vz_{t} + \eps_t \left[ - \big({\mD}(\vz_t) + {\mQ}(\vz_t)\big) \nabla H(\vz_t)
+ \Gamma(\vz_t)\right] \nonumber\\
&+ \eta_t,
\quad
\eta_t \sim \mathcal{N}(0, 2 \eps_t \mD(\vz_t)). \label{Eq:simple-update}
\end{align}
Although \eqref{Eq:simple-update} is in the form of the Euler--Maruyama method, higher order numerical schemes can be used for better accuracy \cite{DingHighOrder,Owhadi,LeimkuhlerHighOrder}.
For example, in HMC, since the diffusion matrix $\mD$ is zero, a neutral integration scheme such as leap frog (a second order integration scheme) is often used for accuracy and stability of integration \cite{NealHMC}.
Other higher order numerical methods such as the splitting scheme \cite{DingHighOrder,Owhadi} and simple modifications to the Euler--Maruyama method \cite{LeimkuhlerHighOrder} can also lead to higher order of accuracy in different scenarios.

The resulting algorithm according to \eqref{Eq:simple-update} is outlined in Algorithm~\ref{alg:SDE_Sampler}.
Note that relying on a sample path from the discretized system of \eqref{Eq:SDE} typically leads to the introduction of bias due to discretization error.
In these cases, the samples only provide unbiased estimates in the limit as $\eps_t \rightarrow 0$ unless further corrections are introduced.
In Sec.~\ref{sec:combined}, we use the dynamics of \eqref{Eq:simple-update} as the proposal distribution inside an irreversible jump process to correct for any potential discretization error.

\begin{algorithm}[t!]
\caption{Continuous Markov Process Sampling Algorithm}
initialize $\vz_0$ \\
 \For{$t=0,1,2\cdots N_{iter}$}{
   \For{$i=1\cdots n$}{$\Gamma_i(\vz) = \sum_j \dfrac{\partial}{\partial \vz_j} \left({\mD}_{ij}(\vz) + {\mQ}_{ij}(\vz)\right)$}
   sample $\eta_t \sim \mathcal{N}(0, 2 \eps_t \mD(\vz_t))$
   \\
   $\vz_{t+1} \leftarrow \vz_{t} +  \eps_t \left[ - \big({\mD}(\vz_t) + {\mQ}(\vz_t)\big) \nabla H(\vz_t)
+ \Gamma(\vz_t)\right] + \eta_t$
 }
 \label{alg:SDE_Sampler}
\end{algorithm}

\subsection{Previous Dynamics in MCMC Algorithms as Special Cases}
\label{sec:previous_continuous}
We explicitly state how some previous continuous-dynamic-based MCMC methods fit within the proposed framework based on specific choices of $\mD(\vz)$, $\mQ(\vz)$ and $H(\vz)$.

\paragraph{Hamiltonian Dynamics}
The key ingredient in HMC~\cite{Duane:1987HMC,NealHMC} is Hamiltonian dynamics, which simulates the physical motion of an object with position $\theta$, momentum $r$, and mass $\mM$ on an frictionless surface as follows:
\begin{align}
\left\{
\begin{array}{l}
      \rd \theta = {\mM}^{-1} r \rd t \\
      \rd r = - \nabla U(\theta) \rd t.
\end{array}
\right. \label{eq:HMC}
\end{align}
%
It can be observed that \eqref{eq:HMC} is a special case of the proposed framework with $\vz = (\theta, r)$, $H(\theta, r) =  U(\theta) + \frac{1}{2} r^T \mM^{-1} r$,  ${\mQ}(\theta, r) =
\left(
\begin{array}{ll}
         0 & -\mI \\
         \mI & 0
\end{array}
\right)$ and $\mD(\theta, r) = \vzero$.
See Sec.~\ref{sec:MALA} for a more complete discussion.

\paragraph{Langevin Dynamics} The Langevin dynamics sampler \cite{MALA,SGLD} proposes to use the following first order (no momentum) Langevin dynamics to
generate samples
\begin{equation}
\rd \theta = - \mD \nabla U(\theta) \rd t + \sqrt{2 \mD} \rd \mW. \label{eq:GLD}
\end{equation}
This algorithm corresponds to taking $\vz=\theta$ with $H(\theta) = U(\theta)$, ${\mD}(\theta) = \mD$, and ${\mQ}(\theta) = \vzero$.

\paragraph{Riemannian Langevin Dynamics} The Langevin dynamics sampler can be generalized to use an adaptive diffusion matrix ${\mD}(\theta)$. Specifically, it is interesting to take ${\mD}(\theta)= \mG^{-1}(\theta)$, where $\mG(\theta)$ is the Fisher information metric \cite{RMALA,SGRLD}. The sampler iterates
\begin{align}
\rd \theta = & [ - {\mG}(\theta)^{-1} \nabla U(\theta)  + \Gamma^{\mD}(\theta)] \rd t
\nonumber\\
&+ \sqrt{2 {\mG}(\theta)^{-1}} \rd \mW.
\label{eq:GRLD}
\end{align}
We can cast this Riemannian Langevin dynamics sampler~\cite{SGRLD} into our framework taking $\mD(\theta) = {\mG}(\theta)^{-1}$, and ${\mQ}(\theta) = \vzero$.  From our framework, we know that here
\begin{align}
\Gamma^{\mD}_i(\theta) = \sum\limits_j \dfrac{\partial {\mD}_{ij} (\theta)}{\partial \theta_j}.
\end{align}
Interestingly, in earlier literature~\cite{RiemannianMALA}, $\Gamma^{\mD}_i(\theta)$ was taken to be $2 \ |\mG(\theta)|^{-1/2} \sum\limits_j \dfrac{\partial}{\partial \theta_j}\left(\mG_{ij}^{-1}(\theta) |\mG(\theta)|^{1/2} \right)$. More recently, it was found that this correction term corresponds to the distribution function with respect to a non-Lebesgue measure~\cite{MALA}.  For the Lebesgue measure, the revised $\Gamma^{\mD}_i(\theta)$ was as determined by our framework~\cite{MALA}.  This is an example of how our framework can provide guidance in devising correct samplers.

\subsection{Irreversibility in continuous dynamics}
In Appendix \ref{appendix:rev}, we show that the continuous stochastic dynamics \eqref{Eq:SDE} decompose into (i) general Riemannian Langevin dynamics and (ii) conserved, deterministic dynamics generalizing Hamiltonian dynamics.
The first component is reversible and is determined by $\mD(\vz)$ while the second part is irreversible and is determined by $\mQ(\vz)$.
The irreversible dynamics generates a circular motion that traverses through the state space.

It has been proven that incorporating this irreversible dynamics (parameterized by $\mQ(\vz)$) can only increase the mixing of the Markov process \cite{hwang1993,hwang2005,Duncan2016,LargeDeviation_Irr}.
The intuition can be drawn from analyzing the nonzero part of the spectrum of the Fokker-Planck operator in \eqref{Eq:FPE}.
It can be shown that the traversing motion brings different eigenvalues closer to each other, making the overall spectrum (except zero) narrower \cite{non_detailedbalance}.
When the spectral gap of the reversible dynamics is nonzero, introducing this narrowing effect only increases the spectral gap.
For a Gaussian target distribution and taking $\mD(\theta)=\mI$, the optimal choice of a constant $\mQ$ to increase the spectral gap has been studied \cite{PavOptimal,HwangOptimal}.

\section{Irreversible Jump Sampler}
\label{sec:jump}
Although irreversible continuous dynamics increase the mixing of the overall stochastic dynamics, the discretized algorithm of \eqref{Eq:simple-update} typically leads to bias due to discretization error as mentioned in Sec.~\ref{sec:discretization}.
If we use Metropolis-Hastings (MH) to correct for this error, the whole process becomes reversible again.
Instead, in this section we propose an irreversible jump sampler (I-Jump) that can be used in place of MH to correct for the discretization error in \eqref{Eq:simple-update} while maintaining irreversibility.

%
Although the focus of this paper is on using this I-Jump algorithm with continuous dynamics proposals---as explored in Sec.~\ref{sec:combined}---the I-Jump algorithm can be used with more traditional proposal distributions as a generic replacement for MH.
Hence, in this section we turn our attention to the jump processes and consider an equivalent representation that enables more ready analysis of the properties of the process, and the development of an efficient irreversible jump process sampler.

\subsection{Irreversible Jump Processes}
A Markov jump process can be defined by the Kolmogorov forward equation
\begin{align}
&\dfrac{\partial}{\partial t} \ptrans{\vy}{\vz}{t} \label{Eq:MJP} \\ \nonumber
&= \int_{\mathbb{R}^d} \rd \vx \Big[ W(\vz|\vx)\ptrans{\vy}{\vx}{t} - W(\vx|\vz)\ptrans{\vy}{\vz}{t} \Big],
\end{align}
where, similar to \eqref{Eq:ItoFPE}, $\ptrans{\vy}{\vz}{t}$ is the probability density function of random vector $\vz$ (parameterized by $t$) taking value $\vz$ at time $t$ conditional on it taking value of $\vy$ at time $0$.
For the process specified by a generic bivariate integrable function $W(\vx|\vz): \mathbb{R}^d\times\mathbb{R}^d\rightarrow\mathbb{R}$, it is challenging to determine which choice of $W(\vx|\vz)$ leads to a jump process with the correct stationary distribution.  Even if one can construct such a $W$, it can be challenging to use $W$ to sample a realization of the jump process.
Instead, one often restricts attention to reversible processes and uses MH.

We revisit Markov jump processes under an equivalent but alternative representation defined in terms of two bivariate functions $S$ and $A: \mathbb{R}^d\times\mathbb{R}^d\rightarrow\mathbb{R}$.
A simple set of constraints on $S$ and $A$ ensures that the target distribution $\pi(\vz)$ is the stationary distribution of the jump process.
%
In particular, we consider
\begin{align}
\dfrac{\partial}{\partial t} \ptrans{\vy}{\vz}{t}
=& \int_{\mathbb{R}^d} \big(S(\vx,\vz)+A(\vx,\vz)\big) \dfrac{\ptrans{\vy}{\vx}{t}}{\pi(\vx)} \rd \vx \nonumber\\
&- \int_{\mathbb{R}^d} S(\vx,\vz) \dfrac{\ptrans{\vy}{\vz}{t}}{\pi(\vz)} \rd \vx,
\label{Eq:Jump_S}
\end{align}
%
where $S(\vx,\vz) = S(\vz,\vx)$ is symmetric representing the reversible part of the process and $A(\vx,\vz) = -A(\vz,\vx)$ is anti-symmetric defining the irreversible part.  Based on the form of \eqref{Eq:Jump_S}, as shown in Appendix~\ref{appnd_NewJump}, we simply have to satisfy the following constraints in order to ensure that $\pi(\vz)$ is the stationary distribution:
\begin{enumerate}
	\item $S(\vx,\vz) \pi^{-1}(\vx)$ and $ A(\vx,\vz) \pi^{-1}(\vx)$ are bounded and integrable
	\item $S(\vx,\vz)+A(\vx,\vz)>0$
	\item $\int_{\mathbb{R}^d} A(\vx,\vz) \rd \vx = 0$.
	\label{Eq:Jump_Constraints}
\end{enumerate}

Discretizing \eqref{Eq:Jump_S} with $\Delta t$ step size gives the following update rule:
\begin{align}
\ptrans{\vy}{\vz}{\Delta t}
=& \dfrac{\Delta t}{\pi(\vy)} \big(S(\vy,\vz)+A(\vy,\vz)\big) \label{Eq:Jump_Sampler} \\ \nonumber
&+  \left[ 1 - \dfrac{\Delta t}{\pi(\vy)} \int_{\mathbb{R}^d} S(\vy,\vx) \rd \vx \right] \delta(\vz-\vy),
\end{align}
which defines a Markov transition kernel entirely by functions $S$ and $A$: $P(\vy, \rd \vz) = p(\vz|\vy;\Delta t) \rd \vz$.
Since the image of $P$ is a probability, it follows from the first constraint of $S$ and $A$ (under Eq.~\eqref{Eq:Jump_S}) that $\Delta t \leq 1/\max\{|| (S+A)/\pi ||_1, || (S+A)/\pi ||_\infty\}$.

Since the jump operator has $\pi(\vz)$ as the stationary distribution assuming the constraints of $S$ and $A$ are satisfied, the transition probability of \eqref{Eq:Jump_Sampler} defines a valid procedure for drawing samples from the target $\pi(\vz)$.  In particular, over time $\Delta t$, state $\vy$ transitions to state $\vz$ with probability
$\Delta t (S(\vy,\vz)+A(\vy,\vz))/\pi(\vy)$, and state $\vy$ remains unchanged with probability
\begin{align}
\left[ 1 - \dfrac{\Delta t}{\pi(\vy)} \int_{\mathbb{R}^d} S(\vy,\vx) \rd \vx \right].
\end{align}
We can further derive from \eqref{Eq:Jump_Sampler} that $\Delta t \cdot A$ is the difference between the probability of a forward path and the backward path in the update procedure:
\begin{align}
A(\vx,\vz) = \dfrac{1}{2 \Delta t}
\big( \pi(\vy) p(\vz|\vy;\Delta t) - \pi(\vz) p(\vy|\vz;\Delta t) \big).
\label{eq:A_irrv}
\end{align}
From this, we clearly see how $A$ determines the irreversibility of the process.
In Sec.~\ref{sec:irr_jump}, we examine a practical algorithm for efficiently implementing such a procedure based on an accept-reject scheme analogous to the MH algorithm outlined in Algorithm~\ref{alg:MH}.  The important challenge we conquer is handling the irreversibility of the process arising from $A \neq 0$.


\subsection{Reversible Samplers as Special Cases ($A=0$)}
\label{sec:jump_previous}

As with past continuous-dynamic-based samplers, we now cast a set of past jump-process-based samplers into our framework.

\paragraph{Direct resampling} Methods that sample directly from $\pi(\vz)$ take $S(\vy,\vz) = \dfrac{1}{\Delta t} \pi(\vy) \pi(\vz)$ and $A(\vy,\vz)=0$.  We can verify this by substituting into \eqref{Eq:Jump_Sampler}.

\paragraph{Metropolis-Hastings} The very popular MH algorithm (Algorithm~\ref{alg:MH}) falls into our framework taking
\begin{align}
S(\vy,\vz) = \dfrac{1}{\Delta t}
\min\big(\pi(\vy) q(\vz|\vy) , \pi(\vz) q(\vy|\vz)\big), \label{Eq:MH_kernels}
\end{align}
where $q(\vz|\vy)$ is the conditional probability of transiting from $\vy$ to $\vz$ and $A(\vy,\vz) = 0$.

Algorithmically, we can define
\begin{align}
	\alpha(\vy,\vz) &= \Delta t \cdot S(\vy,\vz)/(\pi(\vy) q(\vz|\vy)) \nonumber\\
&= \min \left(1,\frac{\pi(\vz)q(\vy|\vz)}{\pi(\vy)q(\vz|\vy)}\right),
	\label{Eq:alpha}
\end{align}
such that both $q(\vz|\vy)$ and $\alpha(\vy,\vz)$ are less than or equal to $1$.
Then the update rule can be expressed as \cite{chib}:
\begin{align}
p(\vz,\Delta t|\vy) =& q(\vz|\vy) \alpha(\vy,\vz) \\ \nonumber
&+ \left[ 1 - \int_{\mathbb{R}^d} q(\vz|\vy) \alpha(\vy,\vz) \rd \vx \right] \delta(\vz-\vy).
\end{align}
When in state $\vy$ at time $t$, we propose to jump to state $\vz$ at $t+\Delta t$ with conditional probability $q(\vz|\vy)$, realized via a random number generator that has a distribution according to $q(\vz|\vy)$; we accept this proposal with probability $\alpha(\vy,\vz)$ to ensure that the target distribution will be preserved under this procedure.
Hence, the total probability of transiting from state $\vy$ to $\vz$ is $q(\vz|\vy) \alpha(\vy,\vz)$. Otherwise, we stay in state $\vy$.
We see that MH restricts our attention to reversible cases as $A(\vy,\vz)$ is always set to be $0$.

\begin{algorithm}[t!]
\caption{Metropolis-Hastings Algorithm}
 \For{$t=0,1,2\cdots N_{iter}$}{
  sample $u \sim \mathcal{U}_{[0,1]}$\\
  propose $\vz(*) \sim q(\vz(*)|\vz{(t)})$\\
  $\alpha\left(\vz^{(t)}, \vz(*)\right)
  = \min\left\{
  1,
  \dfrac{ \pi\left(\vz(*)\right) q(\vz{(t)} | \vz(*))}
  { \pi\left(\vz{(t)}\right) q(\vz(*)|\vz{(t)})}
  \right\}$ \\
  if $u<\alpha\left(\vz{(t)}, \vz(*)\right)$,
  $\vz{(t+1)} = \vz(*)$ \\
  else $\vz{(t+1)} = \vz{(t)}$
 }
 \label{alg:MH}
\end{algorithm}

\paragraph{Summary of past samplers}
In the previously mentioned algorithms, and a majority of those used in practice, only reversible Markov jump processes ($A(\vz,\vy)$ being $0$) are considered. In Sec.~\ref{sec:irr_jump}, we explore the case where the process is irreversible, i.e., $A(\vz,\vy)\neq 0$.

\subsection{Construction of a Practical Irreversible Jump Sampler}
\label{sec:irr_jump}

Analogous to the discussion of Sec.~\ref{sec:framework}, there are two issues with designing samplers using Markov jump processes.
One is the construction of transition probabilities, a task that has been alleviated in part by the new formulation of \eqref{Eq:Jump_Sampler} in terms of $S(\vy,\vz)$ and $A(\vy,\vz)$ with simple constraints, though we still have to construct such probability density functions.
Another is simulating the Markov process of \eqref{Eq:Jump_Sampler}.
In all but the simplest cases, we might not be able to sample from the transition probability $\Delta t \cdot (S(\vy,\vz)+A(\vy,\vz))/\pi(\vy)$.
These two issues are often intertwined posing challenges to the design of samplers.
As mentioned in Sec.~\ref{sec:framework}, the MH algorithm is often resorted to due to its ease of implementation.
%
%
%
It separates the process of proposing a sample into two simple steps: (1) proposing a candidate according to a known conditional probability distribution $q(\vz|\vy)$ and (2)
accepting or rejecting the candidate according to a certain probability.
An important drawback of the vanilla MH sampler, however, is that the reversibility of the jump process being designed can greatly restrict possible ways
to increase the mixing of the Markov chain.

There have been previous efforts to break the restriction of reversibility in different cases.
For example, the {\it non-reversible MH} algorithm adds a vorticity function to the MH procedure \cite{JorisNonReversible}
while the {\it lifting method} makes two replica of the original state space with a skew detailed balance condition to facilitate irreversibility \cite{Lifting_Early,Lifting}.
The authors have shown examples of sampling special distributions, but
it is unclear how to generalize these previous methods to handle a broad set of target distributions.
See Sec.~\ref{sec:relatedwork} for a detailed discussion of these and other methods.
Here, we show how we can devise a practical and efficient irreversible jump process algorithm analogous to MH that can be applied to general targets; this procedure implicitly defines valid functions $S(\vy,\vz)$ and $A(\vy,\vz)$.
In particular, just as MH corresponds to restricting the class of functions $W(\vz|\vy)$, our algorithm also focuses in on particular instances of $A(\vy,\vz)$, but importantly allows $A(\vy,\vz) \neq 0$ (i.e., irreversible processes).
The value of this in practice is demonstrated in the experiments of Sec.~\ref{sec:exp}.

\paragraph{A na\"ive approach}
A straightforward approach to revise the MH algorithm to make the antisymmetric function $A(\vy,\vz)$ nonzero is to utilize different proposal distributions $f(\vz|\vy)$ and $g(\vz|\vy)$, instead of a single $q(\vz|\vy)$.
That is, the transition function of the MH algorithm in \eqref{Eq:MH_kernels} is changed to
\begin{align}
F(\vy,\vz) &= S(\vy,\vz) + A(\vy,\vz) \nonumber\\
&= \dfrac{1}{\Delta t} \min\big(\pi(\vy) f(\vz|\vy) , \pi(\vz) g(\vy|\vz)\big). \label{Eq:naive_kernels}
\end{align}
Here we are considering jump processes with $A(\vy,\vz)=\dfrac12 \big(F(\vy,\vz)-F(\vz,\vy)\big) \neq 0$, in contrast to what we saw for MH.
By adjusting $f$ and $g$, faster mixing rates can possibly be attained while maintaining a simple sampling procedure akin to that of MH (see Algorithm~\ref{alg:MH}, but with $f$ in place of $q$ in the numerator and $g$ in place of $q$ in the denominator of the $\alpha$ calculation).
The more $f$ and $g$ differ, the more irreversibility effect is incorporated in the design of the sampler.
Functions $f$ and $g$ can even be selected to have non-overlapping support in the state space (as is chosen in our experiments), so that new proposals are guided in certain directions until being rejected, encouraging the algorithm to explore farther states.
The primary issue with this construction is that $\int_{\mathbb{R}^d} A(\vy,\vz) \rd \vy \neq 0$ in general, rendering the stationary distribution {\em not} the $\pi(\vz)$ that we desire.
The question is how to design the anti-symmetric function $A(\vy,\vz)$, such that $\int_{\mathbb{R}^d} A(\vy,\vz) \rd \vy = 0$.

\paragraph{Lifting for sampling when $d=1$} A simple modified approach is to follow an adjoint Markov process after being rejected by the original one.
This is inspired by the {\it lifting} idea in discrete spaces \cite{Lifting_Early,Lifting}.
Importantly, this approach has $\pi(\vz)$ as the stationary distribution (marginalized over the auxiliary variable).

Algorithmically, this process introduces a \sloppy one-dimensional, uniformly distributed discrete auxiliary variable $\vy^p \in \{-1,1\}$.
We then define
\begin{align}
\widetilde{f}(\vz,\vz^p|\vy,\vy^p)
= \big( \ind_{\vy^p\geq0} f(\vz|\vy)
 + \ind_{\vy^p<0} g(\vz|\vy) \big) \nonumber\\
\widetilde{g}(\vz,\vz^p|\vy,\vy^p)
= \big( \ind_{\vy^p<0} f(\vz|\vy)
 + \ind_{\vy^p\geq0} g(\vz|\vy) \big) \label{Eq:f&g_1D},
\end{align}
where $f(\vz|\vy)$ and $g(\vz|\vy)$ are different conditional probability distributions, and $\ind_{\mathcal{A}}$ is the indicator function for the set $\mathcal{A}$.

\begin{figure}[t!]
\includegraphics[height=0.25\textheight]{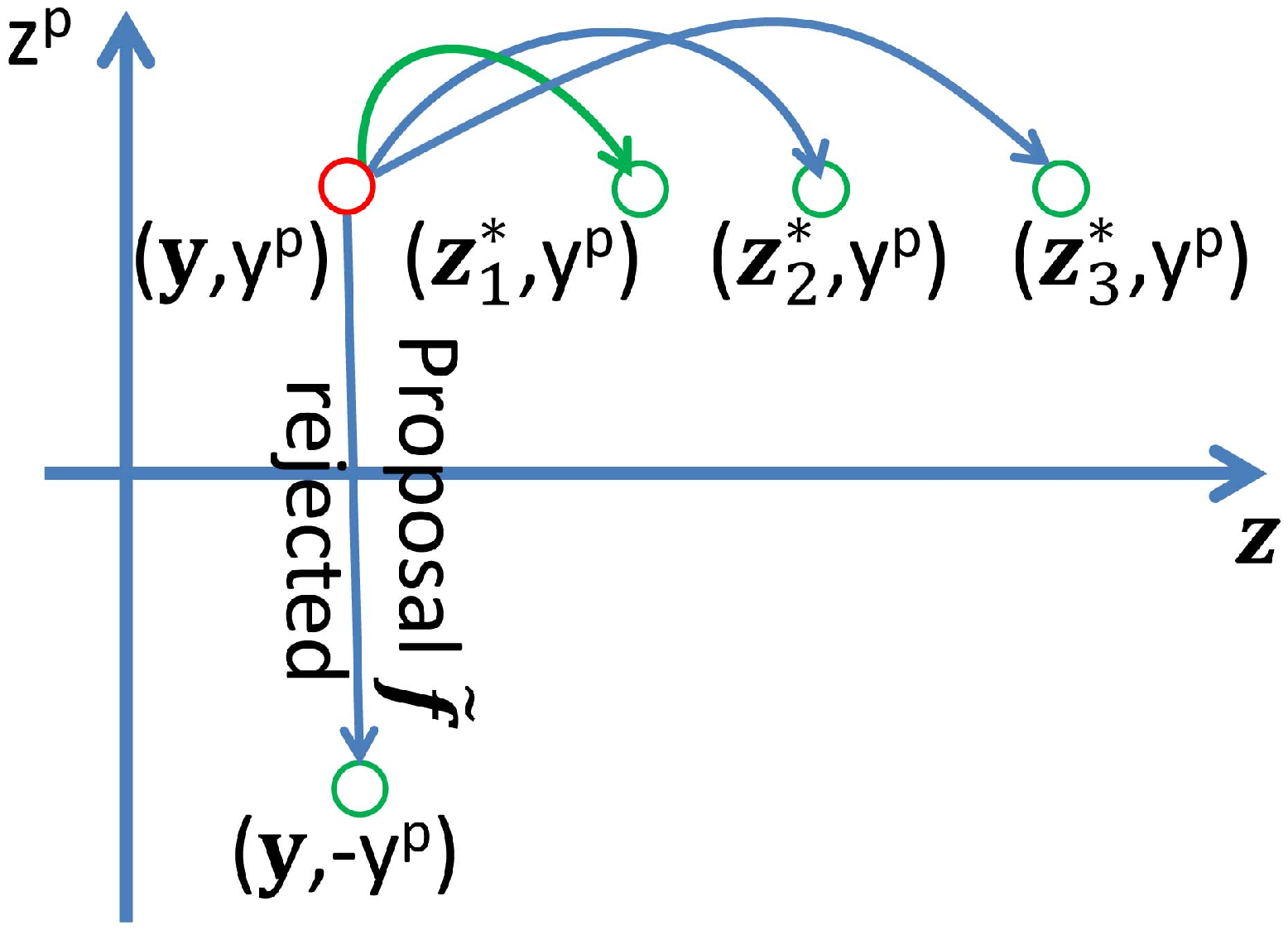}
\hfill
\includegraphics[height=0.25\textheight]{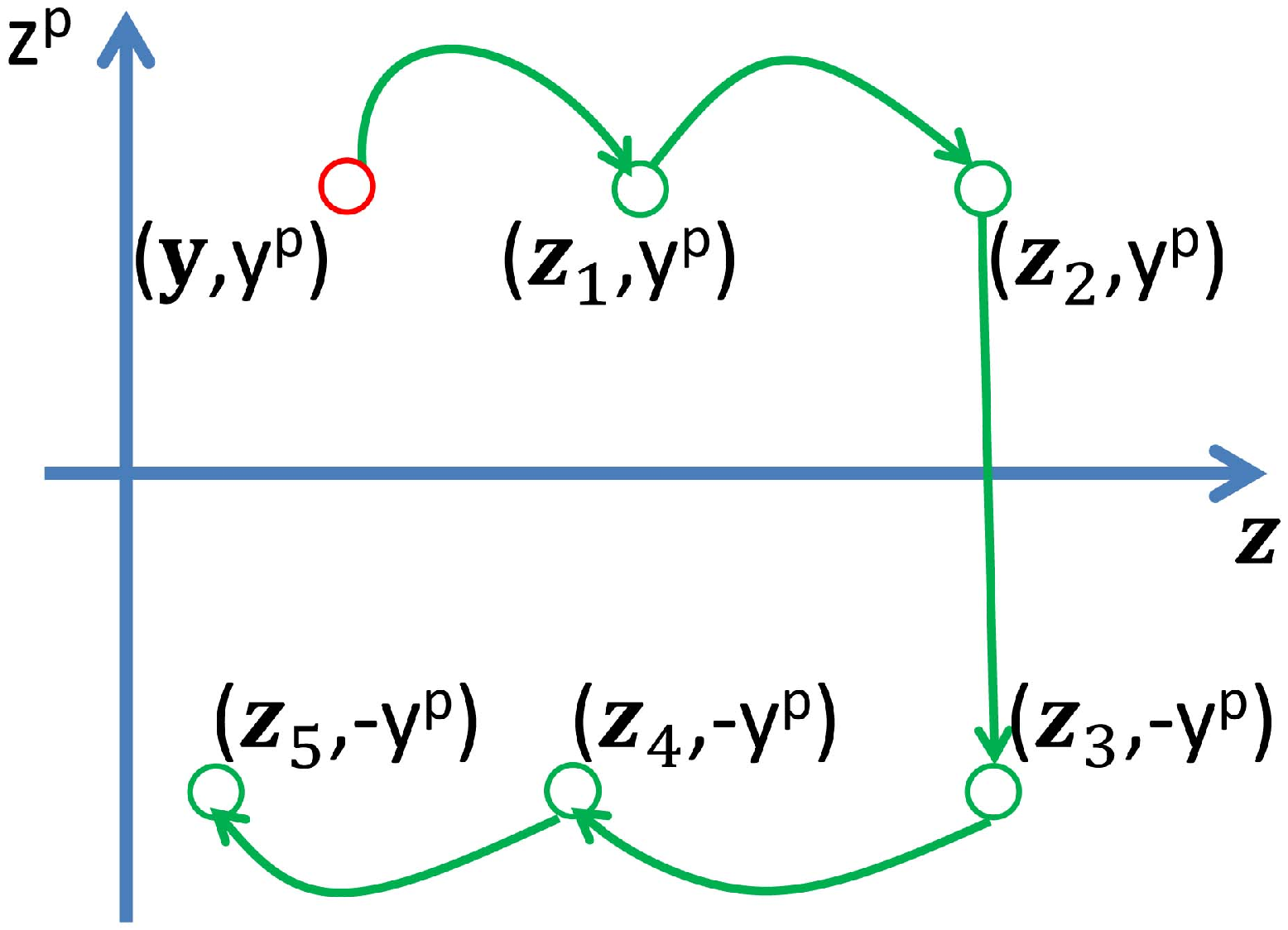}
\caption{
Update rule starting from state $(\vy,\vy^p)$.
\emph{Left:} Several possible states $(\vz^*,\vz^p)$ that the algorithm could visit in the next step.
Without resampling the auxiliary variables, $\vz^p$ can only be $\vy^p$ or $-\vy^p$.
\emph{Right:} Assuming the algorithm visits $(\vz_1,\vy^p)$ as the next state to $(\vy,\vy^p)$ (indicated by the green arrow),
a sample trajectory of states generated.
}
\label{fig:cartoon_update}
\end{figure}

We modify the MH algorithm as described in Algorithm~\ref{alg:Simple},
\begin{algorithm}[t!]
\caption{One-Directional I-Jump Sampler}
randomly pick $z^p$ from $\{1,-1\}$ with equal probability\\
 \For{$t=0,1,2\cdots N_{iter}$}{
  sample $u \sim \mathcal{U}_{[0,1]}$\\
  \If{$z^p>0$}{
  sample $\vz(*) \sim f\left(\vz(*)| \vz{(t)} \right)$\\
  $\alpha\left(\vz{(t)}, \vz(*)\right)
  = \min\left\{
  1,
  \dfrac{ \pi\left(\vz(*)\right) g\left(\vz{(t)} | \vz(*) \right)}
  { \pi\left(\vz{(t)}\right) f\left(\vz(*) | \vz{(t)} \right)}
  \right\}$
  }
  \Else{
  sample $\vz(*) \sim g\left(\vz(*)| \vz{(t)} \right)$\\
  $\alpha\left(\vz{(t)}, \vz(*)\right)
  = \min\left\{
  1,
  \dfrac{ \pi\left(\vz(*)\right) f\left(\vz{(t)} | \vz(*) \right)}
  { \pi\left(\vz{(t)}\right) g\left(\vz(*) | \vz{(t)} \right)}
  \right\}$
  }
  if $u<\alpha\left(\vz{(t)} , \vz(*) \right)$, \\
  $\vz{(t+1)} = \vz(*) $; $z^p{(t+1)} = z^p{(t)} $ \\
  else $\quad$ $\vz{(t+1)} = \vz{(t)}$; $z^p{(t+1)} = - z^p{(t)} $
 }
 \label{alg:Simple}
\end{algorithm}
where we update state $\vy$ and the auxiliary variable $\vy^p$ according to the following transition probability (as in our recipe of \eqref{Eq:Jump_Sampler}):
\begin{align}
p&(\vz,\vz^p | \vy,\vy^p; \Delta t)
\nonumber\\
= &\dfrac{\Delta t}{\pi(\vy)\pi^p(\vy^p)}
\mathfrak{F}(\vy,\vy^p,\vz,\vz^p) \delta(\vz^p-\vy^p) \label{Eq:IrrFlow}
\\ \nonumber
 + &
\left(1 - \dfrac{\Delta t}{\pi(\vy)\pi^p(\vy^p)} \int_{\mathbb{R}^d}\mathfrak{F}(\vy,\vy^p,\vx,-\vz^p) \rd \vx\right) \\ \nonumber
& \cdot \delta(\vz^p+\vy^p) \delta(\vz-\vy),
\end{align}
in which $\mathfrak{F}(\vy,\vy^p,\vz,\vz^p)$ is defined using $\widetilde{f}$ and $\widetilde{g}$:
\begin{align}
\mathfrak{F}(\vy,\vy^p, & \vz,\vz^p)
\nonumber\\
= \min\Big( & \pi(\vy)\pi^p(\vy^p) \widetilde{f}(\vz,\vz^p|\vy,\vy^p), \nonumber\\
& \pi(\vz)\pi^p(\vz^p) \widetilde{g}(\vy,\vy^p|\vz,\vz^p)\Big).
\nonumber
\end{align}
%
This update rule can be understood as follows.  With probability
$\mathfrak{F}(\vy,\vy^p,\vz,\vz^p)/(\pi(\vy)\pi^p(\vy^p))$, state $\vy$ becomes state $\vz$ while the auxiliary state $\vy^p$ remains the same. 
Alternatively, with probability $\\ \displaystyle 
\left[ 1 - \dfrac{1}{\pi(\vy)\pi^p(\vy^p)} \int_{\mathbb{R}^d}\mathfrak{F}(\vy,\vy^p,\vx,-\vz^p) \rd \vx\right]$, no new state $(\vx,\vy^p)$ is accepted conditioning on currently being at state $(\vy,\vy^p)$. Instead,
state $(\vy,\vy^p)$ is directly changed to state $(\vy,-\vy^p)$, leading to a different jump process in $\vy$. An illustration of the update rule is shown in Fig.~\ref{fig:cartoon_update}.
It is worth noting that in the lifted space of $(\vy,\vy^p)$, there are other choices for the probability of transition from $(\vy,\vy^p)$ to $(\vy,-\vy^p)$ (c.f.~\cite{Lifting}).
However, they all lead to a nonzero probability of staying in $(\vy,\vy^p)$, which reduces irreversibility.

From \eqref{eq:A_irrv}, we see that this proposed algorithm takes the anti-symmetric function $A(\vy,\vy^p,\vz,\vz^p)$ to be
\begin{align}
A(\vy,\vy^p, & \vz,\vz^p) \nonumber\\
= \dfrac{1}{2 \Delta t}
\Big(& \pi(\vy)\pi^p(\vy^p) p(\vz,\vz^p | \vy,\vy^p; \Delta t) \nonumber\\
&- \pi(\vz)\pi^p(\vz^p) p(\vy,\vy^p | \vz,\vz^p; \Delta t) \Big) \label{eq:A_alter}
\end{align}
with $p(\vz,\vz^p | \vy,\vy^p; \Delta t)$ as in \eqref{Eq:IrrFlow}.
To ensure correctness of the sampler, $A(\vy,\vy^p,\vz,\vz^p)$ must satisfy (condition~\ref{Eq:Jump_Constraints}):
\begin{align}
\int_{\mathbb{R}^{d+1}} A(\vy,\vy^p,\vz,\vz^p) \ \rd \vy \ \rd \vy^p = 0. \label{eq:A_constraint}
\end{align}
In Appendix~\ref{Appnd_SkewSymm}, we prove that this is indeed the case for $A(\vy,\vy^p,\vz,\vz^p)$ as in \eqref{eq:A_alter}.
The intuition is that the jump in the auxiliary variable introduces a circulative behavior to the whole process (see Fig.~\ref{fig:cartoon_update} for illustration).
This circulation of probability flux is exactly balanced with the jumps in the original variable and the auxiliary variable.
We also see in Fig.~\ref{fig:cartoon_update} that irreversibility introduces a directional effect (just like HMC introduces a direction of rotation).
This algorithm is a generalization of the {\it guided walk Metropolis} method \cite{GUSTAFSON1998} and works well in one dimension, as we demonstrate in Sec.~\ref{sec:1D_Experiment}.
In what follows, we generalize this idea to higher dimensions $d >1$.

\paragraph{Moving to higher dimensions}
An irreversible sampler in $\mathbb{R}^d$ can be constructed as follows.
We expand the state space by introducing a $d^p$-dimensional auxiliary variable $\vy^p \in \mathbb{R}^{d^p}$ in the new state space $(\vy, \vy^p)$.
The total probability can be designated as: $(\vy,\vy^p)\sim\pi(\vy)\pi^p(\vy^p)$.
We further impose symmetry on the auxiliary variables such that $\pi^p(\vy^p)=\pi^p(-\vy^p)$, and let
\begin{align}
\widetilde{f}&(\vz,\vz^p|\vy,\vy^p) \label{Eq:f&g} \\ \nonumber
&=\prod_{i=1}^{n} \big( \ind_{\vy^p_i\geq0} f_i(\vz|\vy, \vy^p)
 + \ind_{\vy^p_i<0} g_i(\vz|\vy, -\vy^p) \big);\\ \nonumber
\widetilde{g}&(\vz,\vz^p|\vy,\vy^p)\\ \nonumber
&=\prod_{i=1}^{n} \big( \ind_{\vy^p_i<0} f_i(\vz|\vy, -\vy^p)
 + \ind_{\vy^p_i\geq0} g_i(\vz|\vy, \vy^p) \big),
\end{align}
where $n$ can be chosen by the user; and $f_i(\vz|\vy, \vy^p)$ and $g_i(\vz|\vy, \vy^p)$ are conditional probability distributions defined by the value of $\vy^p$.
Fitting this definition into the transition probability $p(\vz,\vz^p | \vy,\vy^p; \Delta t)$ in \eqref{Eq:IrrFlow}, the generalized update rule is defined and described in Algorithm~\ref{alg:IrrMC}.
It is worth noting that the accept-reject step in the current setting is the same as in random-walk MH.

This definition of $\widetilde{f}$ and $\widetilde{g}$ is a direct generalization of the definition 
of \eqref{Eq:f&g_1D} to the multi-dimensional case.
Again, we have the anti-symmetric function $\allowbreak A(\vy,\vy^p,\vz,\vz^p)$ as in \eqref{eq:A_alter}.
As we prove in Appendix~\ref{Appnd_SkewSymm}, this construction has
$\int_{\mathbb{R}^{d+{d^p}}} A(\vy,\vy^p,\vz,\vz^p) \ \rd \vy \ \rd \vy^p = 0$ even with our $d^p$-dimensional \emph{continuous} auxiliary variables.

In summary, we can use \eqref{Eq:IrrFlow} to devise a practical algorithm for sampling (I-Jump sampler of Algorithm~\ref{alg:IrrMC}).
In particular, if we define $f_i(\vz|\vy, \vy^p)$ and $g_i(\vz|\vy, \vy^p)$ that are easy to sample from, then we can use the definitions of $\widetilde{f}$ and $\widetilde{g}$ in \eqref{Eq:f&g}
to propose samples in the same way as the MH algorithm.
%
Optionally, we can periodically resample $\vy^p$ according to $\pi^p(\vy^p)$.

One natural choice of $\widetilde{f}$ and $\widetilde{g}$ is to consider the random walk MH algorithm in our framework and add a nonzero $A(\vy,\vz)$ to \eqref{Eq:MH_kernels} while leaving $S(\vy,\vz)$---the reversible part---unchanged.
Then it can be proven that the resulting process will only provide a faster-mixing Markov chain \cite{LargeDeviation_Irr,non_detailedbalance} (Note that this is in contrast to the na\"ive approach of \eqref{Eq:naive_kernels} to modify MH).
This can be achieved by choosing $n=1$; $d^p=d$; $\pi^p(\vy^p)$ equal to a restricted uniform distribution on $\left\{\vy^p \ | \ ||\vy^p||_2 = 1\right\}$; and $f$ and $g$ to be half-space Gaussian distributions.
We write $f(\vz|\vy, \vy^p)$ as:
\[\vz = \vy + \veta \cdot {\rm sgn}(<\veta,\vy^p>), \; \veta\sim\mathcal{N}(0,\sigma^2 I),\]
$g(\vz|\vy, \vy^p)$ as:
\[\vz = \vy - \veta \cdot {\rm sgn}(<\veta,\vy^p>), \; \veta\sim\mathcal{N}(0,\sigma^2 I),\]
where ${\rm sgn}(x) = 2\ind_{x\geq0} - 1$.
In other words,
\begin{align}
f(\vz|\vy, \vy^p) = \ind_{<\vy^p,\vz-\vy> \ \geq \ 0 \ } \dfrac{2}{(2\pi\sigma)^{d/2}}e^{-\frac{1}{2\sigma}||\vz-\vy||^2_2}, \nonumber\\
g(\vz|\vy, \vy^p) = \ind_{<\vy^p,\vz-\vy> \ < \ 0 \ } \dfrac{2}{(2\pi\sigma)^{d/2}}e^{-\frac{1}{2\sigma}||\vz-\vy||^2_2}. \nonumber
\end{align}
Intuitively, we are choosing a random direction $\vy^p$ so that the proposal distribution is restricted in the half space aligned with $\vy^p$.
Then, $\vy^p$ flips signs upon rejection.
One can show that $S(\vy,\vy^p,\vz,\vz^p)$ for the above choice decomposed as $S(\vy,\vz)S(\vy^p,\vz^p)$ and $S(\vy,\vz)$ is indeed $\dfrac{1}{\Delta t} \min(\pi(\vy)q(\vz|\vy), \pi(\vz)q(\vy|\vz))$ for $q(\vz|\vy)$ a Gaussian distribution centered at $\vy$.
That is, the reversible dynamic component determined by $S(\vy,\vz)$ is exactly that of the random walk MH (See \eqref{Eq:MH_kernels}).
However, $A(\vy,\vy^p,\vz,\vz^p)\neq0$ while satisfying \eqref{eq:A_constraint}.
This construction maximizes the irreversibility without changing $S(\vy,\vz)$.
We test this setting in Sec.~\ref{sec:exp_IMALA}.

Alternatively, we can use one-sided distributions for proposals.
For example, we can consider gamma distributions
by setting $n=d^p=d$, and
$f_i\left(\vz| \vy, \vy^p \right)$ as:
\[
\vz_i = \vy_i + \gamma \cdot \vy_i^p, \; \gamma \sim \Gamma(\alpha,\beta);
\]
and $g_i\left(\vz| \vy, \vy^p \right)$ as:
\[
\vz_i = \vy_i - \gamma \cdot \vy_i^p, \; \gamma \sim \Gamma(\alpha,\beta);
\]
and let $\pi^p(\vz^p)$ to be a restricted uniform distribution on the set $\left\{\vy^p \bigg | \dfrac1N ||\vy^p||_1 = 1\right\}$.
We will examine the benefits of this setting in synthetic experiments of Sec.~\ref{sec:jump_exp}.

It can be seen that in both settings there is always a direction of exploration, $\vy^p$, that enjoys most of the benefits from irreversibility.
In multiple dimensions, a favorable direction of exploration is often not clear.
Hence, in experiments, we periodically resample the auxiliary variable $\vy^p$ to explore all directions.
Of course, there are still potential issues when $d$ is large due to the fact that 
resampling of $\vy^p$ may be inefficient for exploring the entire $d$ dimensions.
We leave this as a direction for future research.
In the next Sec.~\ref{sec:combined}, we focus on using the continuous Markov process to provide a favorable direction of exploration and take $n=d^p=1$.
Then $\vz^p$ belongs to a binary set $\{-1,1\}$, rendering Algorithm~\ref{alg:IrrMC} the same as the simpler version, Algorithm~\ref{alg:Simple}, which is the continuous state space generalization of the {\it lifting} method \cite{Lifting_Early,Lifting}.

\begin{algorithm*}[t!]
\caption{I-Jump Sampler}
 \For{$t=0,1,2\cdots N_{iter}$}{
  optionally, periodically resample auxiliary variable $\vz^p$ as ${\vz^p}{(t)}\sim \pi^p(\vz^p)$\\
  sample $u \sim \mathcal{U}_{[0,1]}$\\
  sample $\vz(*) \sim \widetilde{f}\left(\vz(*), {\vz^p}(*) | \vz{(t)}, \vz^p{(t)} \right)$\\
  $\alpha\left(\vz{(t)}, \vz^p{(t)} , \vz(*), \vz^p(*)\right)
  = \min\left\{
  1,
  \dfrac{ \pi\left(\vz(*)\right) \pi^p\left(\vz^p(*)\right) \widetilde{g}\left(\vz{(t)}, \vz^p{(t)} | \vz(*), \vz^p(*) \right)}
  { \pi\left(\vz{(t)}\right) \pi^p\left(\vz^p{(t)}\right) \widetilde{f}\left(\vz(*), \vz^p(*) | \vz{(t)}, \vz^p{(t)} \right)}
  \right\}$ \\
  if $u<\alpha\left(\vz{(t)}, \vz^p{(t)} , \vz(*), \vz^p(*)\right)$, $\quad$
  $\left(\vz{(t+1)}, \vz^p{(t+1)}\right) = \left(\vz(*), \vz^p{(t)}\right)$ \\
  else $\quad$ $\left(\vz{(t+1)}, \vz^p{(t+1)}\right) = \left(\vz{(t)}, - \vz^p{(t)} \right)$
 }
 \label{alg:IrrMC}
\end{algorithm*}

\section{Irreversible Metropolis Adjusted Langevin Algorithm}
\label{sec:combined}
In this section we discuss how to use the continuous dynamics of Sec.~\ref{sec:framework} as a proposal distribution in our I-Jump sampler of Sec.~\ref{sec:jump} (Algorithm~\ref{alg:Simple}), even when the continuous dynamics are not reversible.
This procedure corrects for the discretization error introduced from simulating the continuous dynamics.
Previously, methods such as the Metropolis-adjusted Langevin diffusion (MALA) and Riemannian Metropolis-adjusted Langevin diffusion (RMALA) \cite{MALA,RMALA,RiemannianMALA} have only been proposed for reversible processes.
These methods use one step integration of reversible SDEs to propose samples within an MH algorithm that accepts or rejects the proposal.
We extend these methods to include proposals from any SDE in the form of \eqref{Eq:SDE} (any SDE with a mild integrability condition), without the requirement of reversibility.
Furthermore, our I-Jump sampler allows the resulting overall process to still be irreversible.

In Sec.~\ref{sec:MALA}, we review the (reversible) MALA algorithm.
We then delve into our proposed irreversible MALA (I-MALA) algorithm in two stages.
In Sec.~\ref{Sec:MH_SDE}, we first discuss how one can use the continuous Markov processes of \eqref{Eq:SDE} as a proposal distribution in the I-Jump sampler of Sec.~\ref{sec:jump} and get an acceptance rate equal to $1$ when the continuous Markov process is simulated exactly.
In Sec.~\ref{IrrMALA}, we use a one-step discretized simulation of the continuous Markov process as the proposal distribution
and specify the details of the resulting practical I-MALA algorithm.
In the experiments of Sec.~\ref{sec:exp}, we show that I-MALA can generate better results in terms of rapid and efficient exploration of a distribution than MALA or HMC.

\subsection{Metropolis Adjusted Langevin Algorithm (MALA)}
\label{sec:MALA}

Since the MALA algorithm is a special case of the RMALA algorithm (with $\mD(\vz)$ taken to be constant), we will simply introduce RMALA in this section.
The RMALA algorithm takes $\vz = \theta$ and constructs the proposal distribution $q(\theta(*)|\theta{(t)})$ in the MH algorithm (Algorithm~\ref{alg:MH}) to be the discretized Riemannian Langevin dynamics:
\begin{align}
\theta(*) \leftarrow &\theta{(t)} + \eta(t)
\nonumber\\
+ &\Delta t \cdot [ - {\mG}(\theta{(t)})^{-1} \nabla U(\theta{(t)})  + \Gamma^{\mD}(\theta{(t)})]
,
\nonumber\\
\eta(t) \sim& \mathcal{N}(0, 2 \Delta t {\mG}(\theta{(t)})^{-1}) . \label{eq:MALA_update}
\end{align}
Here, the diffusion matrix $\mD(\theta{(t)}) = \mG(\theta{(t)})^{-1}$ and $\Gamma^{\mD}_i(\theta) = \sum\limits_j \dfrac{\partial {\mD}_{ij} (\theta)}{\partial \theta_j}$.
In the original MALA algorithm \cite{MALA}, $\mG = \mI$, whereas in the RMALA algorithm \cite{RMALA,RiemannianMALA}, $\mG$ is taken to be the Fisher information metric.
Therefore, the resulting transition probability $q(\theta(*)|\theta{(t)})$ in the MH Algorithm~\ref{alg:MH} is:
\begin{align}
q(\theta(*)|\theta{(t)})
= \mathcal{N}\big\{\theta(*) \big| \mu(\theta{(t)}, \Delta t), 2 \Delta t {\mG}(\theta{(t)})^{-1}\big\},
\label{eq:MALA_prob}
\end{align}
where
\begin{align}
&\mu(\theta{(t)}, \Delta t)
\nonumber\\
&= \theta{(t)}
+ \Delta t \cdot [ - {\mG}(\theta{(t)})^{-1} \nabla U(\theta{(t)}) + \Gamma^{\mD}(\theta{(t)})] \nonumber.
\end{align}
This algorithm provides a sampling procedure to exactly simulate the reversible continuous Markov dynamics.
And in doing so, gradient information is used to help the sampler efficiently explore the target distribution.

Use of the MH procedure inevitably restricts the sampler to be reversible.
From \eqref{eq:MALA_prob}, we also observe that only reversible Langevin dynamics are used in the MALA algorithm.
Although irreversible dynamics can be used in \eqref{eq:MALA_update}, as will be discussed in the Sec.~\ref{Sec:MH_SDE}, the acceptance rate would decrease with the increase of irreversibility since irreversibility increases the difference between $q(\theta(*)|\theta{(t)})$ and the reverse proposal $q(\theta{(t)}|\theta(*))$.
We explore this further in Sec.~\ref{Sec:MH_SDE} (see \eqref{eq:accept_irr_MH}).

\paragraph{Comparison with HMC}
HMC expands the sampling space to $\vz=(\theta,r)$ and simulates the Hamiltonian dynamics \eqref{eq:HMC} discussed in Sec.~\ref{sec:previous_continuous} with a neutral integration scheme.
When the mass matrix $\mM$ in \eqref{eq:HMC} is constant, a leap-frog step integration is often applied:
\begin{align}
\left\{
\begin{array}{l}
      r_{t+1/2} \leftarrow r_t - \eps_t/2 \nabla U(\theta_t) \\
      \theta_{t+1} \leftarrow \theta_t + \eps_t{\mM}^{-1} r_{t+1/2} \\
      r_{t+1} \leftarrow r_{t+1/2} - \eps_t \nabla U(\theta_{t+1}).
\end{array}
\right. \label{eq:HMC_update}
\end{align}
When $\mM = \mM(\theta)$ is adaptive, a more involved symplectic integrator is often needed to preserve volume in $\vz$.
To obtain valid samples, the momentum variable $r$ is resampled (an example of a jump process sampler described in ``direct resampling" of Sec.~\ref{sec:jump_previous}) for ergodicity.
An MH accept-reject step is applied with acceptance rate calculated using the volume preserving property of the Hamiltonian dynamics.
While MALA quickly converges towards a mode and diffusively explores around it, HMC excels at deterministically traversing along level sets of the Hamiltonian
\[
H(\theta, r) =  U(\theta) + \frac{1}{2} r^T \mM^{-1} r.
\]


\subsection{General SDE Proposals under Small Step Size Limit}
\label{Sec:MH_SDE}
Our ultimate goal is to use the stochastic dynamics of \eqref{Eq:SDE} to propose samples in the framework of Algorithm~\ref{alg:Simple}.  In practice, we need to simulate from the discretized SDE of \eqref{Eq:simple-update}.  Before analyzing this case, we first examine what would happen if we could \emph{exactly} simulate the SDE of \eqref{Eq:SDE}.

Here, we imagine using the transition probability density $P\big(\vz | \vy; t \big)$ of the continuous Markov process to construct a particular case of Algorithm \ref{alg:Simple}.
We take $f(\vz|\vy,\vy^p)$ in \eqref{Eq:f&g} to be equal to $P\big(\vz | \vy; t\big)$ defined via an exact solution to the SDE, starting at $\vy$:
\begin{align}
\rd \vz =& \Big[ - \big(\mD(\vz)+\mQ(\vz)\big) \nabla H(\vz) + \Gamma(\vz)\Big] \rd t
\nonumber\\
&+ \sqrt{2 \mD(\vz)} \rd \mW,
\label{Eq:forwardSDE}
\end{align}
where $\Gamma_i(\vz) = \sum_j \dfrac{\partial}{\partial \vz_j} \left({\mD}_{ij}(\vz) + {\mQ}_{ij}(\vz)\right)$.

For the reverse proposal $g(\vz|\vy,\vy^p)$ in \eqref{Eq:f&g}, we use the adjoint process $P^\dag\big(\vz | \vy; t\big)$, inverting the irreversible dynamics via $\mQ(\vz)\rightarrow-\mQ(\vz)$ \cite{MaQian}:
\begin{align}
\rd \vz =& \Big[ - \big(\mD(\vz)-\mQ(\vz)\big) \nabla H(\vz) + \widetilde{\Gamma}(\vz)\Big] \rd t
\nonumber\\
&+ \sqrt{2 \mD(\vz)} \rd \mW,
\label{Eq:adjointSDE}
\end{align}
where $\widetilde{\Gamma}_i(\vz) = \sum_j \dfrac{\partial}{\partial \vz_j} \left({\mD}_{ij}(\vz) - {\mQ}_{ij}(\vz)\right)$.


\begin{theorem}
\label{thm:adjoint}
For the Markov processes
\[
{P\big( \vz | \vy; t \big)} {\rm \ and \ } {P^\dag\big( \vz | \vy; t \big)}
\]
defined by the SDEs of \eqref{Eq:forwardSDE} and \eqref{Eq:adjointSDE} through It\^o integration, the following equality holds:
\begin{equation}
\dfrac{P\left( \vz | \vy; t \right)}{P^\dag\left( \vy | \vz; t \right)} = \dfrac{\pi\left(\vz\right)}{\pi\left(\vy\right)}.
\end{equation}
\end{theorem}
The proof is in Appendix~\ref{append_adjoint}.  Using Theorem~\ref{thm:adjoint}, we have
\begin{align}
\alpha\left(\vy, \vz\right)
  &= \min\left\{
  1,
  \dfrac{\pi\left(\vz\right) P^\dag\left( \vy | \vz; t \right)}
  {\pi\left(\vy\right) P\left(\vz | \vy; t \right)}\right\}
  \nonumber\\
  &= 1.
  \label{eq:accept_conv}
\end{align}
%
%
Even though in Sec.~\ref{sec:cont} we saw that SDEs of the form in \eqref{Eq:SDE} have $\pi(\vz)$ as the invariant distribution, it is not immediately obvious that using this SDE as a proposal in Algorithm~\ref{alg:Simple} would lead to an acceptance rate of 1.
In fact, it might be tempting to directly plug $P\left(\vz | \vy; t \right)$ into the MH Algorithm~\ref{alg:MH}.
However, that would result in a MH acceptance rate:
\begin{align}
\alpha^{MH}\left(\vy, \vz\right)
  &= \min\left\{
  1,
  \dfrac{\pi\left(\vz\right) P\left( \vy | \vz; t \right)}
  {\pi\left(\vy\right) P\left(\vz | \vy; t \right)}\right\}
  \nonumber\\
  &\neq 1.
  \label{eq:accept_irr_MH}
\end{align}
And the more irreversibility is introduced, the less the acceptance rate $\alpha^{MH}$ will be in the MH algorithm.
This gap between $\alpha^{MH}$ and $1$ was first discovered in the statistical mechanics literature and relates to the ``house keeping heat" \cite{Crooks,Sasa,MaQian}.

\eqref{eq:accept_conv} also gives us insight into the fact that using more accurate numerical integrators could lead to higher acceptance rates.
In Sec.~\ref{IrrMALA}, we analyze the accept-reject scheme for the simple first-order integration of \eqref{Eq:simple-update} with finite step size $\Delta t$.

\subsection{I-MALA via I-Jump Correction}
\label{IrrMALA}
Since in practice we rely on finite step sizes $\Delta t>0$, there will be numerical error resulting in
$\\ \dfrac{P\left(\vz(*) | \vz{(t)}; \Delta t\right)}{P^\dag\left(\vz{(t)} | \vz(*); \Delta t\right)}$ differing from $\dfrac{\pi\left(\vz(*)\right)}{\pi\left(\vz{(t)}\right)}$.
We now propose an irreversible generalization of the MALA algorithm to correct for these errors.
We make use of Algorithm~\ref{alg:Simple} and take a general SDE and its adjoint process defined in Sec.~\ref{Sec:MH_SDE} to propose samples using a one-step numerical integration (as in MALA).  Because we have the local gradient information in the SDEs to guide us, the direction of the exploration is determined.
So, we simply use a 1-dimensional discrete auxiliary variable $\vy^p$, and thus rely on Algorithm~\ref{alg:Simple} instead of the more general Algorithm~\ref{alg:IrrMC}. We call the resulting algorithm the \emph{I-MALA} method.

Assuming a one-step numerical integration using a $\Delta t$ period of time, the discretization of the SDE of \eqref{Eq:forwardSDE} leads to
\begin{align}
{P(\vz | \vy; \Delta t)}
= \mathcal{N}\big\{\vz | \mu(\vy, \Delta t), 2 \Delta t \cdot \mD(\vy)\big\}
\label{Eq:Prob_forwardSDE},
\end{align}
where
\begin{align}
\mu(\vy, \Delta t) = \vy
+ \Big[& - \big(\mD(\vy)+\mQ(\vy)\big) \nabla H(\vy)
\nonumber\\
&+ \Gamma(\vy)\Big]\Delta t,
\nonumber\\
\Gamma_i(\vz) = \sum_j \dfrac{\partial}{\partial \vz_j} \big( & {\mD}_{ij}(\vz) + {\mQ}_{ij}(\vz)\big).
\end{align}
Importantly, this allows us to compute $f\left(\vz(*)| \vz{(t)} \right) = P(\vz(*) | \vz{(t)}; \Delta t)$ in Algorithm~\ref{alg:Simple}.  The corresponding calculation for the adjoint process with the SDE in \eqref{Eq:adjointSDE} is:
\begin{align}
{P^\dag(\vz | \vy; \Delta t)}
= \mathcal{N}\big\{\vz | \mu^\dag(\vy, \Delta t), 2 \Delta t \cdot \mD(\vy)\big\}
\label{Eq:Prob_backwardSDE},
\end{align}
where
\begin{align}
\mu^\dag(\vy, \Delta t) = \vy
+ \Big[& - \big(\mD(\vy)-\mQ(\vy)\big) \nabla H(\vy) \nonumber\\ &+ \widetilde\Gamma(\vy)\Big]\Delta t,
\nonumber\\
\widetilde{\Gamma}_i(\vz) = \sum_j \dfrac{\partial}{\partial \vz_j} \big( & {\mD}_{ij}(\vz) - {\mQ}_{ij}(\vz)\big).
\end{align}
This allows us to compute
\[
g\left(\vz(*)| \vz{(t)} \right) = P^\dag(\vz(*) | \vz{(t)}; \Delta t).
\]
The resulting I-MALA algorithm is summarized in Algorithm~\ref{alg:Irr_MALA}.

We know from Sec.~\ref{Sec:MH_SDE} that in the small $\Delta t$ limit,
\begin{align}
\alpha\left(\vz{(t)}, \vz(*)\right)
&= \min\left\{1,
\dfrac{P^\dag\left( \vz{(t)} | \vz(*) \right)}{P\left(\vz(*) | \vz{(t)} \right)} \cdot \dfrac{\pi\left(\vz(*)\right)}{ \pi\left(\vz{(t)}\right) }
\right\}
\nonumber\\
&\rightarrow 1.
\end{align}
From this result, we see that there seems to be a step-size/acceptance-rate tradeoff.  As mentioned in Sec.~\ref{Sec:MH_SDE}, a higher-order numerical scheme could potentially increase the acceptance rate with the same step size \cite{DingHighOrder,Owhadi,LeimkuhlerHighOrder}.
We leave this as a direction for future research.

\begin{algorithm}[t!]
\caption{I-MALA}
randomly pick $z^p$ from $\{1,-1\}$ with equal probability\\
 \For{$t=0,1,2\cdots N_{iter}$}{
  sample $u \sim \mathcal{U}_{[0,1]}$\\
  \If{$z^p>0$}{
  sample $\eta_t \sim \mathcal{N}(0, 2 \eps_t \mD(\vz_t))$\\
  $\vz(*) \leftarrow \vz_{t} -  \eps_t \left[\big({\mD}(\vz_t) + {\mQ}(\vz_t)\big) \nabla H(\vz_t)
+ \Gamma(\vz_t)\right] + \eta_t$\\
  $\alpha\left(\vz{(t)}, \vz(*)\right)
  = \min\left\{
  1,
  \dfrac{ \pi\left(\vz(*)\right) P^\dag(\vz{(t)} | \vz(*); \Delta t) }
  { \pi\left(\vz{(t)}\right) P(\vz(*) | \vz{(t)}; \Delta t) }
  \right\}$
  }
  \Else{
  sample $\eta_t \sim \mathcal{N}(0, 2 \eps_t \mD(\vz_t))$\\
  $\vz(*) \leftarrow \vz_{t} -  \eps_t \left[\big({\mD}(\vz_t) - {\mQ}(\vz_t)\big) \nabla H(\vz_t)
+ \widetilde\Gamma(\vz_t)\right] + \eta_t$
  $\alpha\left(\vz{(t)}, \vz(*)\right)
  = \min\left\{
  1,
  \dfrac{ \pi\left(\vz(*)\right) P(\vz{(t)} | \vz(*); \Delta t)}
  { \pi\left(\vz{(t)}\right) P^\dag(\vz(*) | \vz{(t)}; \Delta t) }
  \right\}$
  }
  if $u<\alpha\left(\vz{(t)} , \vz(*) \right)$, \\
  $\vz{(t+1)} = \vz(*) $; $z^p{(t+1)} = z^p{(t)} $ \\
  else $\quad$ $\vz{(t+1)} = \vz{(t)}$; $z^p{(t+1)} = - z^p{(t)} $
 }
 \label{alg:Irr_MALA}
\end{algorithm}


\section{Related Work}
\label{sec:relatedwork}
There have been previous efforts to construct irreversible Markov processes for sampling.
One example is using continuous dynamics to achieve this goal, which has been studied extensively.
Theoretically, one can make use of Hamiltonian or generalized Hamiltonian dynamics to introduce irreversibility into the sampling procedure \cite{hwang1993,hwang2005,Kostas,Duncan2016}.
There have been samplers that utilize specific irreversible continuous dynamics stemming from physical systems, such as underdamped Langevin \cite{SGHMC} and Nos\'e-Hoover \cite{SGNHT,Shang,Thermostat_Math} dynamics and their generalizations \cite{relat_HMC,geod_HMC}, 
although irreversibility was not the emphasis in these works.
As described in Sec.~\ref{sec:cont}, any dynamic process that has a nonzero $\mQ$ matrix can be used to devise an irreversible sampler within our framework.
The problem, however, is that simulating the continuous Markov processes using the discretized system typically leads to the introduction of bias due to discretization error (see Sec.~\ref{sec:discretization}).
One option to correct for this bias is to introduce a MH step, but of course this yields the whole process reversible again.
In all the previously mentioned works,
no MH correction is used and instead a small, finite stepsize is used and some resulting bias tolerated.
Our I-MALA method provides a mechanism for handling the discretization error while maintaining an overall irreversible process.

One alternative way of introducing irreversibility into samplers without causing additional bias is to combine Hamiltonian dynamics with an ergodic stochastic dynamics that can be exactly simulated.
For Hamiltonian dynamics with quadratic kinetic energy, flipping the sign of the momentum variable is equivalent to following the adjoint process ($\mQ\rightarrow-\mQ$) in Algorithm~\ref{alg:Simple} because of its special structure.
This is a special case of the I-Jump algorithm where the proposal is via simulating Hamiltonian dynamics.
Therefore, one can simulate the ergodic dynamics exactly, simulate the Hamiltonian dynamics, then change the sign of the momentum variable after rejection (instead of resampling it) to form an irreversible sampler.
For example, in \cite{Horowitz:1991,ottobre2016}, Langevin dynamics over the momentum variable is integrated exactly and combined with the Hamiltonian dynamics to form an irreversible sampler (SOL-HMC) \cite{ottobre2016}.
During the review of this paper, it came to our attention that a recently released manuscript \cite{poncet} proposes two new irreversible algorithms.
The first uses a special case of I-MALA with discussions on novel implicit integration schemes.
The second one (Hybrid MALA) uses MALA to simulate reversible dynamics exactly and combine with Hamiltonian dynamics to obtain an irreversible sampler.
In this version of our manuscript, we compare to both SOL-HMC and hybrid MALA in the experiments of Sec.~\ref{sec:exp_IMALA}.

%

Only recently have researchers constructed irreversible jump processes that form valid sampling procedures.
In the {\it non-reversible MH} algorithm \cite{JorisNonReversible}, a vorticity function (or matrix) is added to the MH procedure.
Then, the difficulty of construction is translated to defining a valid vorticity function, similar to the difficulty of defining the antisymmetric function $A(\vy,\vz)$.
For the multivariate Gaussian distribution, the author discretized an irreversible Ornstein-Uhlenbeck process to obtain a suitable vorticity function.
The {\it lifting method} \cite{Lifting_Early,Lifting} makes a replica of the original state space ($\mathbb{R}^d \times \{-1,1\}$) to facilitate irreversibility in the sampling procedure.
A skew detailed balance condition is imposed to ensure a valid antisymmetric function $A(\vy,\vz)$ in the expanded state space.
The authors showed an example of applying the method to spin models.
For both the {\it non-reversible MH} and {\it lifting} methods, it has not been clear
how to come up with practical, easy-to-construct algorithms to handle a broad set of target distributions.
Our I-Jump sampler incorporates both ideas: lifting the state space (to $\mathbb{R}^d \times \mathbb{R}^{d^p}$) and using an irreversible accept-reject procedure similar to the non-reversible MH algorithm.
Combination of the two ideas yields a simple MCMC procedure that generalizes to arbitrary target distributions.

The combined approach of using both continuous dynamics and jump processes has recently been proposed for constructing irreversible samplers.
The {\it bouncy particle} \cite{BouncyParticle} and {\it Zig-Zag} \cite{ZigZag,ZigZag2} samplers use deterministic dynamics (irreversible in nature) combined with a Poisson process to create valid MCMC procedures.
These two methods use continuous dynamics to guide a Poisson jump process with an inhomogeneous rate (or intensity) to ensure the invariance of the target distribution.
In practice, a Poisson thinning step is often required to generate Poisson processes with inhomogeneous rates, hence posing further constraints on the target distribution (e.g., prior knowledge of global lower bounds for the norm of the gradient or Hessian of the negative log posterior).
Our I-MALA algorithm avoids the difficulty of sampling from a Poisson process.
Additionally, we end up with an algorithm that is a simple modification of vanilla MH, making it straightforward to use and plug in to existing algorithmic frameworks.

\section{Experiments}
\label{sec:exp}

In this section, we first examine the correctness and attributes of our I-Jump sampler (Algorithm~\ref{alg:IrrMC}).
We consider various simulated scenarios, including the challenging cases of heavy tailed, multimodal, and correlated distributions.

We then explore the I-MALA algorithm (Algorithm \ref{alg:Irr_MALA}) and compare it against numerous baselines.


\subsection{Visual Comparison of Samplers}
\begin{figure*}[!t]
\centering
\begin{tabular}{cc}
    \hspace{-0.25in}
    \begin{tabular}{c}
        \rotatebox{90}{\textbf{50 Steps}}
        \vspace{0.9in}\\
        \rotatebox{90}{\textbf{1000 Steps}}
    \end{tabular}
    \begin{tabular}{ccc}
        \hspace{-0.3in}
        \vspace{-0.05in}
        \includegraphics[width = 1.8in]{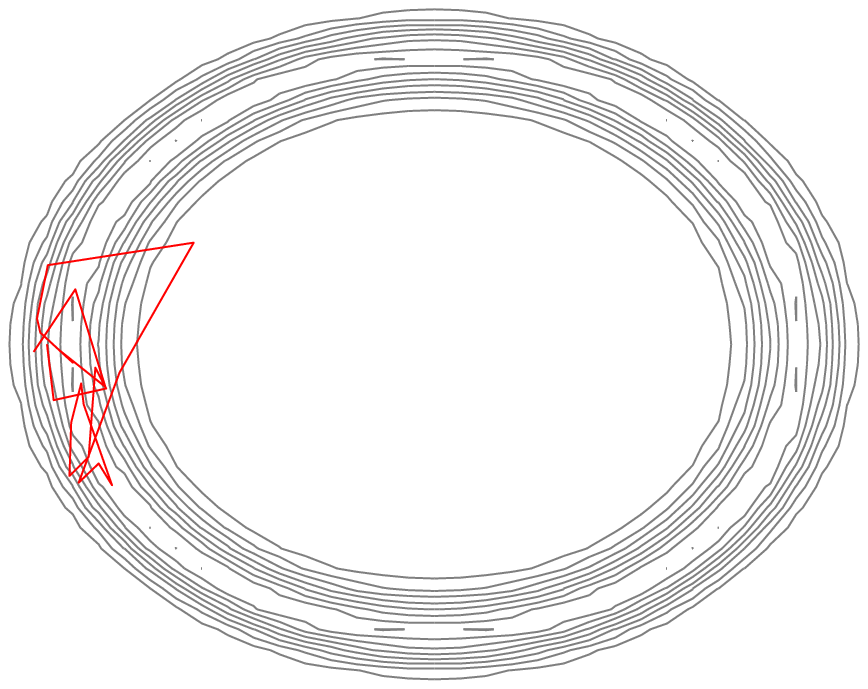} & \hspace{-0.2in}
        \includegraphics[width = 1.8in]{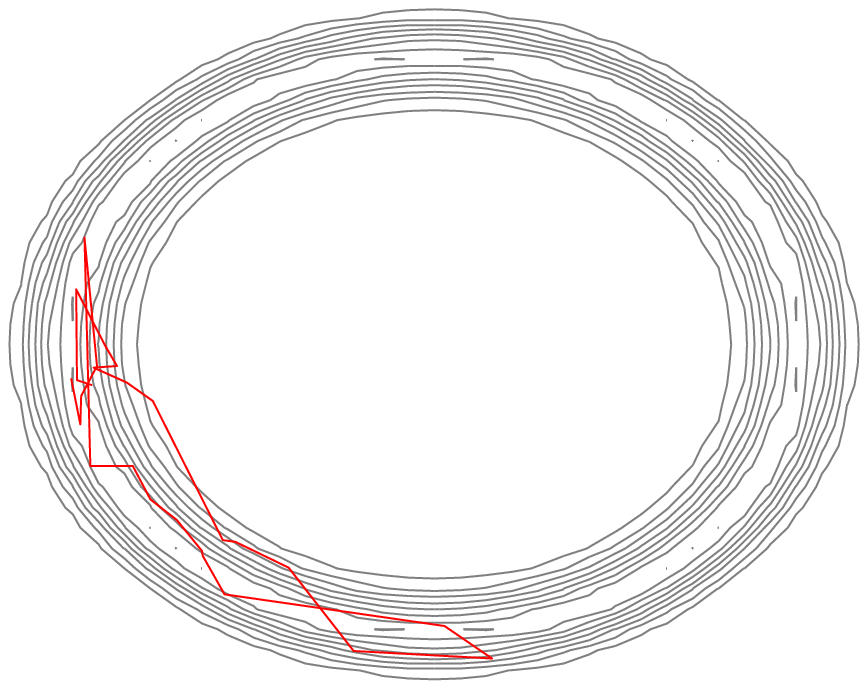} & \hspace{-0.2in}
        \includegraphics[width = 1.8in]{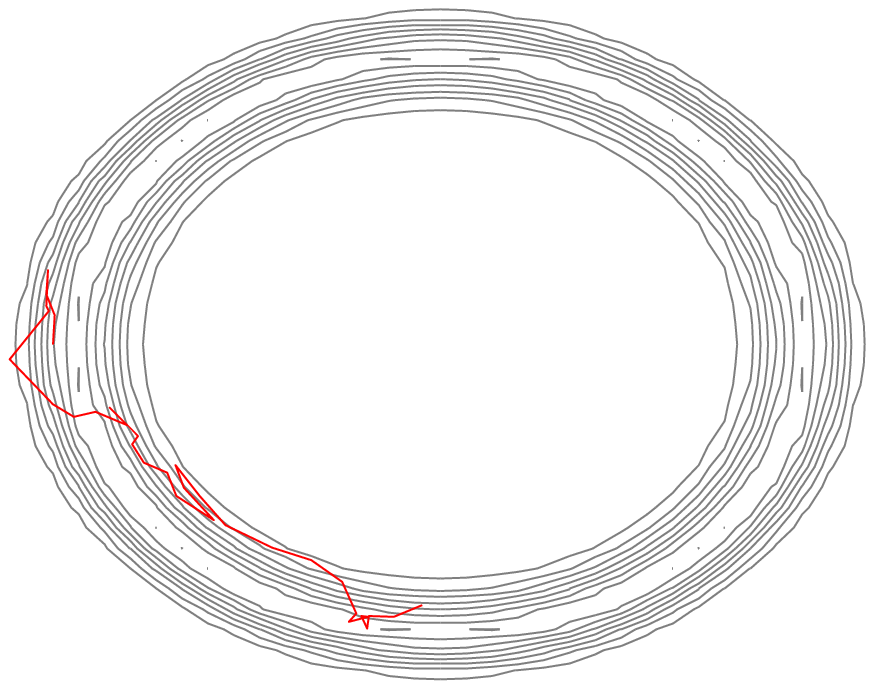} \\ \hspace{-0.2in}
        \includegraphics[width = 1.8in]{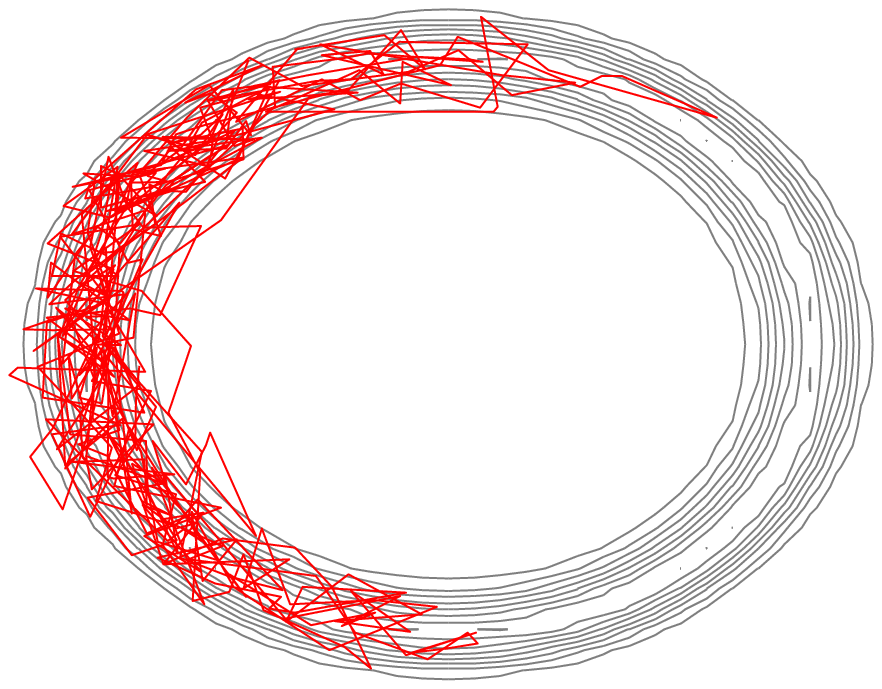} & \hspace{-0.2in}
        \includegraphics[width = 1.8in]{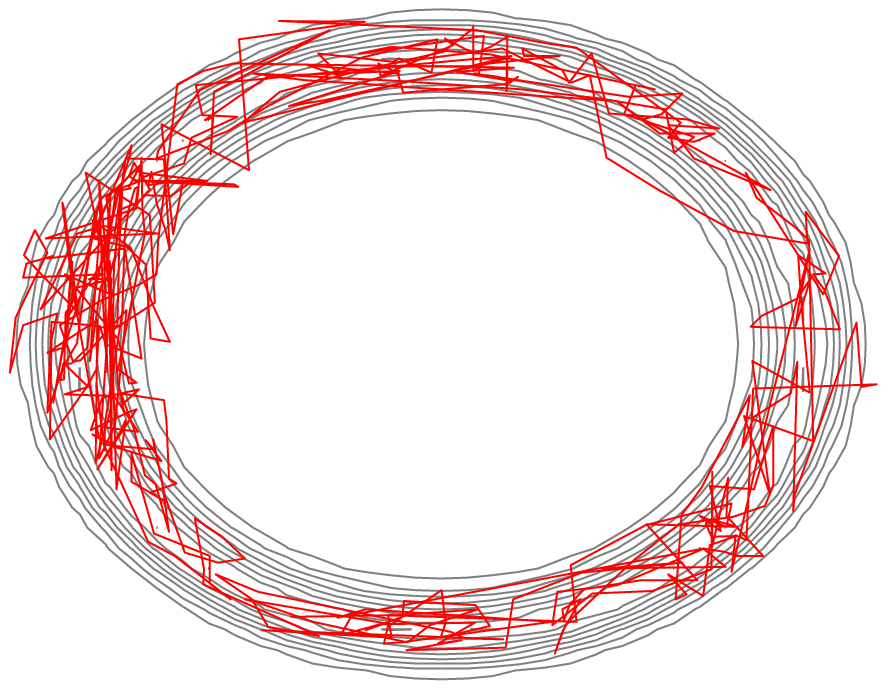} & \hspace{-0.2in}
        \includegraphics[width = 1.8in]{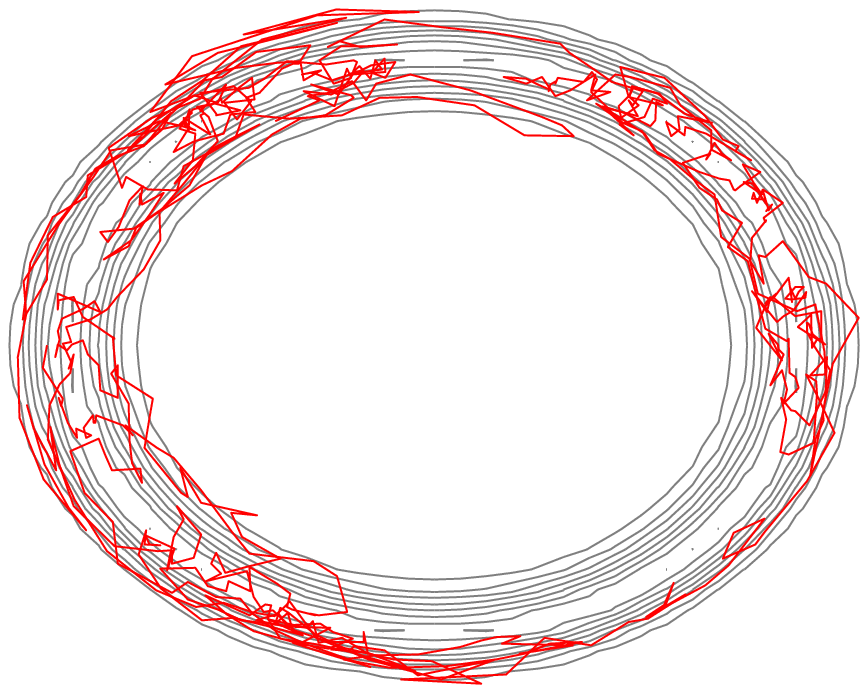} \\
        \hspace{0.25in}
        \textbf{MH} &
        \hspace{0.2in}
        \textbf{I-Jump} &
        \hspace{0.3in}
        \textbf{I-MALA}\\
    \end{tabular}
\end{tabular}
\caption{
\emph{Top row:} Trajectory of first $50$ steps of (left) MH algorithm using Gaussian random walk proposals, (middle) I-Jump algorithm with gamma proposals and (right) I-MALA algorithm.
\emph{Bottom row:} Similarly for the first $1000$ steps of the algorithms.
}
\label{fig:TracePlot}
\end{figure*}

We first perform a qualitative comparison between the random walk MH algorithm, our I-Jump sampler, and the I-MALA algorithm to provide insights into their differences.
It is demonstrated in Fig.~\ref{fig:TracePlot} that the standard MH sampler jumps around randomly, but does so within a local region of the previous sample and irrespective of previous (directions of) jumps, leading to slow exploration of the distribution.
In contrast, our irreversible counterpart (here using gamma proposals) more rapidly traverses the distribution by following the direction of the previous jump, until being rejected.
Finally, the I-MALA algorithm provides an even smoother trajectory by using continuous dynamics in place of independent gamma proposals.

Having visually examined the differences between the samplers to gain intuition, in what follows we provide a more quantitative analysis of the proposed samplers.



\subsection{Synthetic Experiments for I-Jump Sampler}
\label{sec:jump_exp}

In the following experiments we examine the I-Jump sampler on various challenging synthetic distributions.
For this section, we consider gamma proposal.
That is, as mentioned in Sec.~\ref{sec:irr_jump}, we take
$f_i\left(\vz(*)| \vz{(t)}, \vz^p{(t)} \right)$ according to $\vz_i(*)=\vz_i(t) + \gamma \vz_i^p{(t)}$ with $\gamma \sim \Gamma(\alpha,\beta)$ and
$g_i\left(\vz(*)| \vz{(t)}, \vz^p{(t)} \right)$ according to $\vz_i(*)=\vz_i(t) - \gamma \vz_i^p{(t)}$.
We set $\pi(\vz^p)$ to be a restricted uniform distribution on the set $\left\{\vz^p | \dfrac1N ||\vz^p ||_1 = 1\right\}$.

The hyperparameters $\alpha$ and $\beta$ are chosen using a generic procedure.
We take the gamma shape parameter to be $\alpha=1.1$,
and change the rate parameter $\beta$ approximately as $\beta \propto \sqrt{V}$ with $V$ is the volume of the region we would like to explore.
%
Further details are in Appendix~\ref{appnd_Experiment}.

\subsubsection{1D Heavy-tailed Distribution}
\label{sec:1D_Experiment}
\begin{figure*}[!t]
\centering
\includegraphics[width=0.25\textheight]{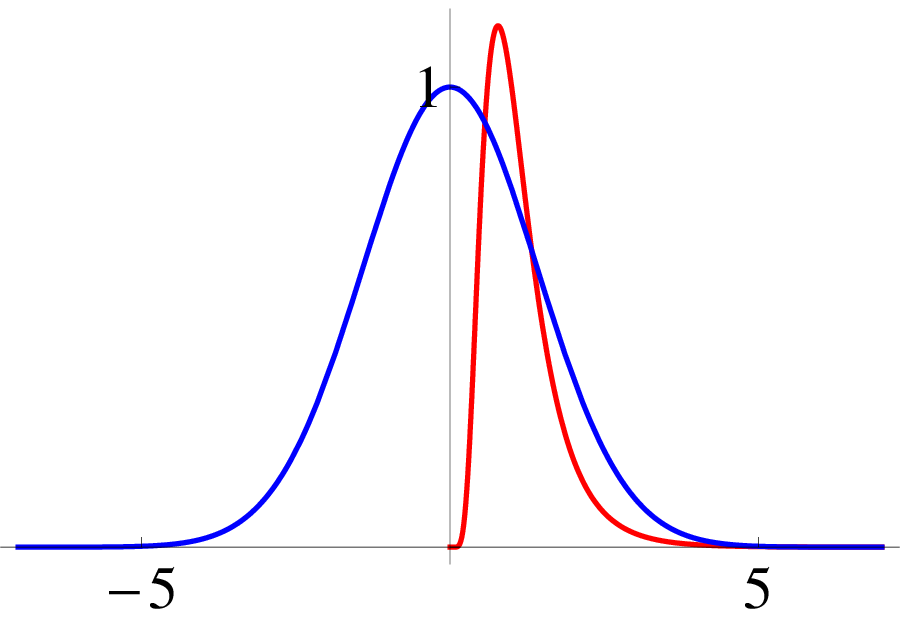}
\includegraphics[width=0.25\textheight]{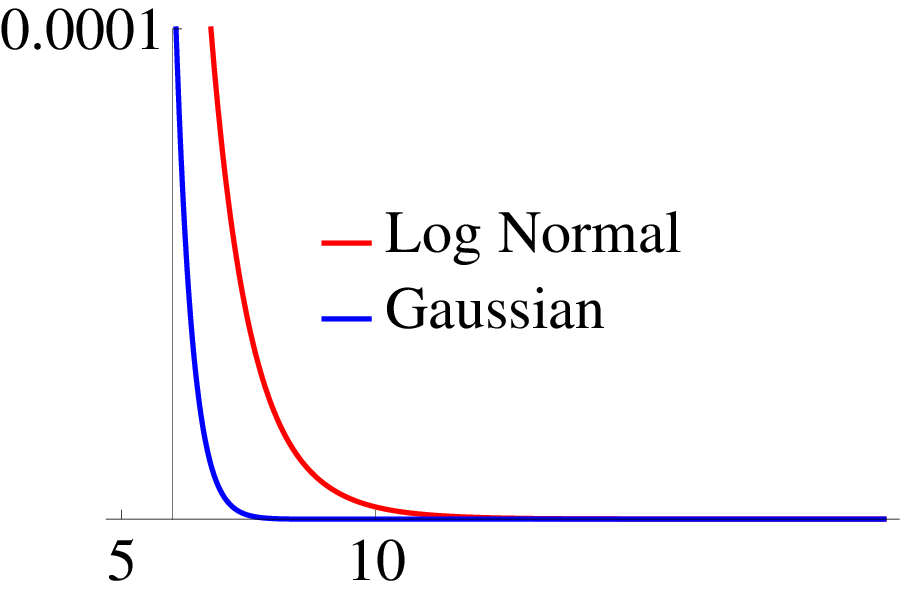}
\\
\begin{minipage}[b]{0.7\textheight}
\centering
\vspace{1pt}
\includegraphics[width=0.3\textheight]{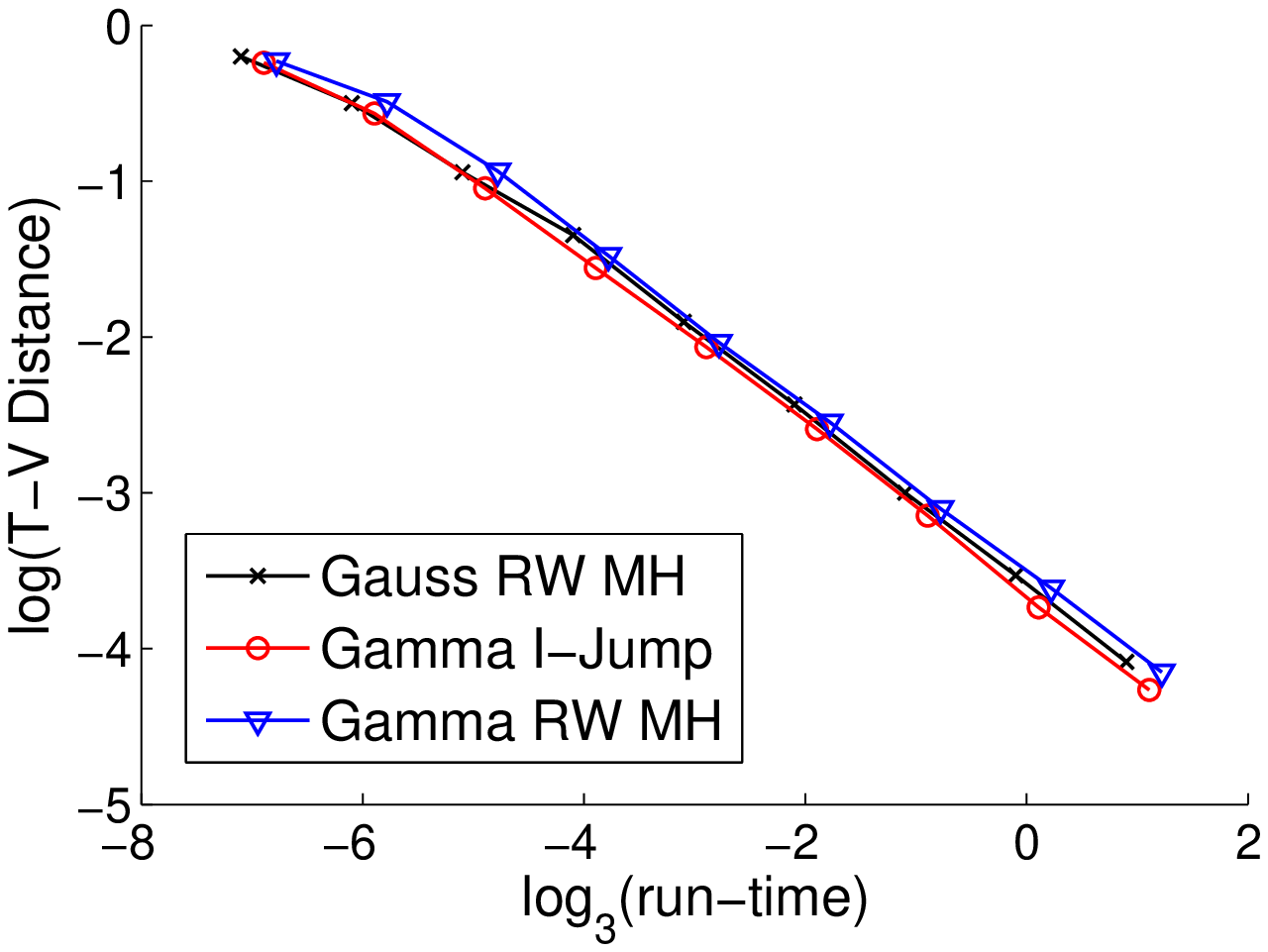}
\includegraphics[width=0.3\textheight]{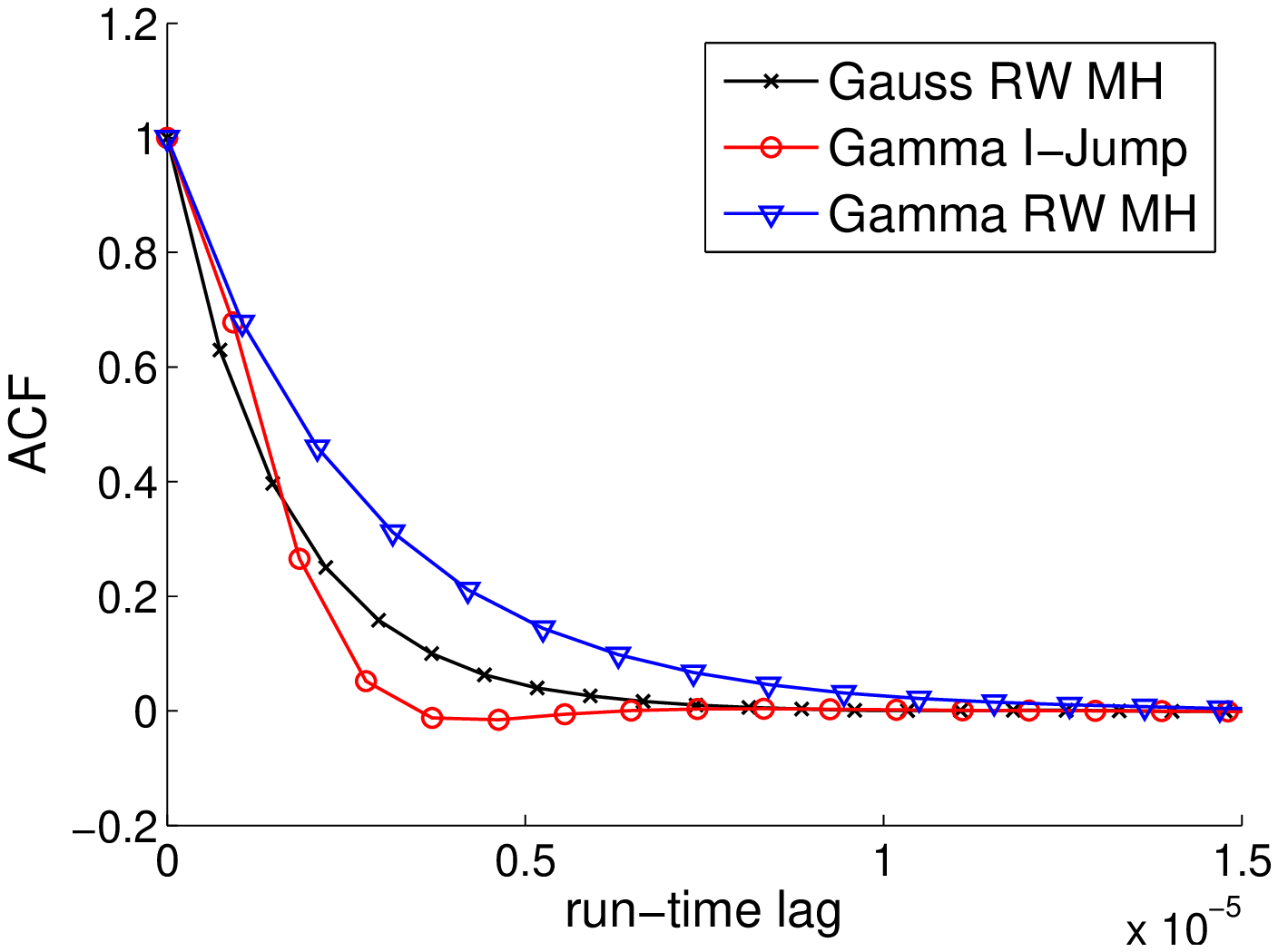}
\end{minipage}
\postcap
\\
\caption*{Normal}
\begin{minipage}[b]{0.7\textheight}
\centering
\includegraphics[width=0.3\textheight]{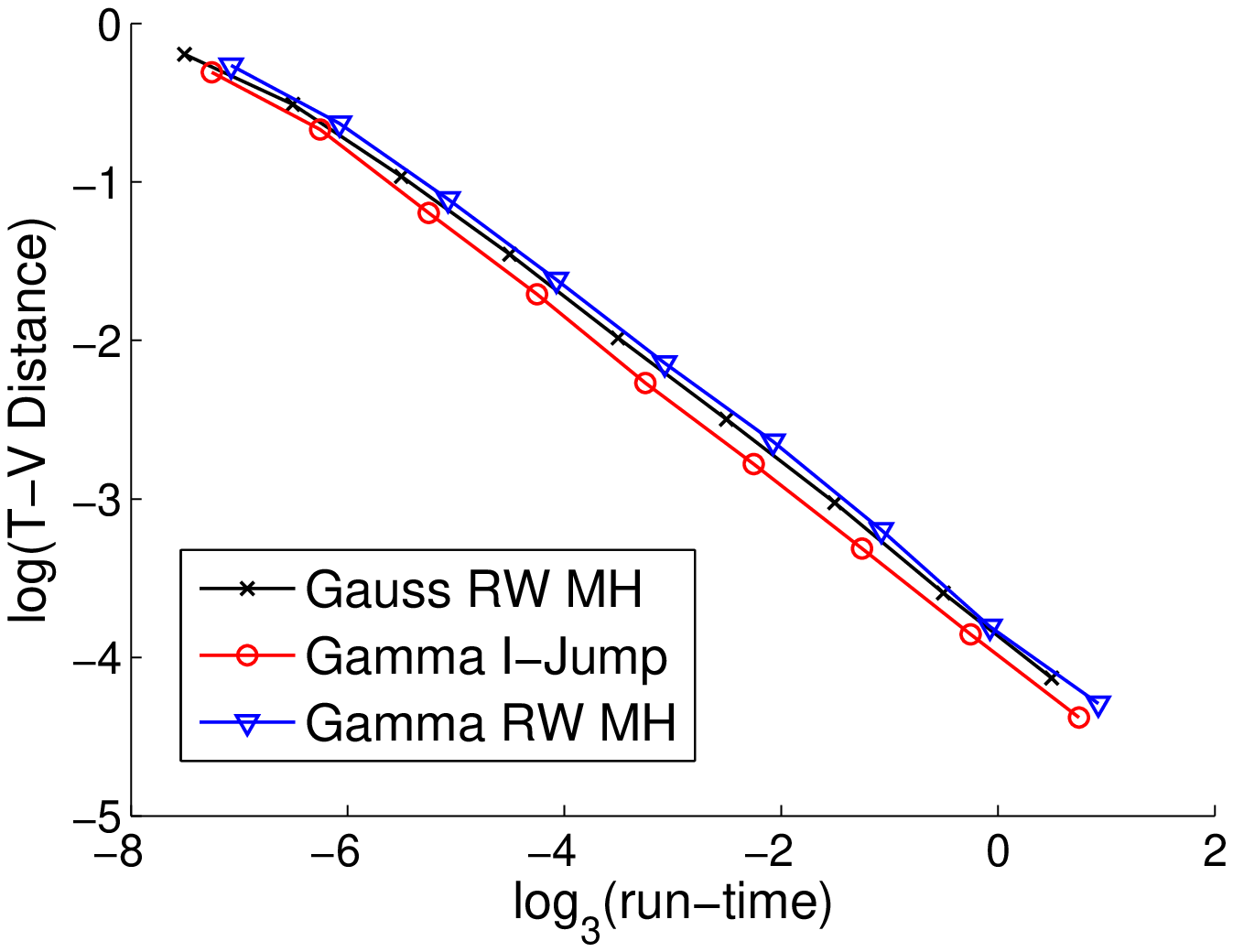}
\includegraphics[width=0.3\textheight]{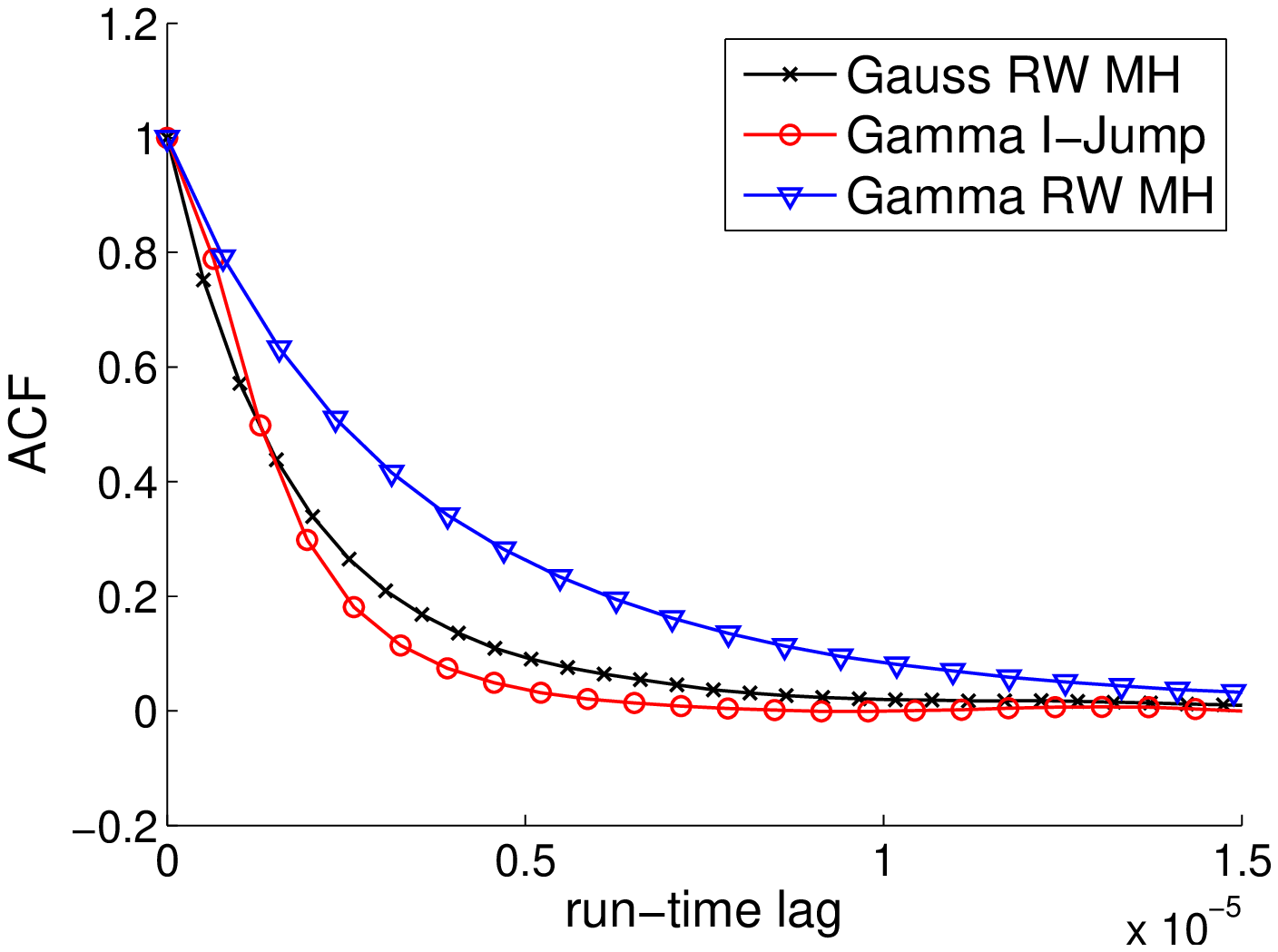}
\end{minipage}
\postcap
\caption*{Log-normal}
\caption{
\emph{Top row:} (Left) Normal and log-normal target distributions, and (right) zoom in of the tails.
\emph{Middle row:} Results for normal target in terms of log total variation distance (T-V distance) vs.~log runtime (left) and ACF vs.~lag in runtime (right).
\emph{Bottom row:} Analogous plots for log normal target.
Comparisons are made among the I-Jump sampler with gamma proposals (Gamma I-Jump), random walk MH algorithm with Gaussian proposals (Gauss RW MH), and random walk MH algorithm with symmetrized gamma proposals (Gamma RW MH).
Runtime is measured in seconds.
}
\label{fig:1D_Experiment}
\end{figure*}

We start by considering the task of sampling from 1D normal and log-normal distributions, the latter of which is a heavy-tailed distribution.
The motivation for considering the simple 1D normal distribution is to validate the correctness of the sampler and to serve as a comparison relative to the heavy-tailed setting.
We compare performance to a Gaussian random walk MH algorithm.
The results are shown in Fig.~\ref{fig:1D_Experiment}.
We also compare against an MH algorithm using a symmetrized gamma proposal distribution:
\begin{align}
(\vz(*)-\vz{(t)}) \sim \dfrac12 \left( f\left(\vz(*)| \vz{(t)}\right) + g\left(\vz(*)| \vz{(t)}\right) \right)
\nonumber
\end{align}
to more closely match the gamma proposal of the I-Jump sampler.

We compute the decrease of total variation distance and autocorrelation function to compare the performance of the samplers in Fig.~\ref{fig:1D_Experiment}.
The total variation distance is computed through discretizing the space and comparing the histogram of the samples and the true target distribution.
We find the I-Jump sampler to have the best performance even in the simple normal target distribution case.
In particular, I-Jump can decrease autocorrelation without increasing the rejection rate (the rejection rate of all three methods are similar).
Intuitively, this result can be understood from Fig.~\ref{fig:cartoon_update}: the irreversible algorithm leads to further exploration in one direction before circling back. (Also, see Fig.~\ref{fig:TracePlot}.)

For the heavy-tailed distribution, similar behavior is observed: I-Jump converges to the desired distribution faster because its samples decorrelate more rapidly as a function of runtime.

\subsubsection{Multimodal Distributions}
\begin{figure*}[p]
\centering
\includegraphics[width=0.23\textheight]{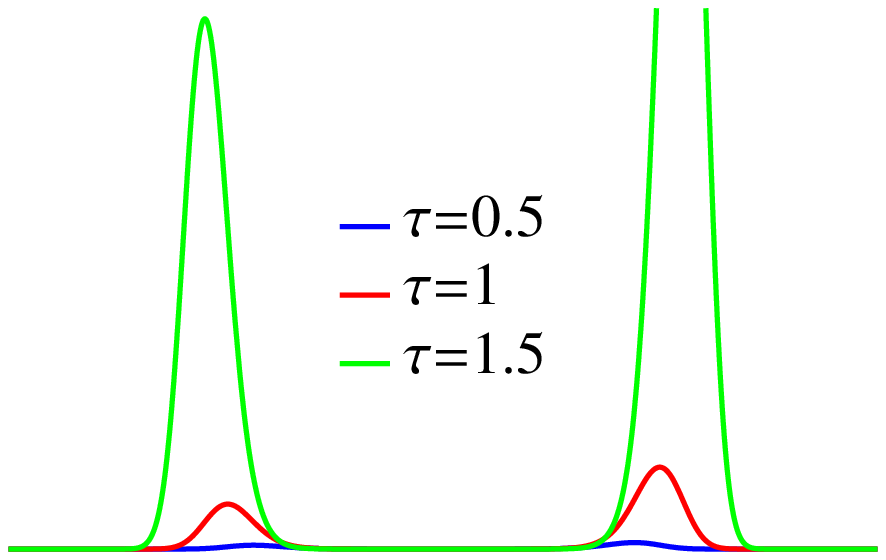}
\includegraphics[height=0.12\textheight]{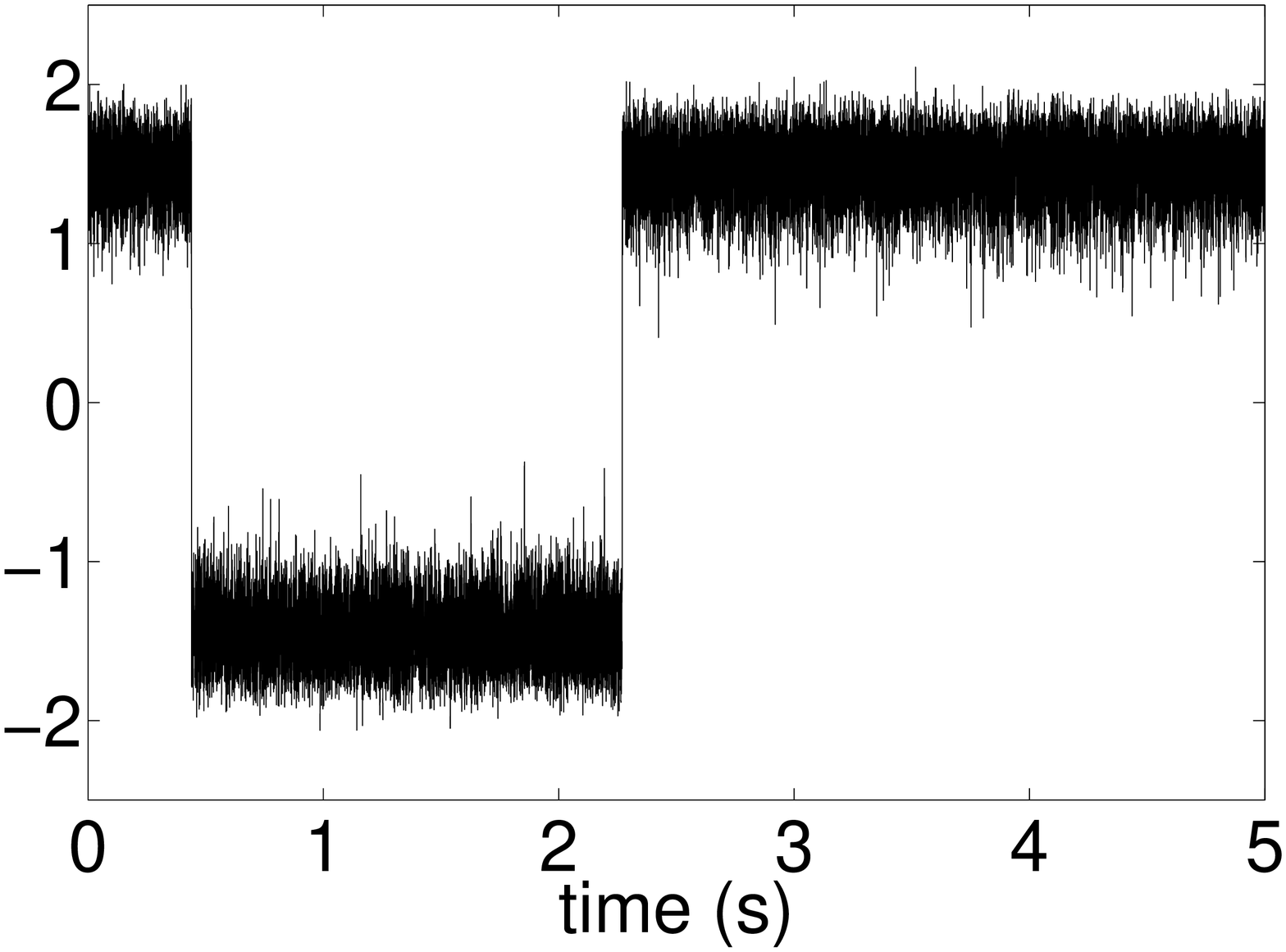}
\includegraphics[height=0.12\textheight]{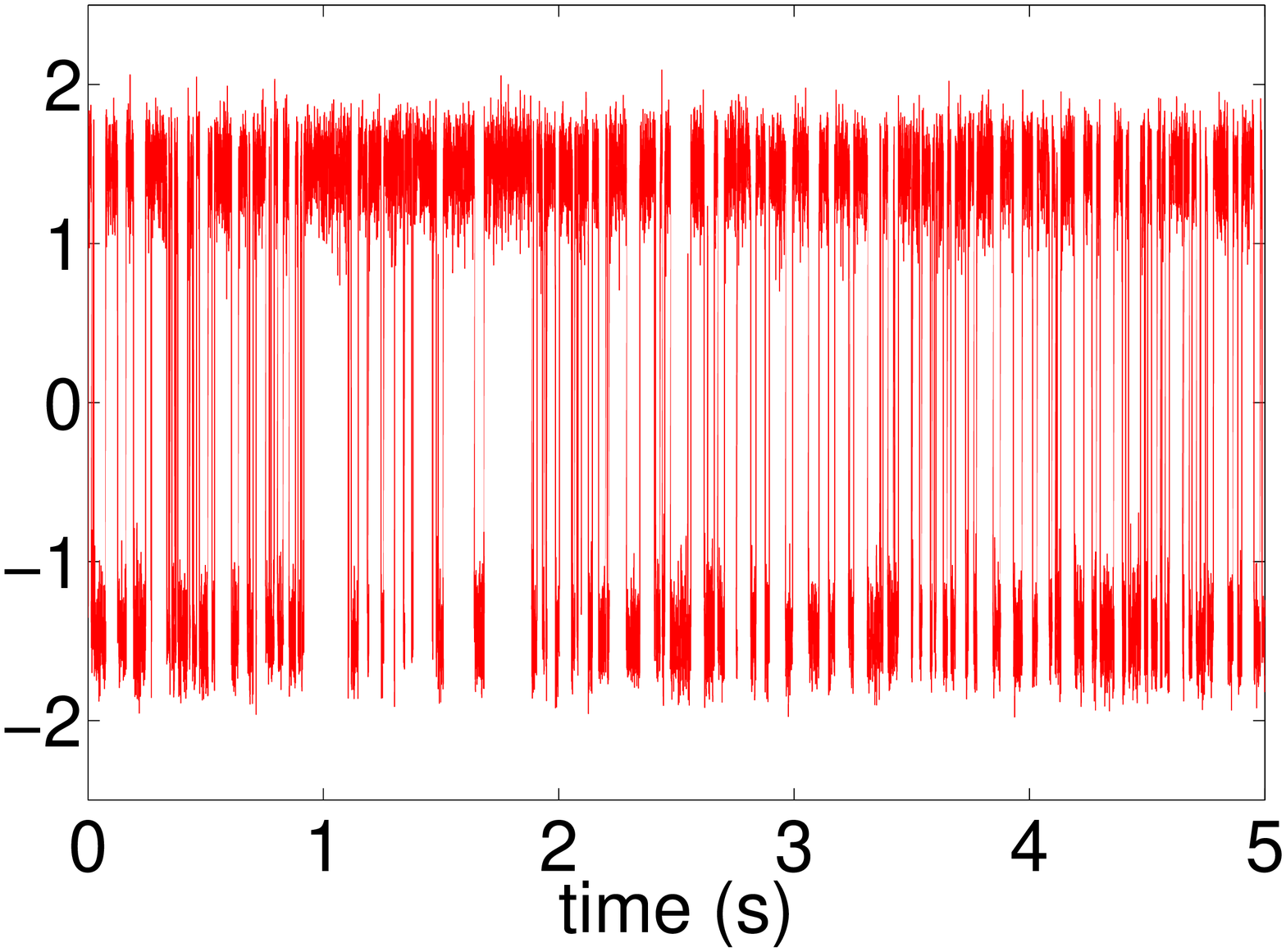}
\\
\begin{minipage}[b]{0.7\textheight}
\centering
\includegraphics[width=0.3\textheight]{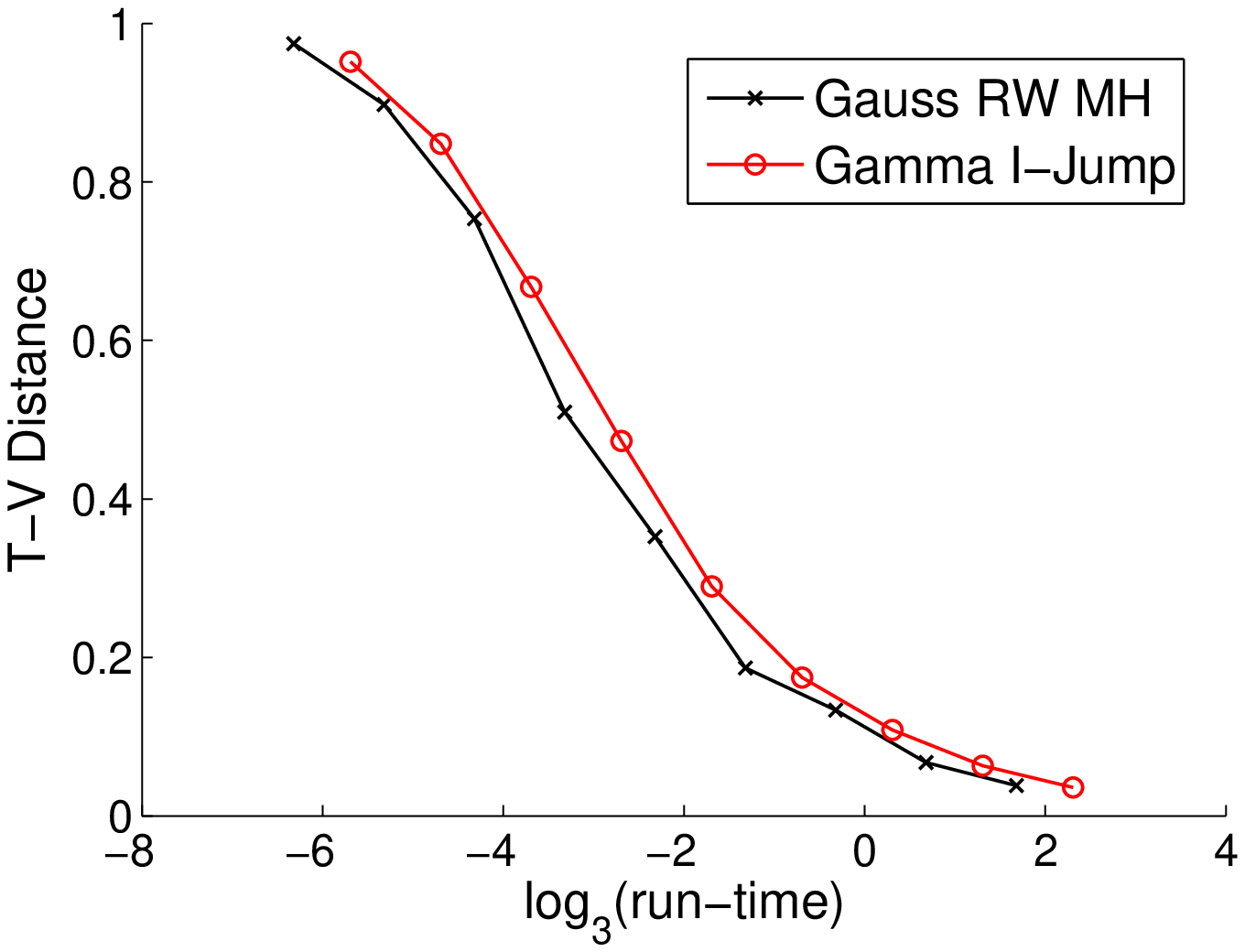}
\includegraphics[width=0.3\textheight]{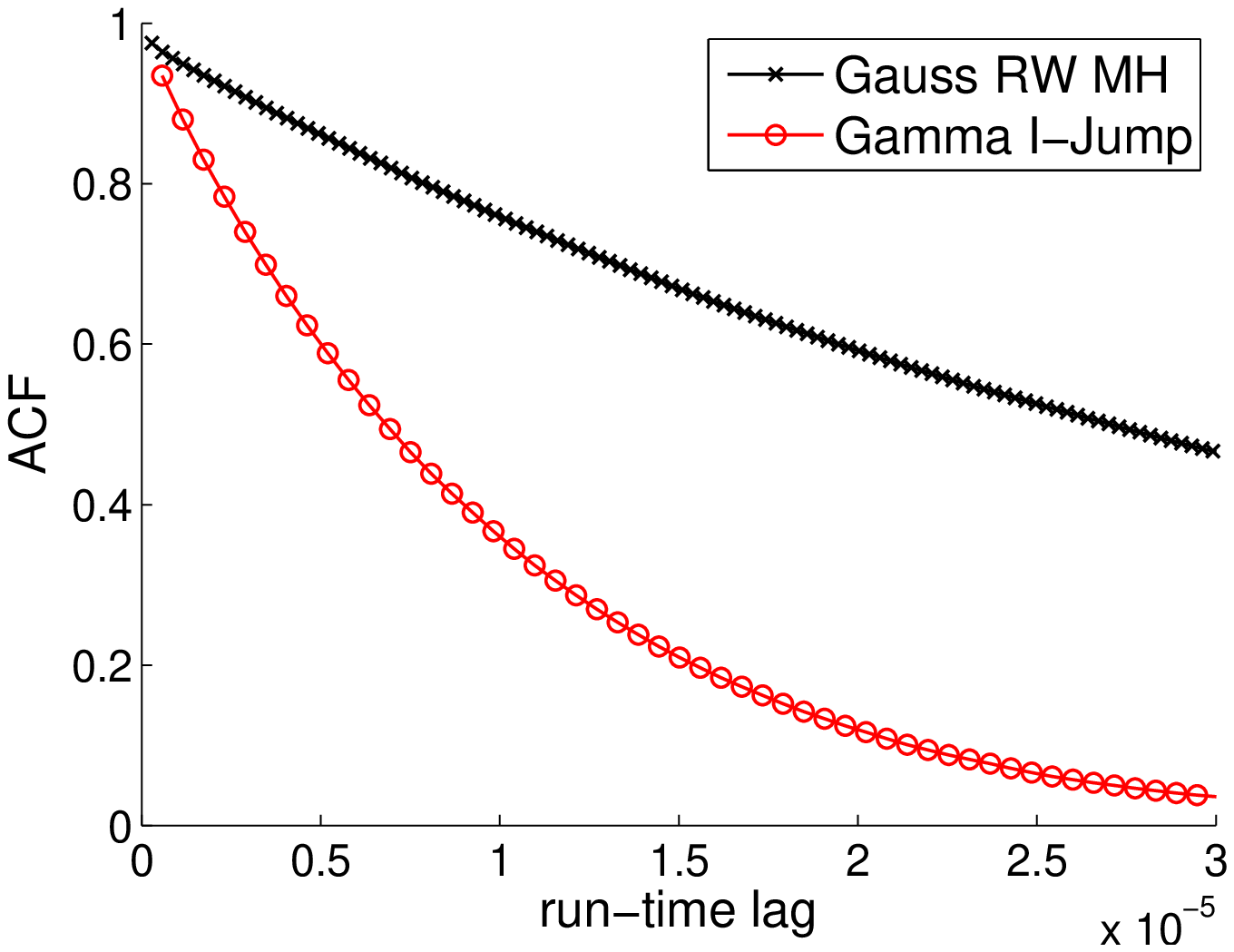}
\end{minipage}
\postcap
\caption*{$\tau=0.5$}
\begin{minipage}[b]{0.7\textheight}
\centering
\includegraphics[width=0.3\textheight]{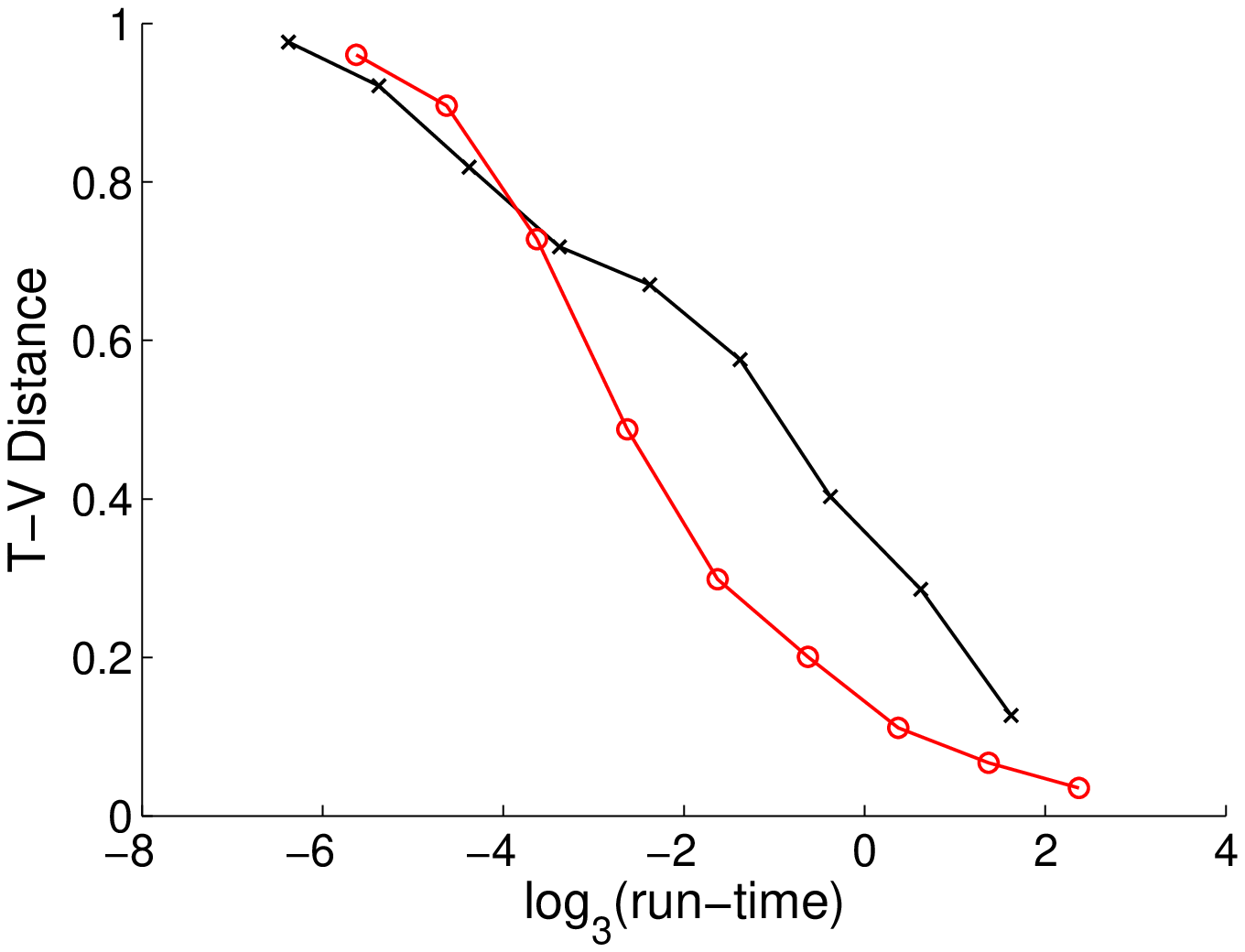}
\includegraphics[width=0.3\textheight]{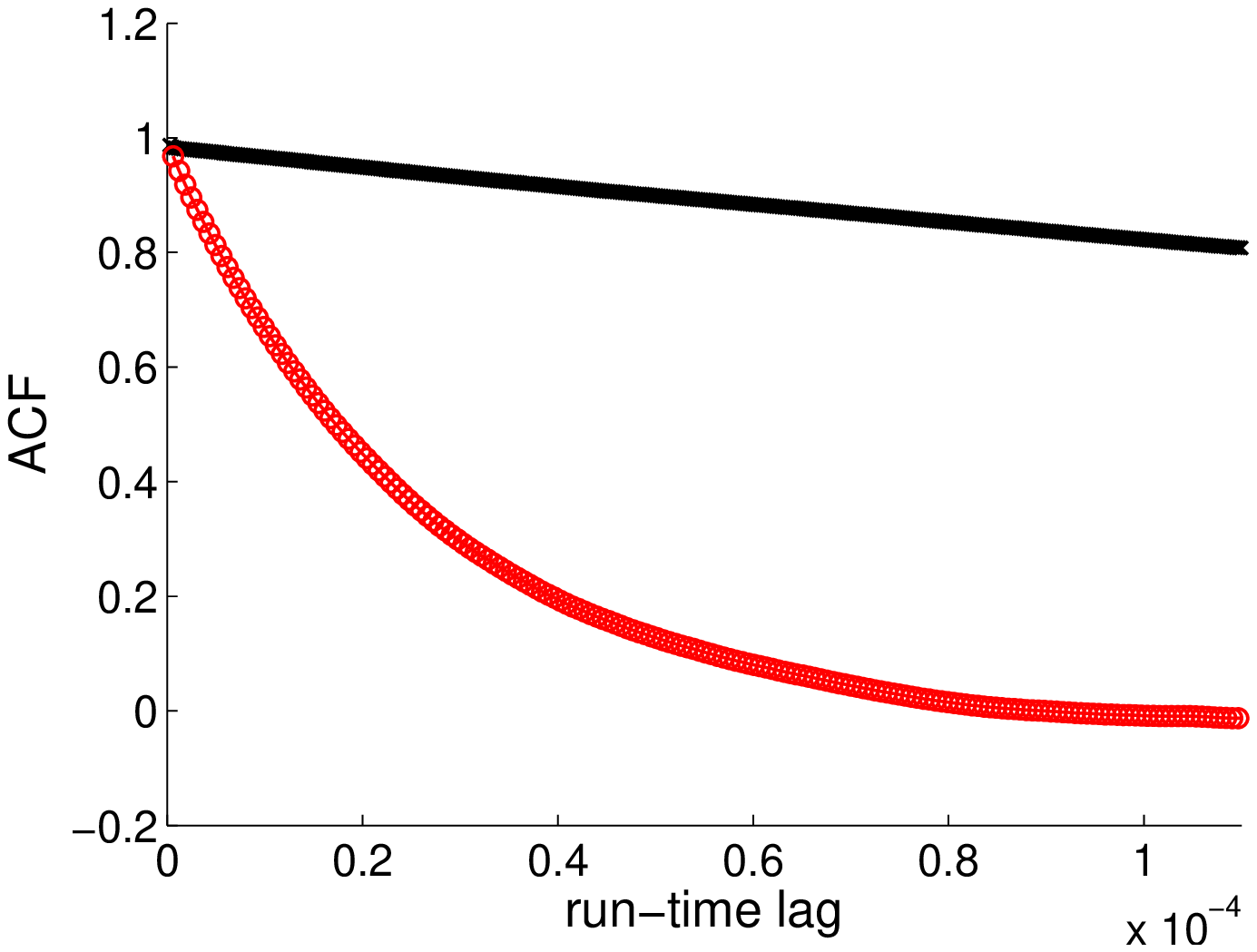}
\end{minipage}
\postcap
\caption*{$\tau=1$}
\begin{minipage}[b]{0.7\textheight}
\centering
\includegraphics[width=0.3\textheight]{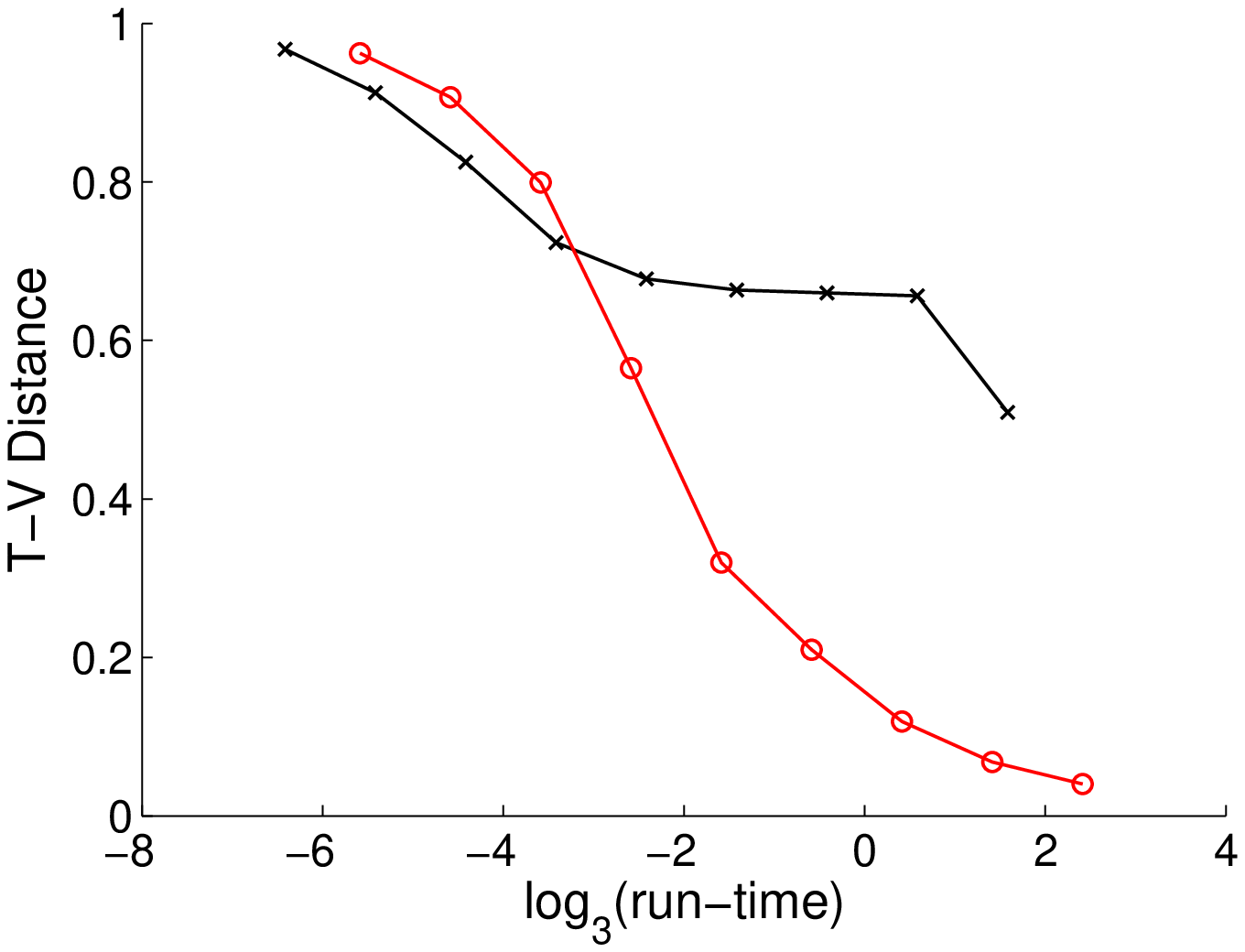}
\includegraphics[width=0.3\textheight]{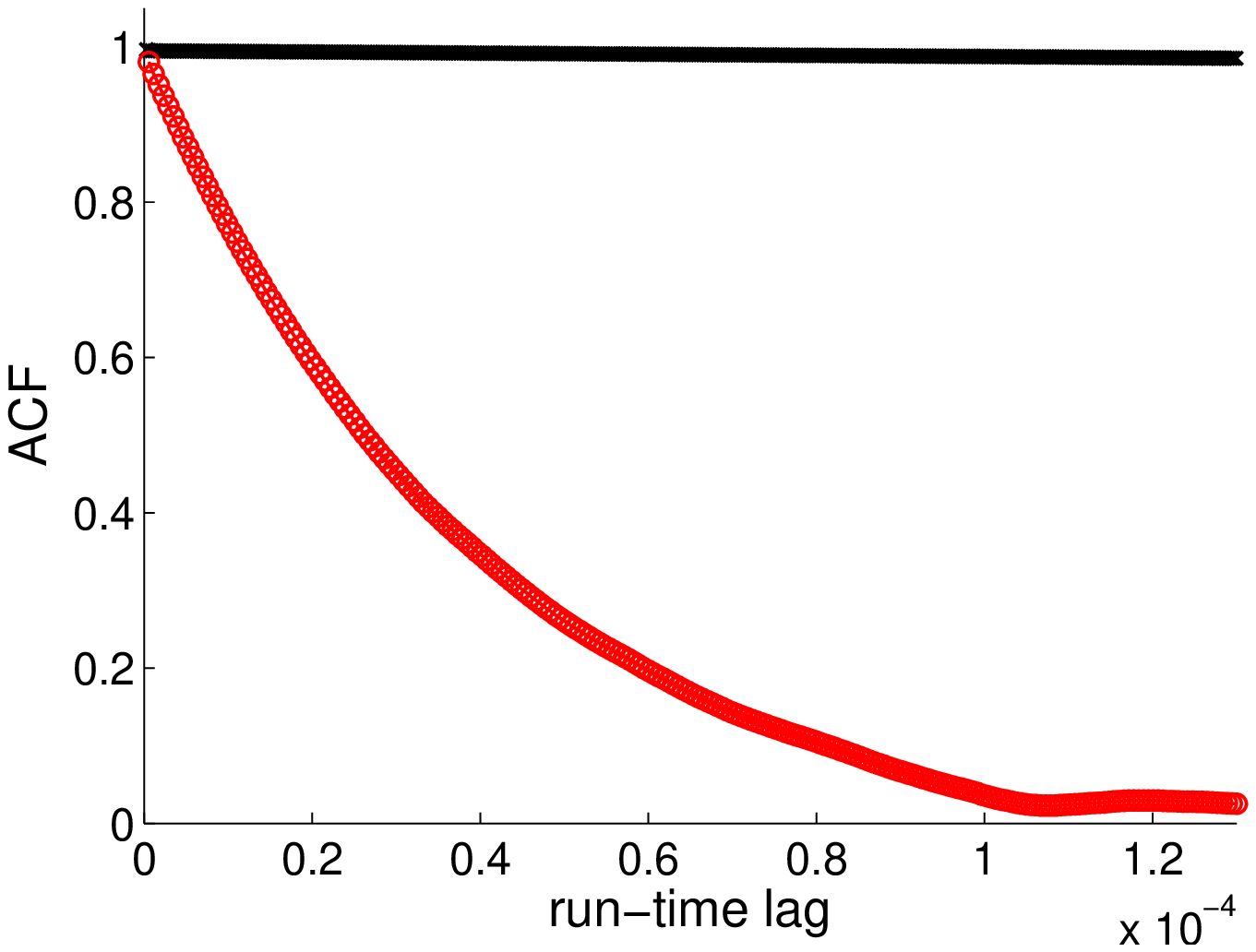}
\end{minipage}
\postcap
\caption*{$\tau=1.5$}
\caption{
\emph{Top row:} (Left) Bimodal targets, $\pi(z_1,z_2) = 2(z_1^2 - \tau)^2 - 0.2 z_1 - 5 z_1^2 + 5 z_2^2$, for various values of $\tau$.
Here we demonstrate a 1D cross section of the 2D distribution. (Middle) Sample state trajectories for Gauss RW MH and (right) Gamma I-Jump for $\tau=1$.
\emph{Bottom rows:} Total variational distance vs. log runtime (left) and ACF vs. lag in runtime (right),
with each row corrresponding to a specific choice of $\tau$.
}
\label{fig:2D_Bimodal}
\end{figure*}

\paragraph{2D Bimodal distributions}
We use our I-Jump sampler to sample increasingly challenging bimodal distributions in 2D, $\pi(z_1,z_2) = 2 (z_1^2 - \tau)^2 - 0.2 z_1 - 5 z_1^2 + 5 z_2^2$ for $\tau=0.5, 1.0$, and $1.5$, as displayed in Fig.~\ref{fig:2D_Bimodal}.
Based on the results of Sec.~\ref{sec:1D_Experiment}, we simply compare against the Gaussian random walk MH sampler and drop the symmetrized gamma proposal case.
In Fig.~\ref{fig:2D_Bimodal} we see that the I-Jump sampler significantly outperforms the random walk MH algorithm.
Intuitively, this is facilitated by the greater traversing ability of the I-Jump sampler so that with the same acceptance rate, the I-Jump sampler can explore more possible states than the reversible sampler, and have greater chance of transiting into another mode.

One way to quantify this effect is in terms of {\em escape time} from local modes, which we summarize in Table \ref{table1}.
We see that the I-Jump jump sampler has escape times orders of magnitude lower. Furthermore, these escape times increase at a much smaller rate as the local modes become more concentrated, indicating much more rapid mixing between modes.

\begin{table*}
  \centering
  \begin{tabular}{ c | c | c} 
{\bf $\tau$} & Avg. Escape Time for I-Jump Sampler & Avg. Escape Time for MH Sampler\\ \hline
0.5 & $1.94\times10^2$ & $1.06\times10^3$ \\ \hline
1   & $4.64\times10^2$ & $2.47\times10^4$ \\ \hline
1.5 & $9.06\times10^2$ & $7.89\times10^5$ \\ \hline
2   & $2.41\times10^3$ & N/A \\
\end{tabular}
  \caption{Comparison of average escape time from one local mode to another between Gamma I-Jump and Gauss RW MH.
  The distribution in 2D becomes more challenging with larger values of $\tau$ (plotted in Fig.~\ref{fig:2D_Bimodal}).
  ``N/A" in the last entry means that the escape time is so long that an accurate estimate is not available.
  }\label{table1}
\end{table*}

\begin{figure*}[!t]
\centering
\includegraphics[height=0.17\textheight]{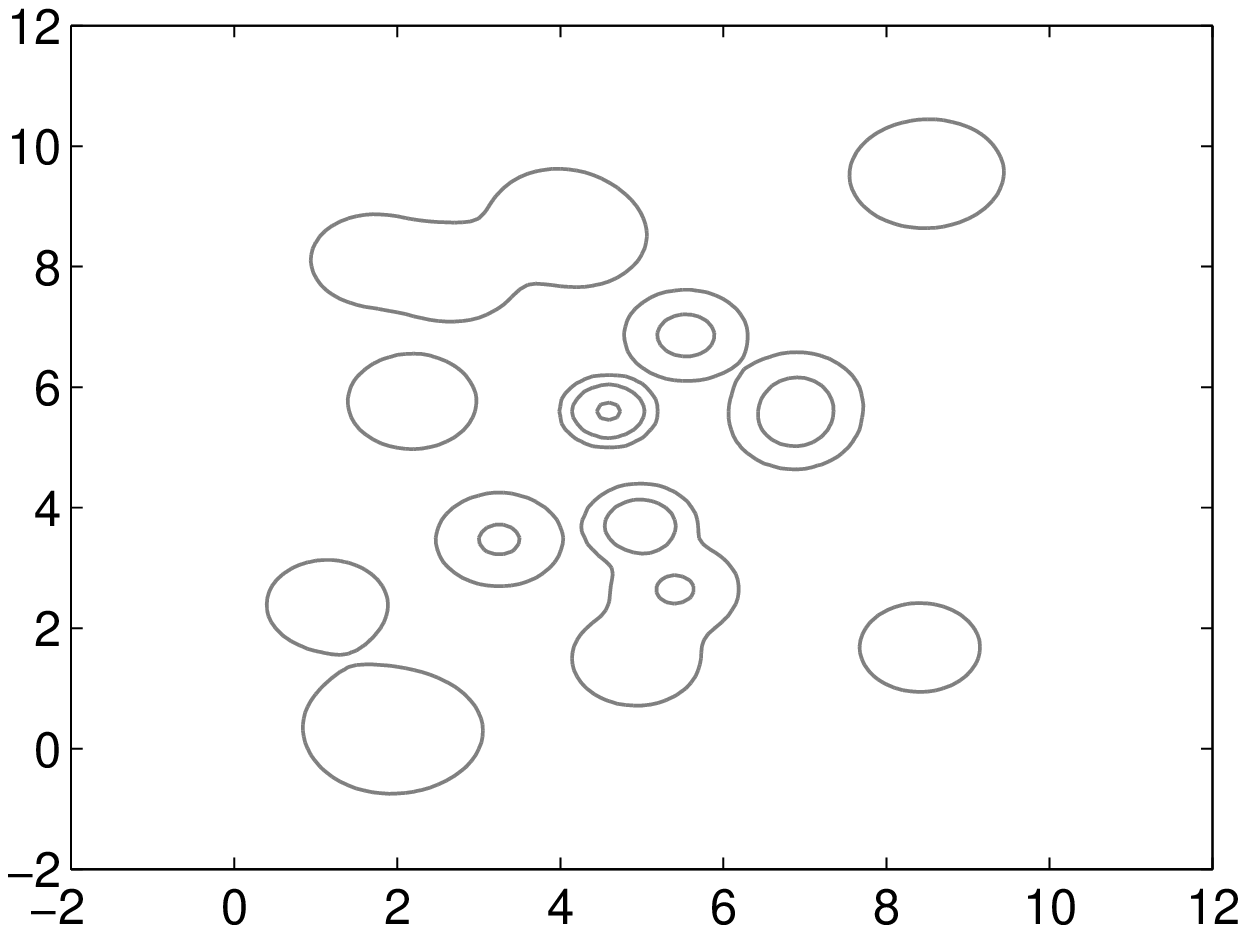} \hspace{-0.2in} 
\includegraphics[height=0.17\textheight]{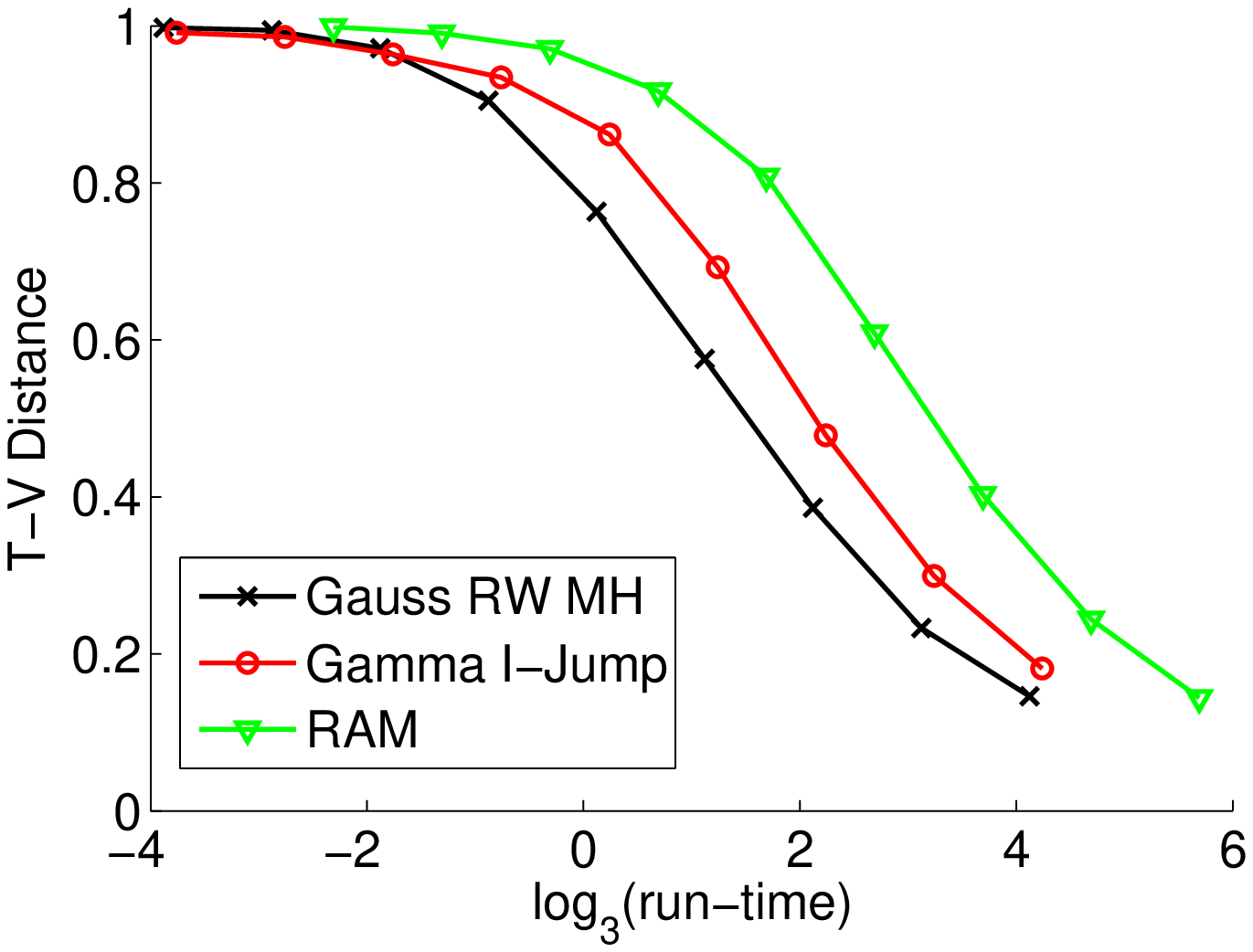} \hspace{-0.2in}
\includegraphics[height=0.17\textheight]{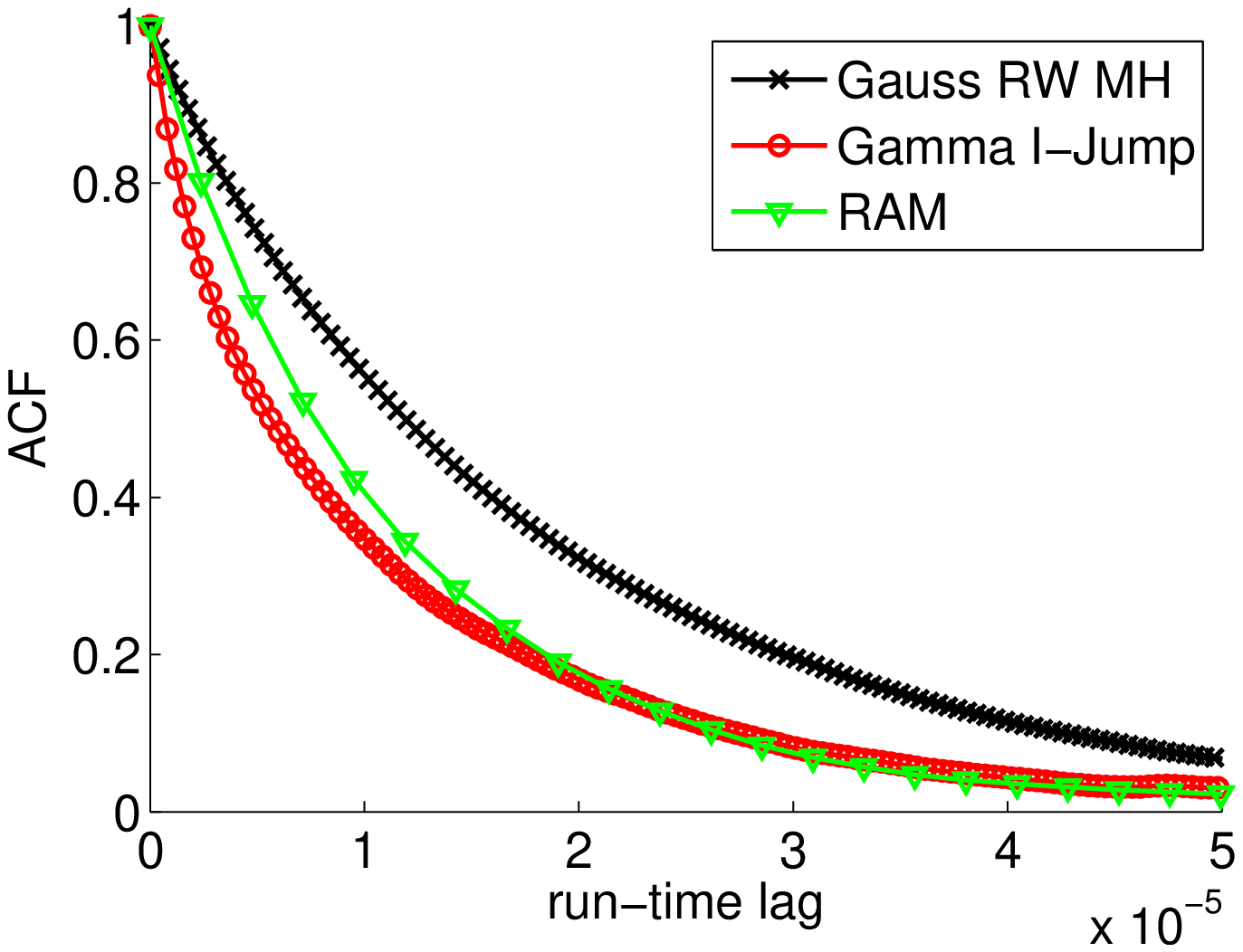}\\
\includegraphics[height=0.17\textheight]{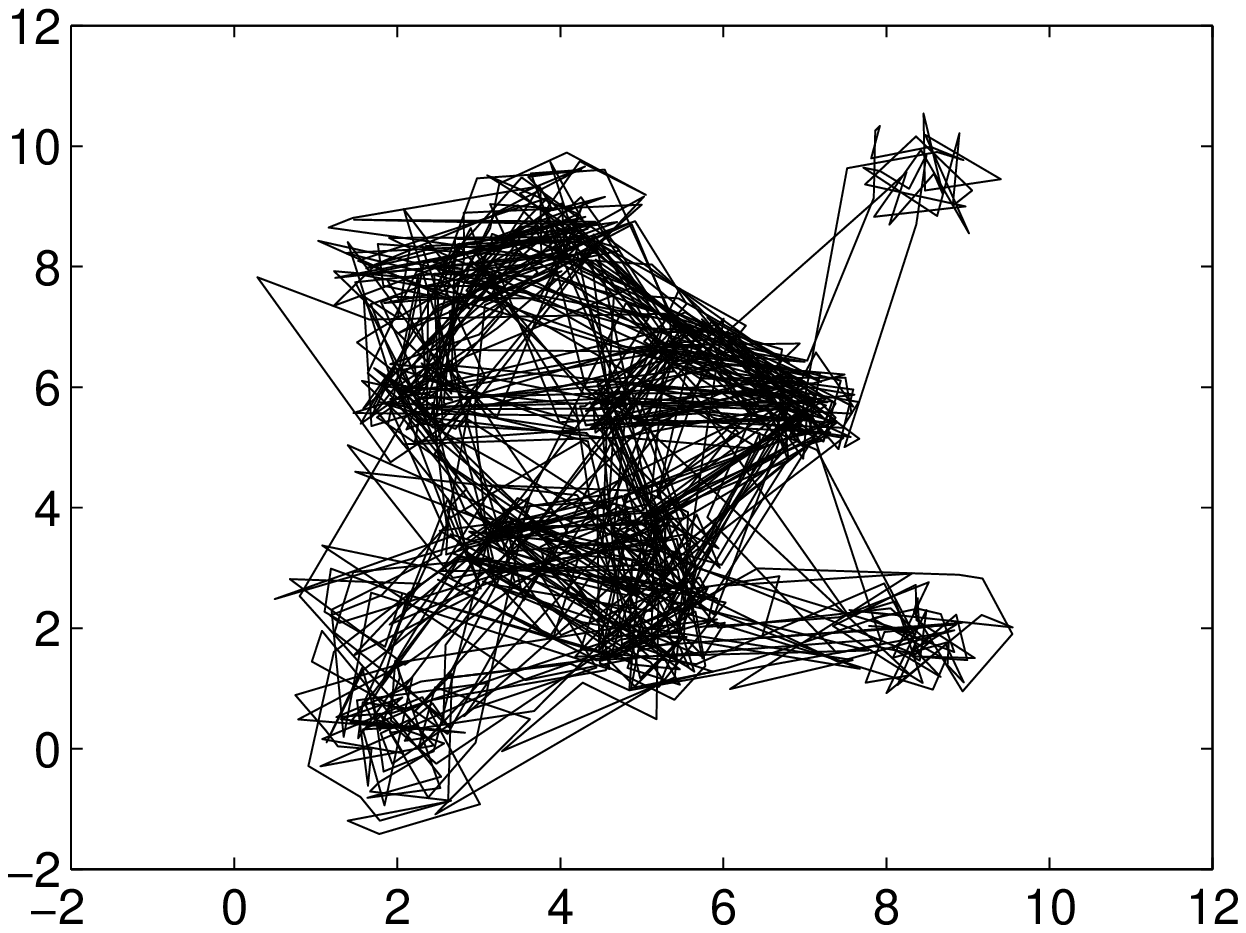} \hspace{-0.2in}
\includegraphics[height=0.17\textheight]{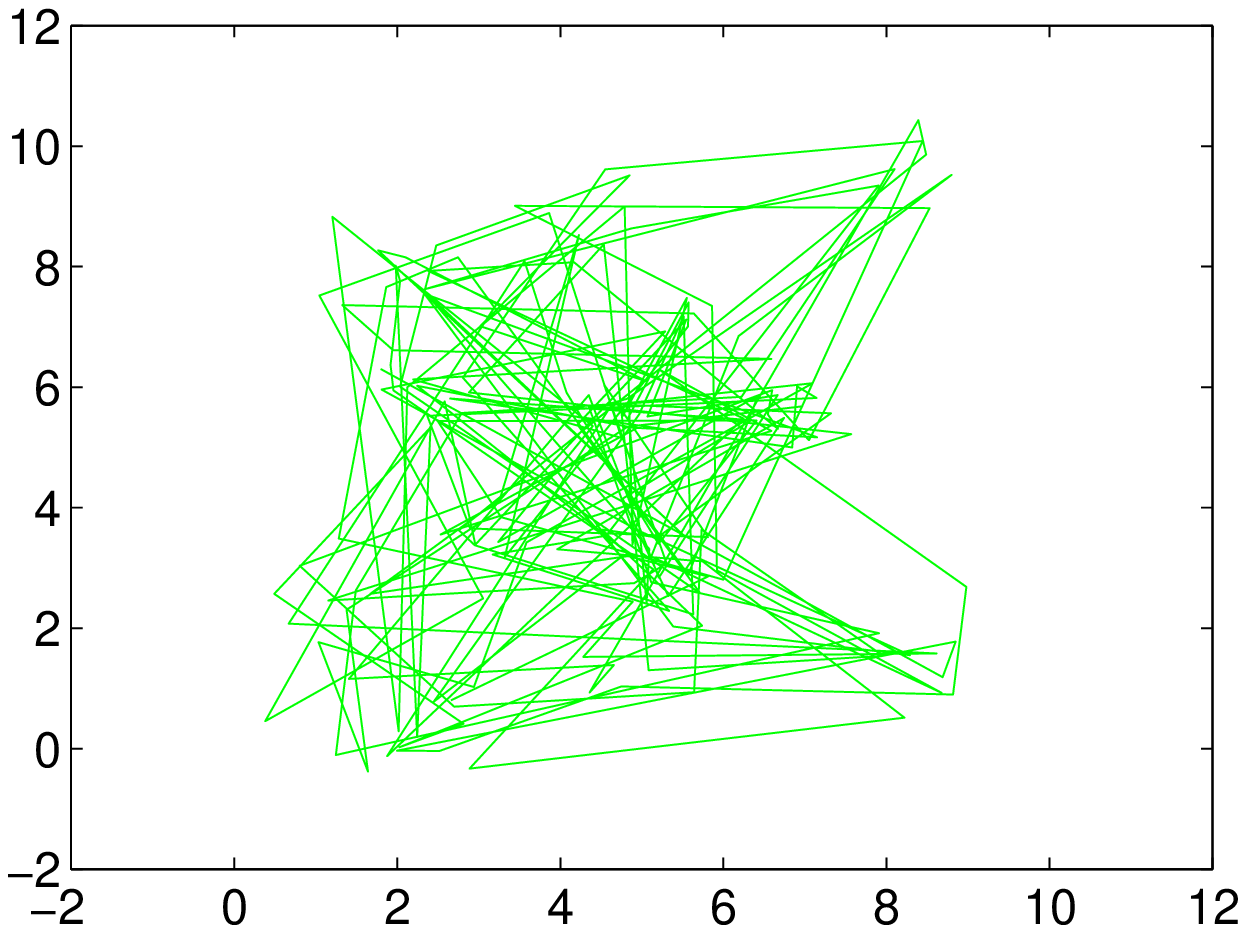} \hspace{-0.2in}
\includegraphics[height=0.17\textheight]{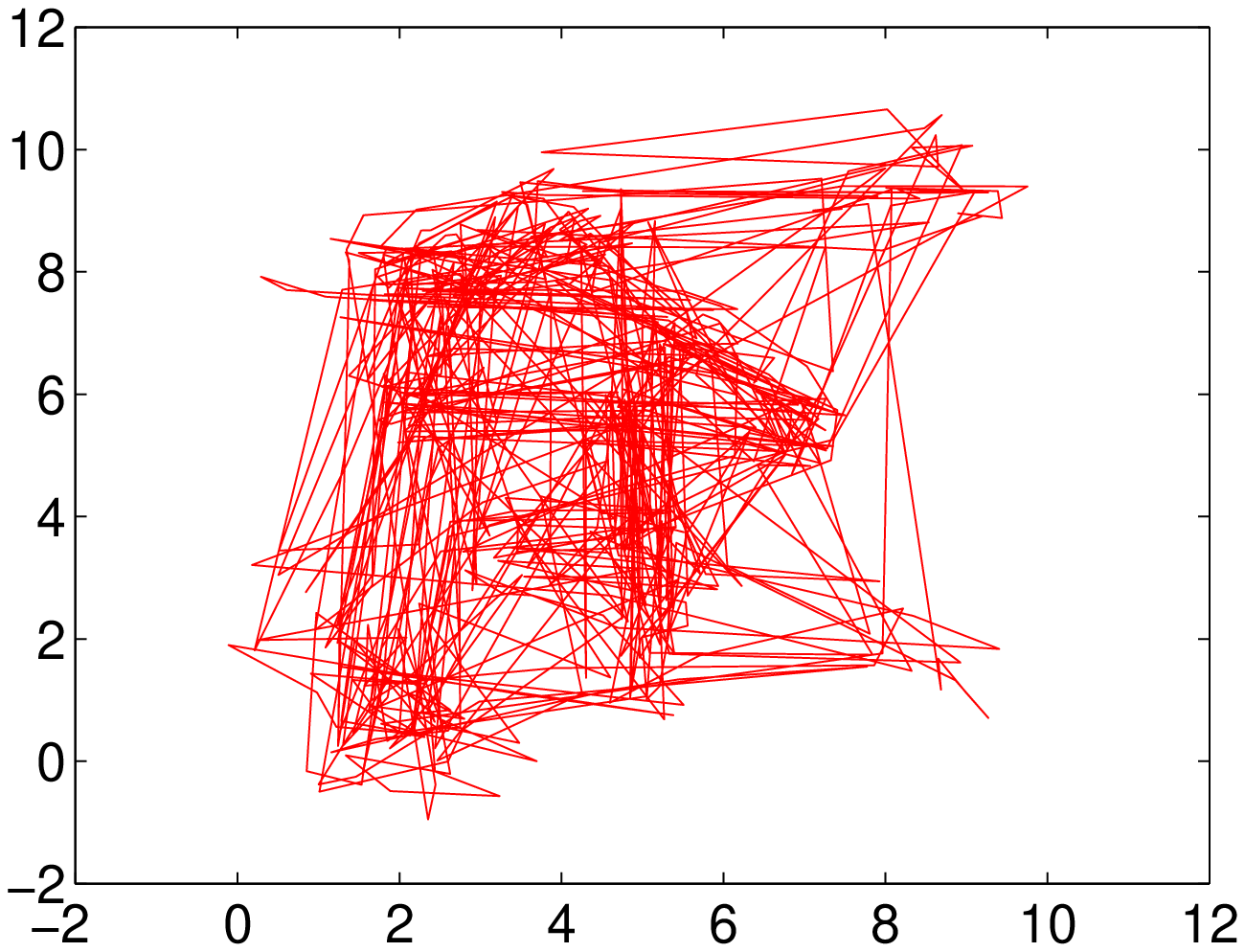}
\caption{
\emph{Top row:} (Left) Contour plot of a challenging multimodal probability density function; (middle)
T-V distance and ACF comparisons among Gauss RW MH, Gamma I-Jump, and the recently proposed repulsing-attracting Metropolis (RAM) sampler \cite{RAM}.
\emph{Bottom row:} A sample run of all three samplers, respectively.
}
\label{fig:2D_Multimodal_easy}
\end{figure*}

\begin{figure*}[!t]
\centering
\includegraphics[height=0.17\textheight]{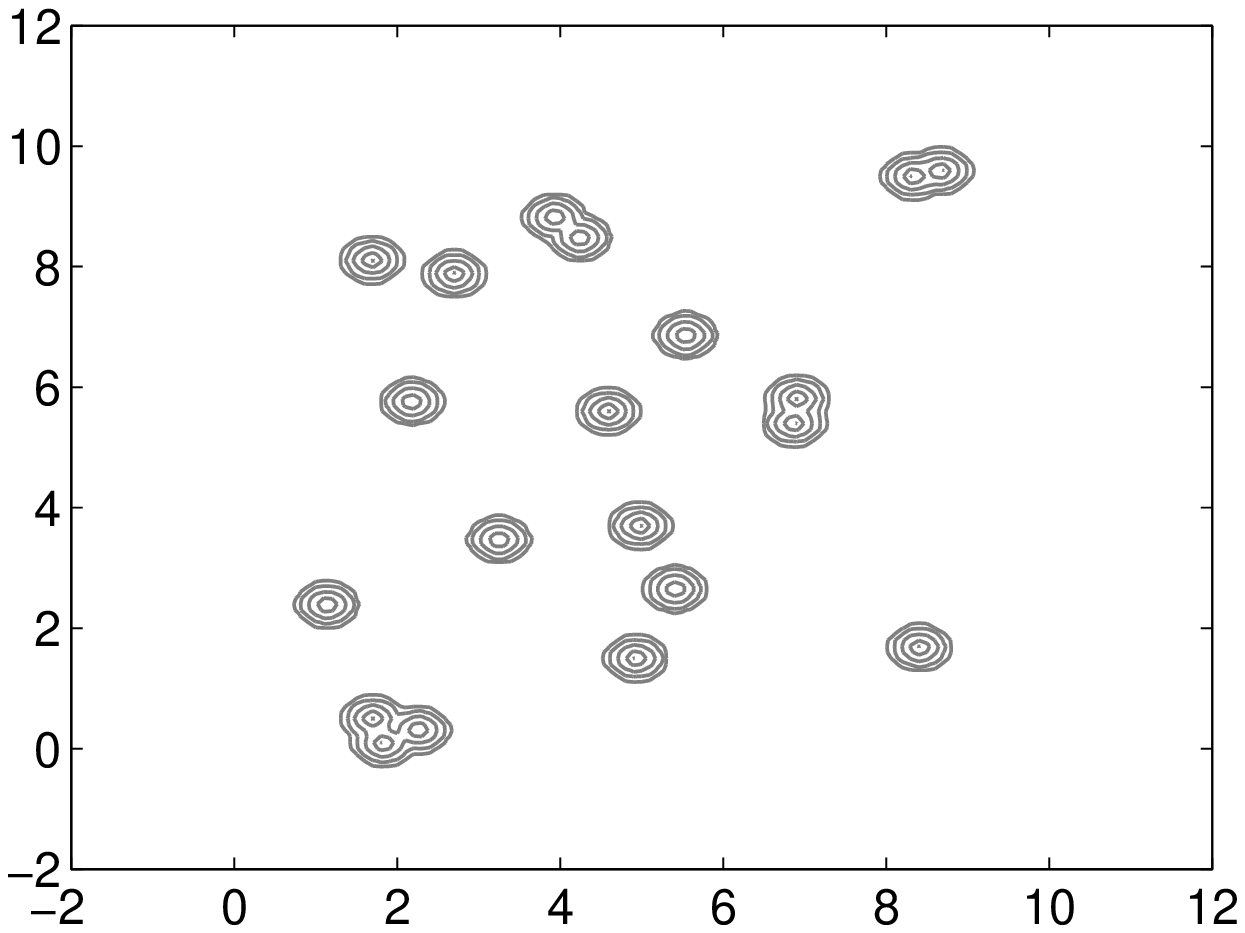} \hspace{-0.1in}
\includegraphics[height=0.17\textheight]{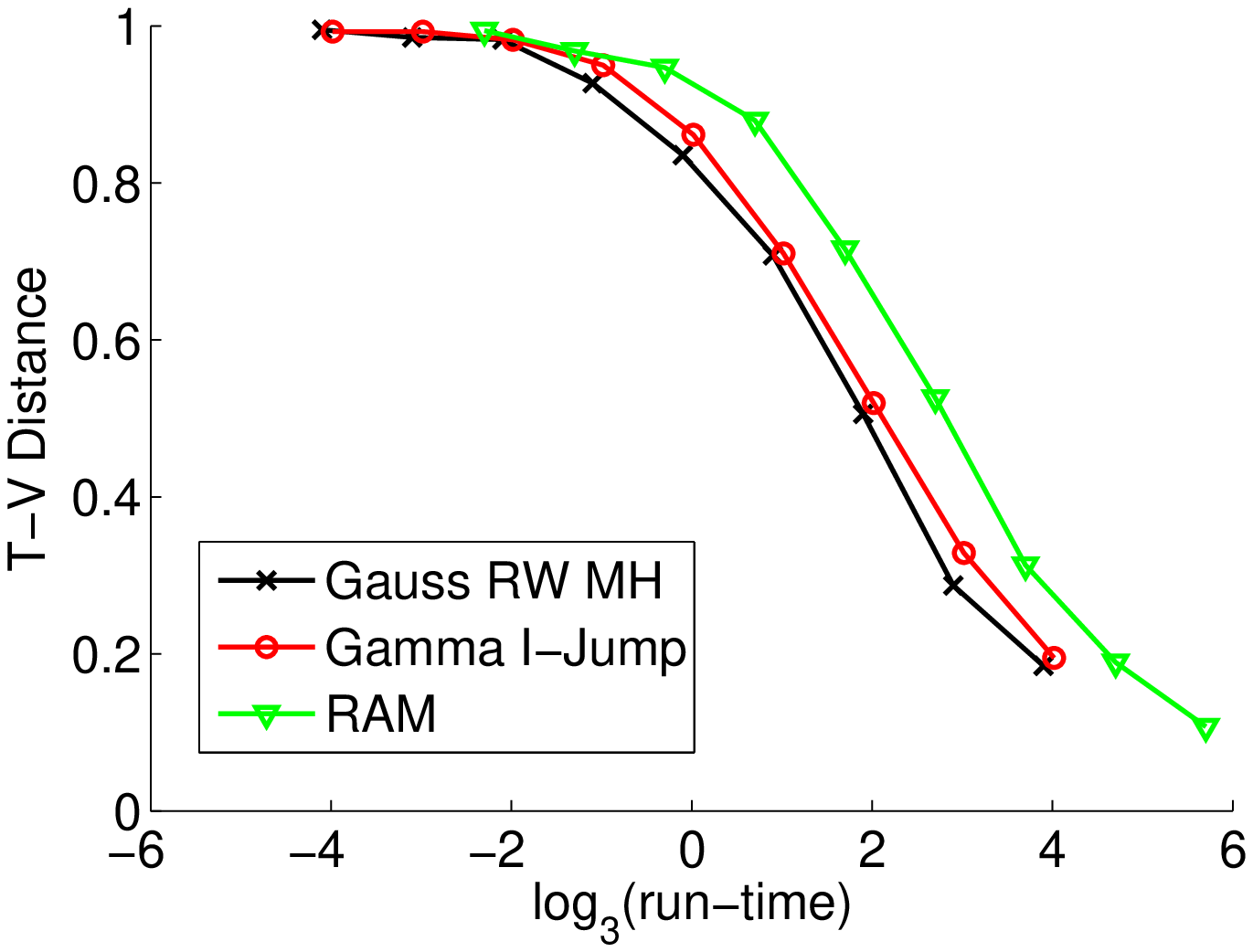} \hspace{-0.2in}
\includegraphics[height=0.17\textheight]{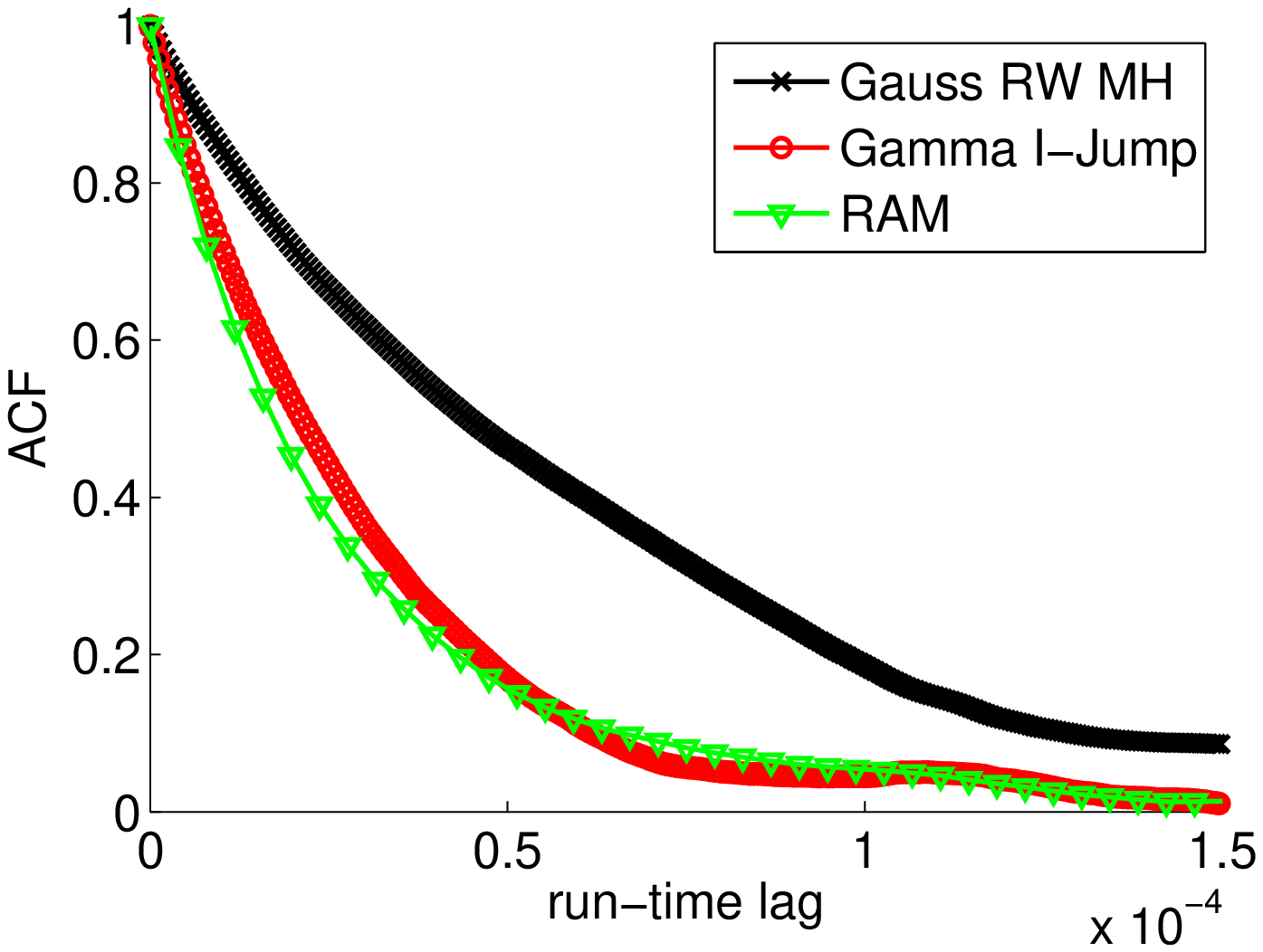}\\
\includegraphics[height=0.17\textheight]{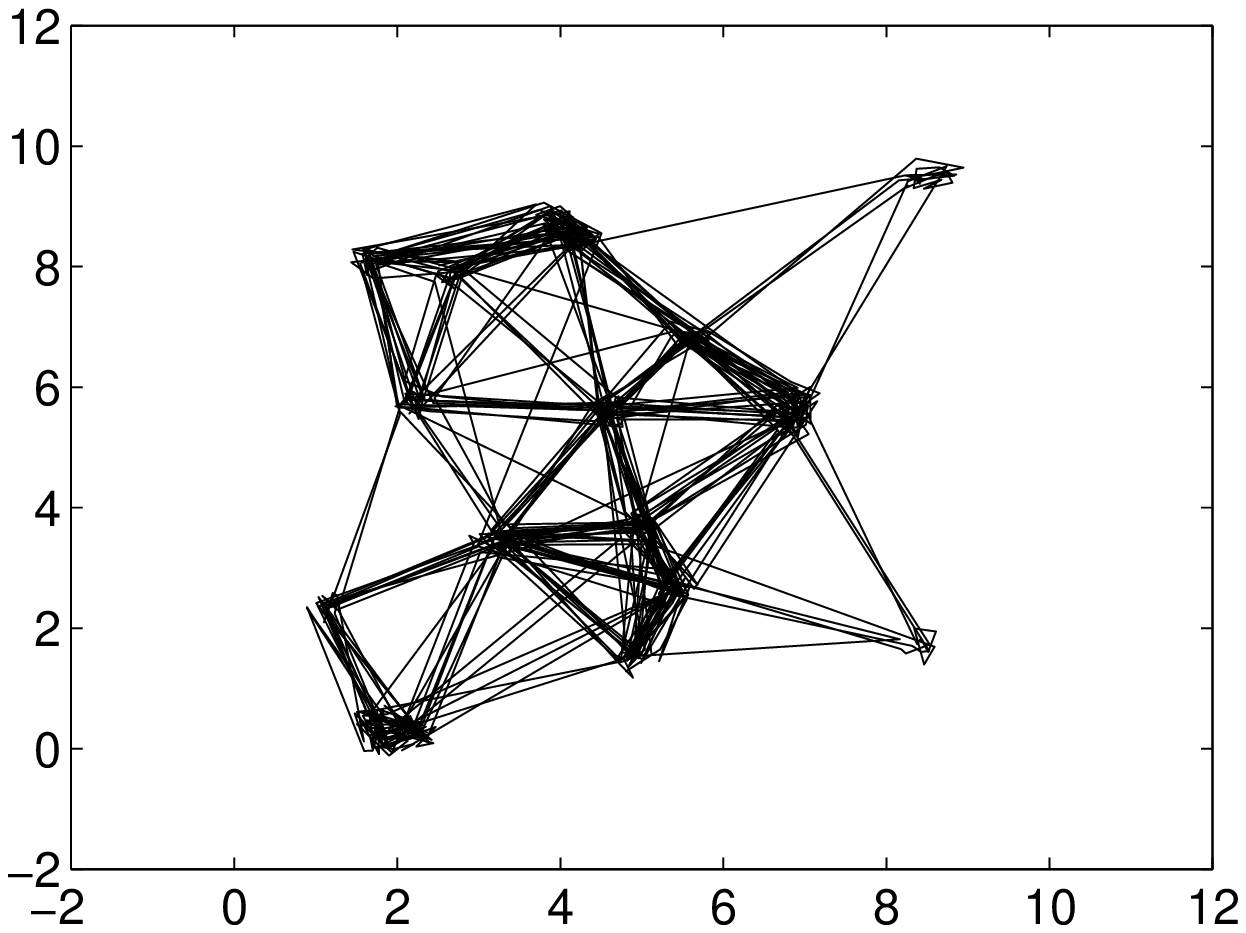} \hspace{-0.2in}
\includegraphics[height=0.17\textheight]{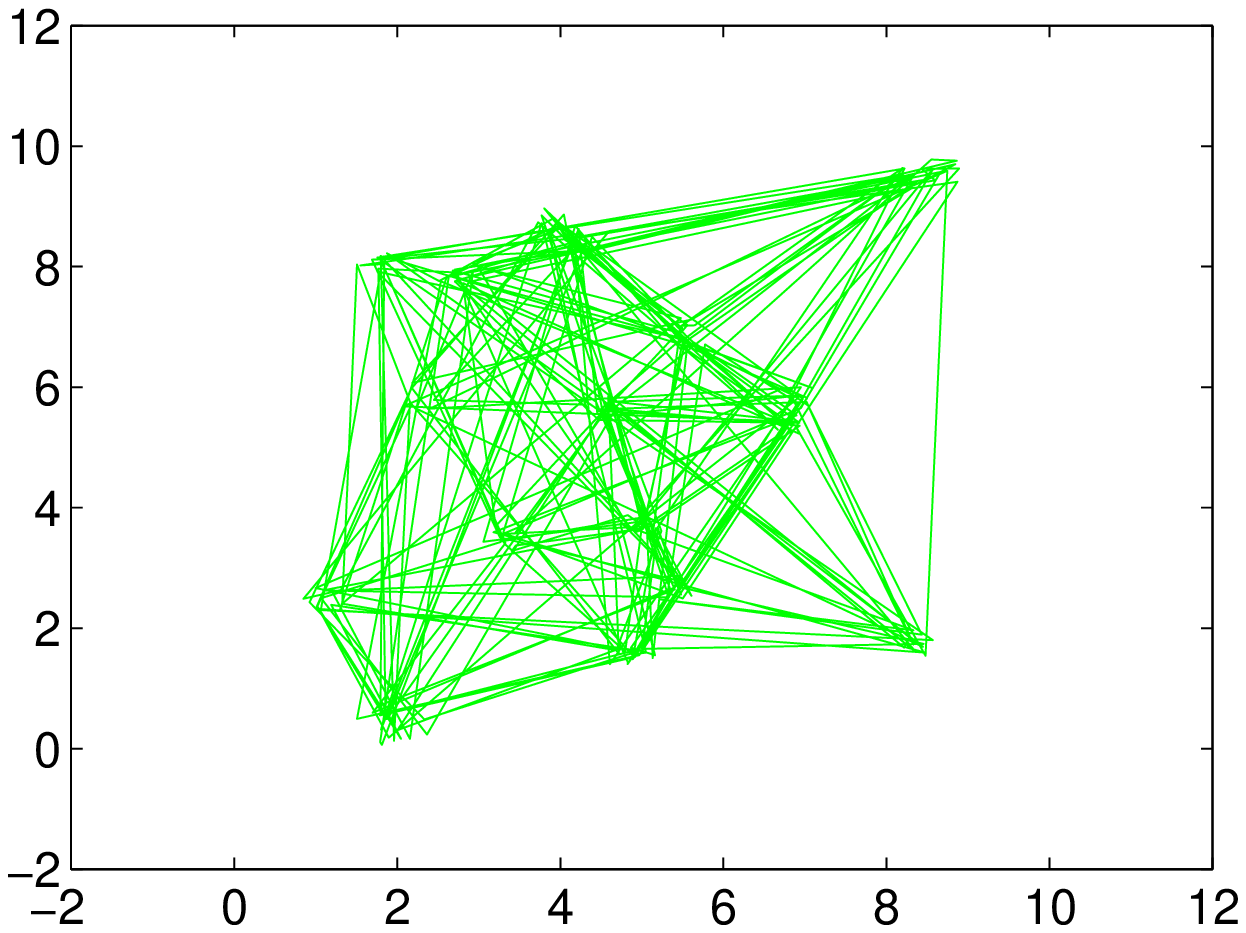} \hspace{-0.2in}
\includegraphics[height=0.17\textheight]{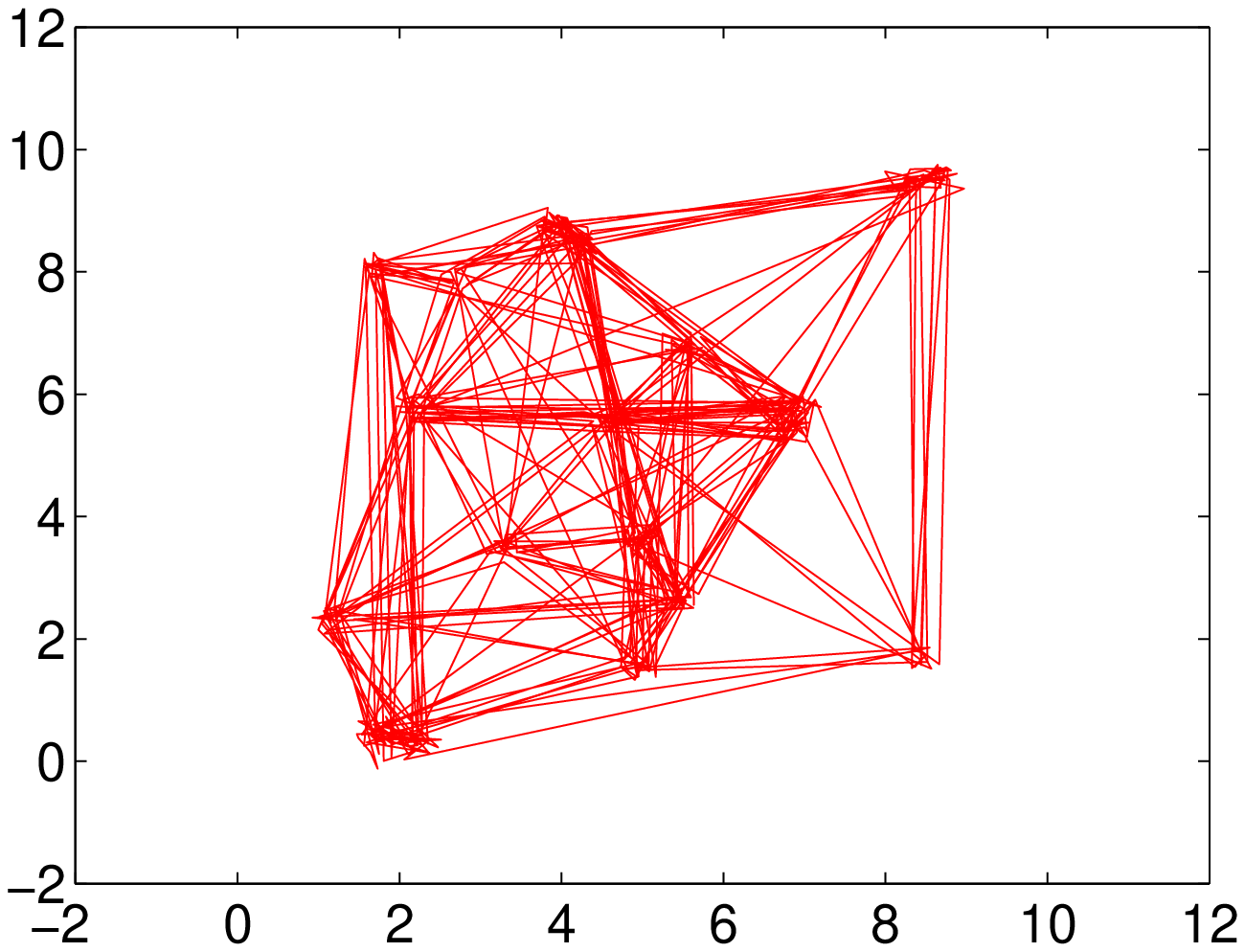}
\caption{
Plots as in Fig.~\ref{fig:2D_Multimodal_easy}, but for an even more challenging multimodal case where the modes are very concentrated and well separated.
}
\label{fig:2D_Multimodal_hard}
\end{figure*}

\paragraph{2D Multimodal distributions}
We also tested our method against a recently considered multimodal setting \cite{RAM}.
In the first setting considered, the target distribution is highly multimodal in 2D with unevenly distributed modes.
Furthermore, the high mass modes have smaller radii of variation.
In the second setting considered, these modes are highly concentrated and well separated, which is an extremely challenging setting for most samplers.
See Figs.~\ref{fig:2D_Multimodal_easy} and \ref{fig:2D_Multimodal_hard}.
In \cite{RAM}, a {\em repulsing-attracting Metropolis (RAM)} sampler was proposed with a structure specifically designed to efficiently handle these types of multimodal distributions.
We use this as a gold-standard comparison, since this method was already shown to outperform parallel tempering and alternatives \cite{Kou} in this setting.

We focus our performance analysis on the decay speed of the autocorrelation function (ACF).
This can be understood by taking the Gaussian random walk MH algorithm as an example: Although the Gaussian random walk MH algorithm seems to perform well in terms of convergence of total variation distance, this effect is based on exploring one mode really well in a short period of time, instead of making more distant moves to explore other modes.
In contrast, the ACF better characterizes the exploration of the samples through the whole space.

Our results are summarized in Figs. \ref{fig:2D_Multimodal_easy} and \ref{fig:2D_Multimodal_hard} for each of the two simulated multimodal scenarios.
In the first scenario, our sampler outperforms both MH and RAM.
In the second scenario, where we have highly concentrated and separated modes, the RAM method tailored to this scenario slightly outperforms our approach.
Overall, however, the I-Jump sampler provides surprisingly good performance despite not having been designed specifically for this setting.

\subsubsection{Correlated Distribution}
\label{sec:exp_corr}
\begin{figure*}[!t]
\centering
\includegraphics[height=0.17\textheight]{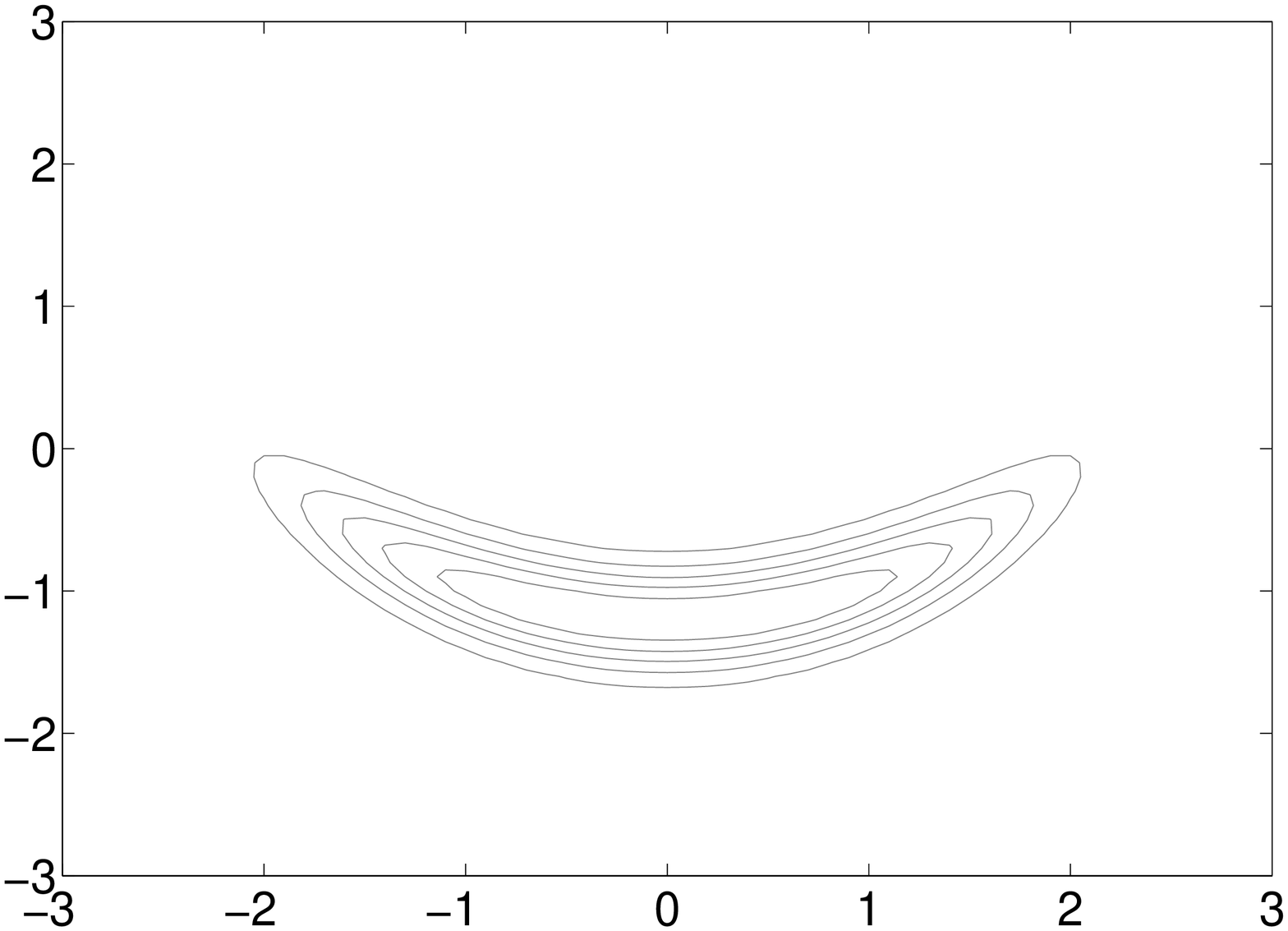}\hspace{-0.2in}
\includegraphics[height=0.17\textheight]{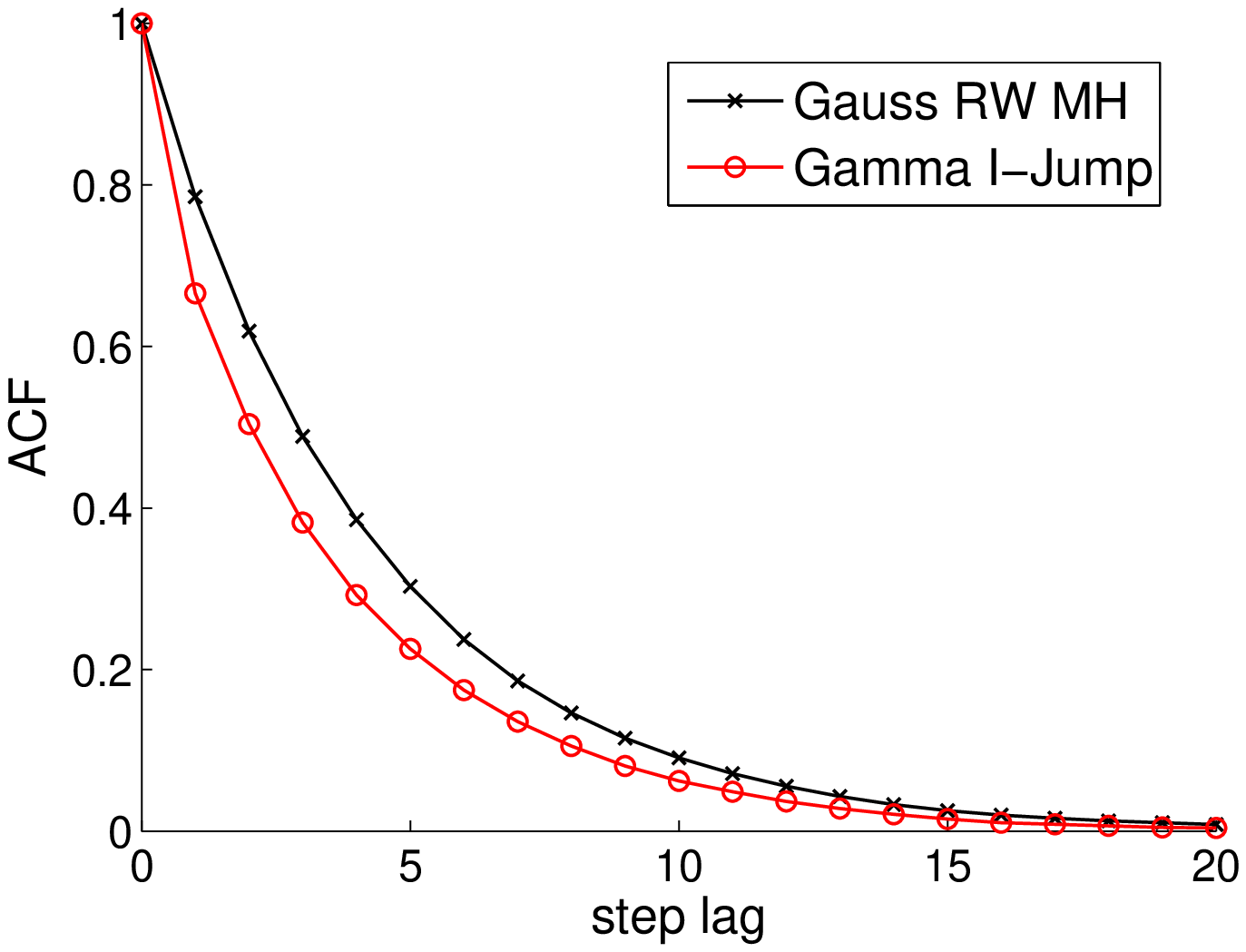}\hspace{-0.2in}
\includegraphics[height=0.17\textheight]{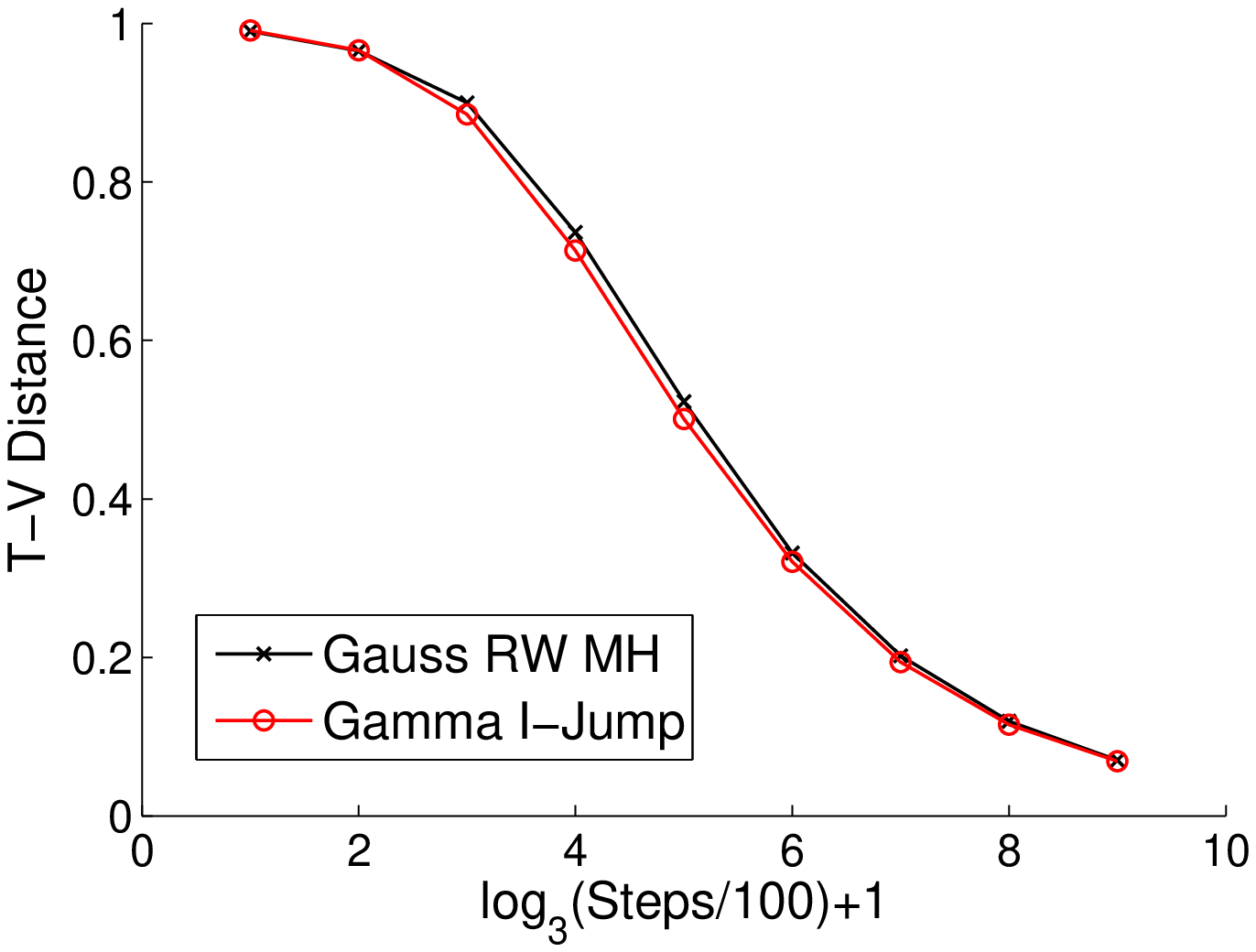}
\\
\includegraphics[height=0.17\textheight]{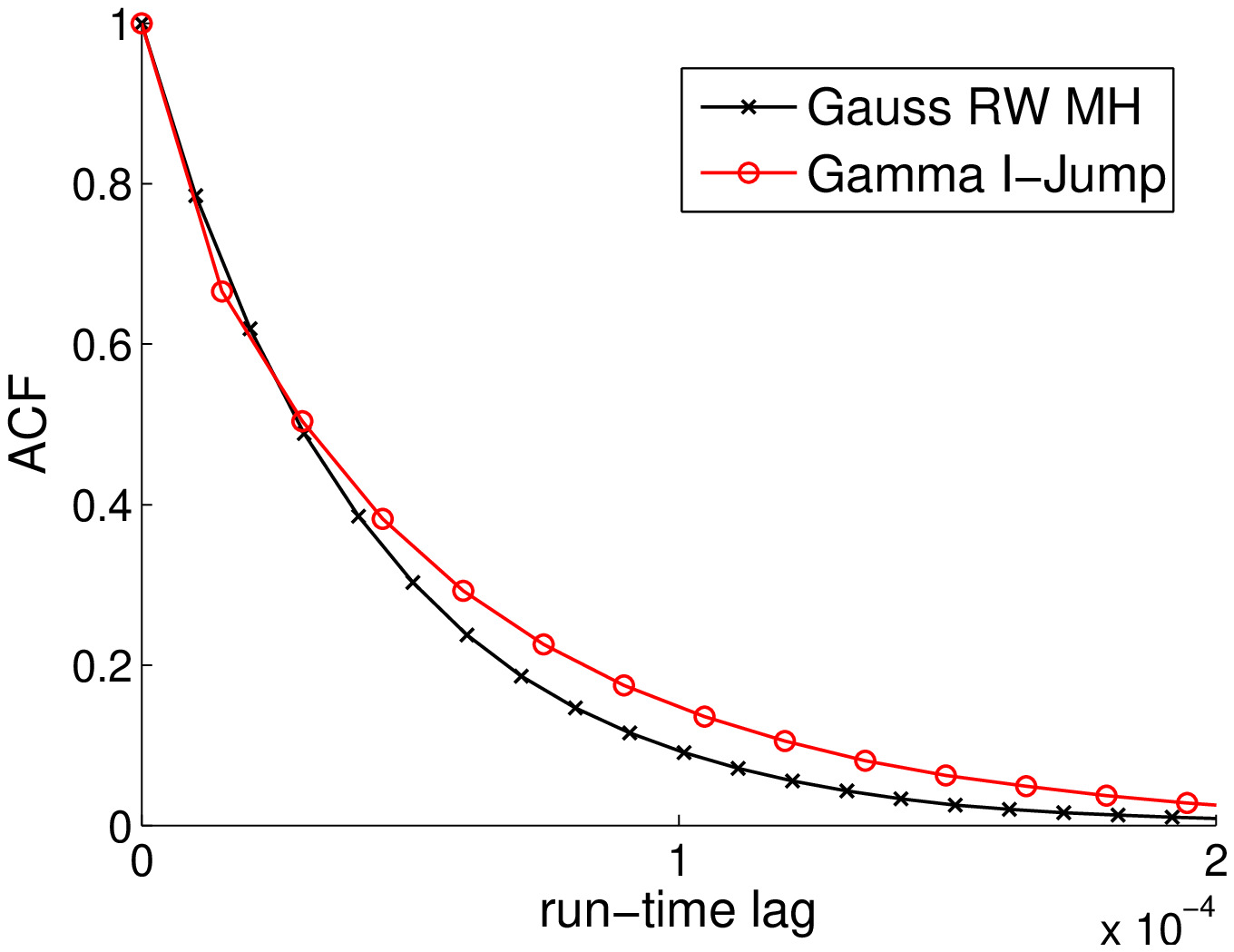}\hspace{-0.2in}
\includegraphics[height=0.17\textheight]{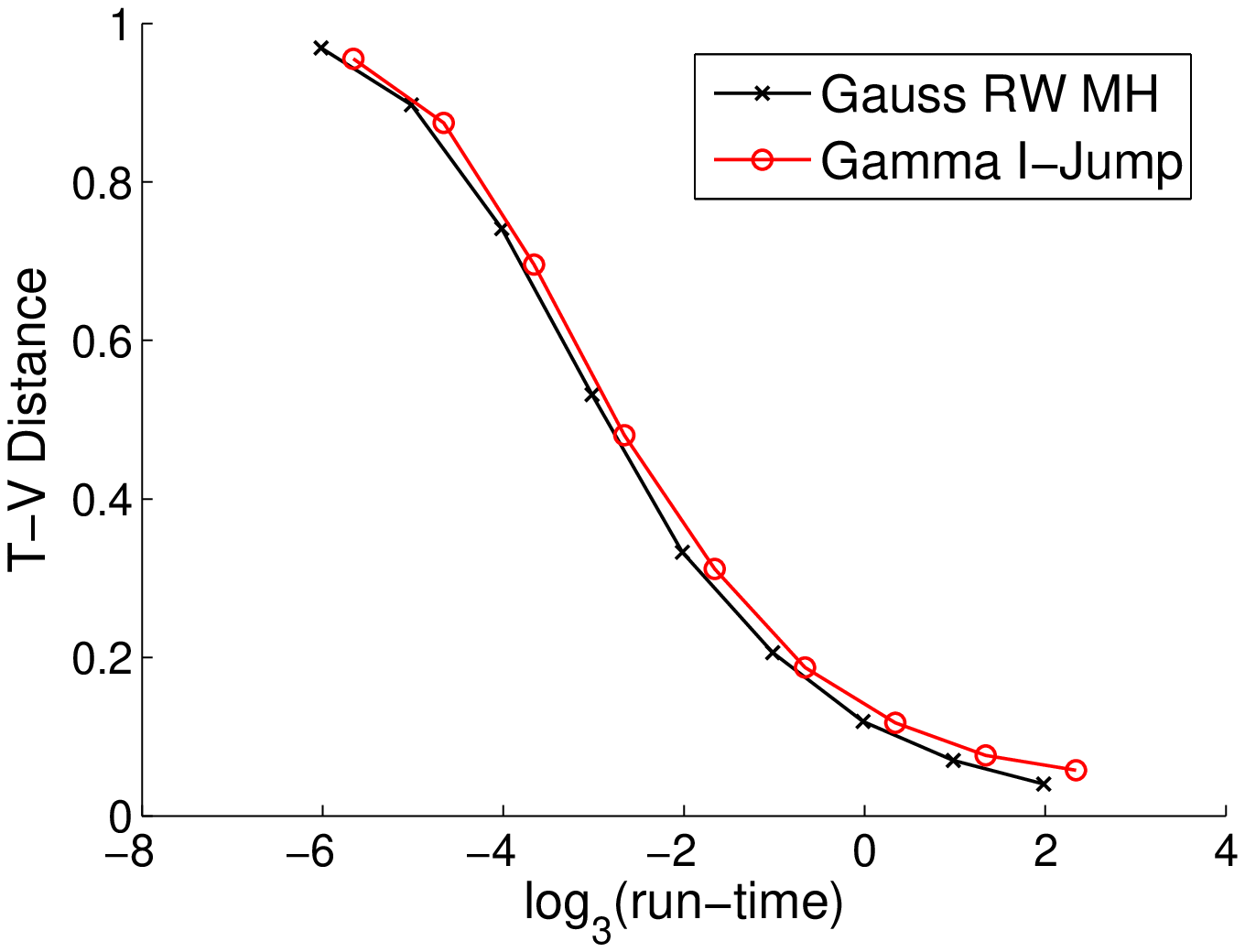}\hspace{-0.2in}
\includegraphics[height=0.17\textheight]{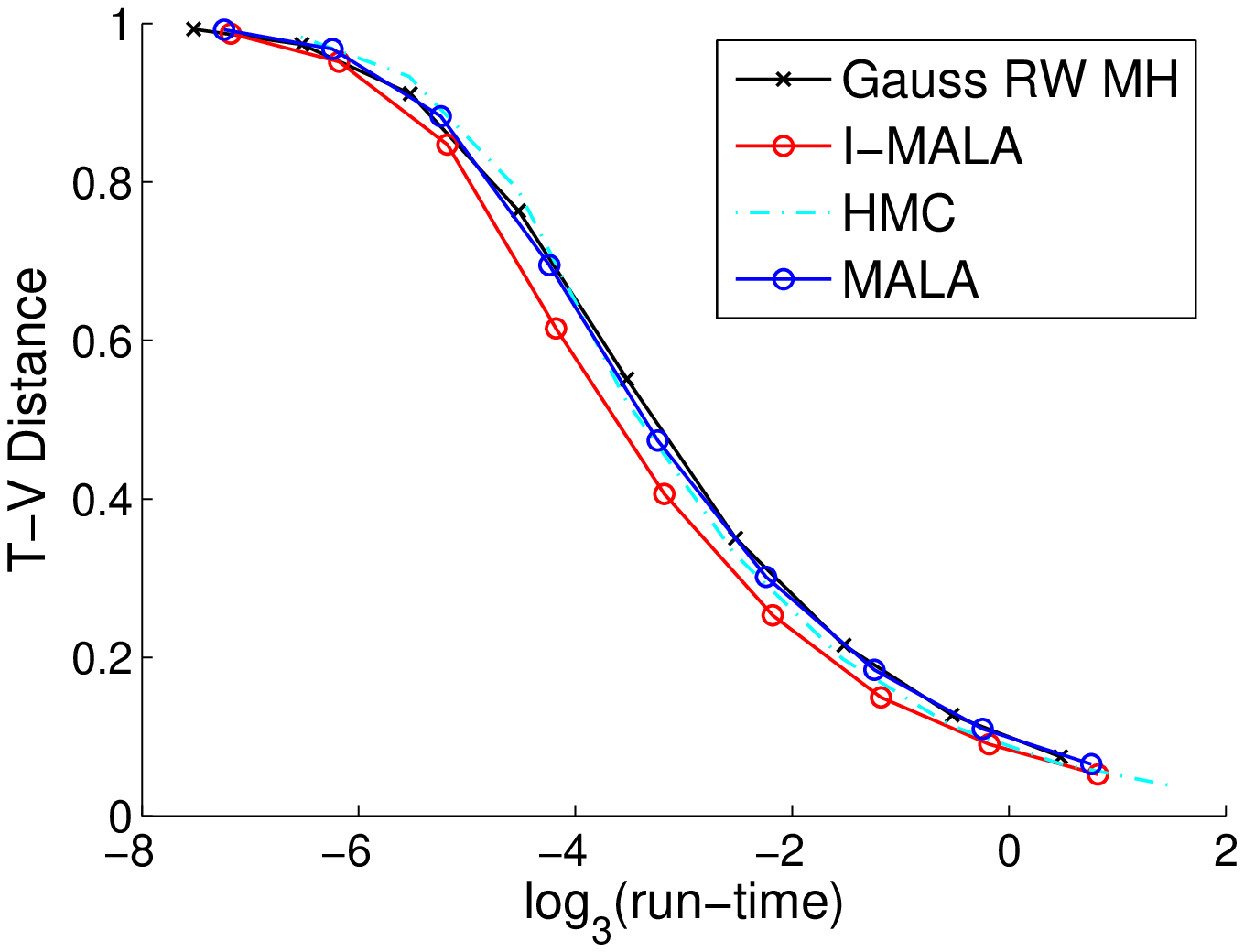}
\caption{
\emph{Top row:} Correlated distribution with complex geometry in 2D, $\pi(z_1,z_2) = z_1^4/10 + (4(z_2+1.2) - z_1^2)^2/2$ (left), ACF vs. lag in steps (middle), and T-V distance vs. log steps of Gamma I-Jump against Gauss RW MH (right).
\emph{Bottom row:}  ACF vs. lag in runtime of Gamma I-Jump against Gauss RW MH (left), T-V distance vs. log runtime between Gauss RW MH and Gamma I-Jump (middle) and Gauss RW MH, I-MALA, HMC, and MALA (right).
}
\label{fig:2D_MoonJump}
\end{figure*}

We now test the correctness and attributes of our algorithm on a highly correlated (moon-shaped) target distribution, where $\pi(z_1,z_2) = z_1^4/10 + (4(z_2+1.2) - z_1^2)^2/2$.
In terms of number of iterations, the I-Jump sampler decorrelates and converges to the target distribution faster.
However, in terms of runtime, the I-Jump sampler does not perform as well as random walk MH, as explored in Fig.~\ref{fig:2D_MoonJump}.
The reason is that the correlated distribution has a complex geometry.
Faster exploration in random directions, as provided by our I-Jump sampler with independent proposals, only marginally increases the mixing effect in each step relative to the reversible independent proposals of MH.
Since the calculation of the distribution is not demanding in this case, the small overhead of the irreversible sampler 
(generating gamma proposals and periodically resampling the direction of exploration)
actually makes a difference and results in slightly worse performance in terms of runtime.
%

To improve the performance of our I-Jump sampler further in this correlated target case, it would be appealing to take the geometric information about the level sets---including the higher mass regions---into account.
Indeed, we are able to do this by replacing the independent gamma proposals with proposals from our continuous dynamics sampler---our I-MALA algorithm (Algorithm \ref{alg:Irr_MALA})---as described in Sec.~\ref{sec:combined}.
To demonstrate the effect of irreversibility, we choose $\mD(\vz) = \left(
\begin{array}{ll}
         1 & 0 \\
         0 & 1
\end{array}
\right)$ and $\mQ(\vz) = \left(
\begin{array}{ll}
         0 & -1 \\
         1 & 0
\end{array}
\right)$
in Eqs.~\eqref{Eq:forwardSDE}, \eqref{Eq:adjointSDE}, \eqref{Eq:Prob_forwardSDE}, and \eqref{Eq:Prob_backwardSDE}.
In this case, I-MALA significantly outperforms random walk MH.
Because the target distribution has complex geometry, the continuous dynamics can provide guidance on locating the higher mass regions and exploring the contours rapidly with the gradient information.
HMC and MALA also exploit this effect, so we provide comparisons to these methods as well.
From these comparisons, we see the benefits of the irreversibility itself of the I-MALA algorithm.

This experiment demonstrates the gains that are possible by combining our continuous dynamics and jump process frameworks, beyond what either can provide individually.
We explore this I-MALA algorithm more extensively on a set of real-world scenarios in the following section.


\subsection{Experiments for I-MALA}
\label{sec:exp_IMALA}
We test our I-MALA algorithm in sampling both Bayesian logistic regression and stochastic volatility models.
In these experiments, we compare to numerous baselines.
The comparison with I-Jump allow us to disentangle the importance of our SDE proposal in place of independent proposals.
We used half-space Gaussian proposals since it is guaranteed to only increase the mixing of random walk MH.
The comparison with MALA provides insights into the importance of an irreversible accept-reject correction to underlying irreversible SDE proposals.
The vanilla random walk MH and HMC algorithms provide other popular choices to serve as baselines.
We also provide comparison with some of the recently proposed irreversible samplers like Zig-Zag, SOL-HMC, and hybrid MALA algorithms.
\subsubsection{Bayesian Logistic Regression}
\label{sec:BayesLogisticRegression}
In this section, we demonstrate results from sampling a Bayesian logistic regression model.
Similar to the setting in \cite{RiemannianMALA}, we consider an $N \times D$ design matrix ${\bf X}$ comprised of $N$ samples each with $D$ covariates and a binary response variable $\vy\in\{0,1\}^{N}$.
If we denote the logistic link function by $s(\cdot)$, a Bayesian logistic regression model of the binary response is obtained by introducing regression coefficients $\beta\in\mathbb{R}^D$ with an appropriate prior \cite{Gelman,JunLiu}; for illustration we take $\beta\sim\mathcal{N}(\vzero,\alpha \mI)$, where $\alpha=100$ in the experiments, employing a wide Gaussian prior.

We make use of three datasets available at the STATLOG project: https://archive.ics.uci.edu/ml/machine-learning-databases/statlog/.
The first two datasets describe the connections between credit card approval and various attributes of the applicants in Australia and Germany.
The third dataset is about the connection between the absence or presence of heart disease and various patient-specific covariates.

\begin{table*}
  \centering
  \begin{tabular}{ c | c | c | c }
Sampler $\setminus$ Dataset & Australian Credit ($D=15$) & German Credit ($D=25$) & Heart ($D=14$)\\
\hline
MH & $4.30$, \;$4.29$ & $2.82$, \;$2.51$ & $22.92$, \;$10.76$\\
\hline
I-Jump & $4.93$, \;$5.29$ & $2.94$, \;$2.93$ & $24.32$, \;$12.35$\\
\hline
Zig-Zag & $5.24$, \;$5.65$ & $3.09$, \;$3.07$ & $25.82$, \;$11.68$\\
\hline
MALA & $9.05$, \;$5.17$ & $3.67$, \;$3.65$ & $29.26$, \;$13.29$\\
\hline
HMC & $10.96$, \;$4.36$ & $3.73$, \;$2.89$ & $30.83$, \;$18.04$\\
\hline
SOL-HMC & $13.29$, \;$8.40$ & $3.75$, \;$2.92$ & $32.71$, \;$25.83$\\
\hline
Hybrid MALA & $11.01$, \;$5.24$ & $4.08$, \;$3.63$ & $39.54$, \;$26.69$\\
\hline
I-MALA & $\bf{15.95}$, \;$\bf{8.86}$ & $\bf{4.47}$, \;$\bf{4.03}$ & $\bf{41.23}$, \;$\bf{27.27}$\\
\end{tabular}
  \caption{Comparison of $\widehat{ESS}$ per second of runtime ({\it Left}: using $\widehat{ESS}^{BW}$ of \eqref{eq:ess} and {\it Right}: using $\widehat{ESS}^{MBM}$ of \eqref{eq:ess_batch}) for various samplers on three datasets described in Sec.~\ref{sec:BayesLogisticRegression}.
  }\label{table:Bayes_logistic}
\end{table*}
\paragraph{Performance metric} We measure performance in terms of the per second effective sample size (ESS/runtime),
where ESS is calculated as the number of steps $N$ divided by the integrated autocorrelation time $\tau_{int}$:
$ESS = \dfrac{N}{\tau_{int}}$ \cite{JunLiu,XiAn,Andrieu2008,Neal:1996}.
In \cite{RiemannianMALA}, $\tau_{int}$ is estimated through the initial positive sequence estimator: ${\tau_{int}} = {1 + 2 \sum_k \gamma(k)}$, where $\gamma(k)$ is the $k$-lagged autocorrelations and the sum is over the $K$ monotone sample autocorrelations \cite{Geyer}.
The initial positive sequence estimator assumes the Markov chain is reversible.
Hence to include irreversible chains, we use the \emph{Bartlett window estimator} for the integrated autocorrelation time $\tau_{int}$ \cite{Priestley,Geyer,Bartlett} so that:
\begin{equation}
\widehat{ESS}^{BW}
= \dfrac{N}{1 + 2 \sum_{k=1}^M \left(1-\dfrac{k}{M}\right) \gamma(k)}, \label{eq:ess}
\end{equation}
where $M$ is a large number (taken to be $3000$ in the experiments).
We also use a more robust estimator, the \emph{multivariate batch mean estimator} \cite{Mult_BatchMean,Consist_BatchMean}, to estimate the effective sample size:
\begin{equation}
\widehat{ESS}^{MBM} = K \left(\dfrac{|\widehat{\Lambda}|}{|\widehat{\Sigma}|}\right)^D, \label{eq:ess_batch}
\end{equation}
where $K=\sqrt{N}$ is the number of batches, $\widehat{\Sigma}$ is the estimated covariance of generated samples, $\widehat{\Lambda}$ is the estimated covariance of sample batch means, and $|\cdot|$ denotes determinant of the covariance matrices.
(When the dimension $D$ is very large, e.g., $D=2000$ in the stochastic volatility model, we calculate $\widehat{ESS}^{MBM}$ as the median of the batch mean estimator over each dimension to circumvent the rank deficiency issue in $\widehat{\Lambda}$).


\paragraph{Optimal hyperparameters} We select hyperparameters for each method via a grid search.
For MH, we corroborate that the selected hyperparameters are indeed those obtained by tuning the acceptance rate between $20\%$ and $40\%$ at stationary.
We also compare to the Zig-Zag sampler (without subsampling).  
One option is to use a lower bound on the Hessian of the negative log posterior (see Section \ref{sec:relatedwork}), which in this case is $\dfrac14 {\bf X}{\bf X}^T + \alpha^{-1} \mI$.  
However, as discussed in the experimental results below, we found this bound to be too loose to be of use.  
Instead, we combine the bound on the gradient of the log likelihood (a constant term) with the Poisson rate for the Gaussian prior, which results in a tighter bound on the Poisson rate of the overall posterior.
To derive the Poisson rate function bound, we used the recently proposed idea of superposition of Poisson rates for decomposable posteriors \cite{BouncyParticle}.
For the I-Jump sampler, we find that taking
the same parameters as in MH already generates better performance than MH.
Taking a slightly smaller variance in the half-space Gaussian proposal leads to even higher $\widehat{ESS}$ per second with an acceptance rate between $30\%$ and $50\%$.

For HMC, we find that using $10$ leapfrog steps to generate a sample is most efficient in terms of $\widehat{ESS}$ per second (as opposed to the commonly used $50$ to $100$ steps).
The acceptance rate is around $90\%$.
For MALA, an acceptance rate between $40\%$ and $60\%$ generates the best $\widehat{ESS}$ per second.
For SOL-HMC and hybrid MALA, we use the optimal hyperparameters found for HMC and MALA to center our grid search.
It seems that the hybrid MALA is more sensitive to the hyperparameters when numerical stability is concerned, possibly because of the combination of simulating Lagenvin and Hamiltonian dynamics.
For the I-MALA algorithm, we take $\mD = \mI$,
and
$\mQ(\vz) = \left(
\begin{array}{ll}
         0 & -\mI_{d\times d} \\
         \mI_{d\times d} & 0
\end{array}
\right)$
where $d = \lfloor (D + 1) / 2 \rfloor$, just as in Sec.~\ref{sec:exp_corr} to combine benefits of Langevin diffusion and Hamiltonian dynamics.
The acceptance rates are between $40\%$ and $60\%$.

\paragraph{Experimental results} Our results are summarized in Table~\ref{table:Bayes_logistic}.
We see that I-Jump provides a gain over random walk MH in all datasets.
The Zig-Zag sampler outperforms both MH and I-Jump.  
However, it is important to note that having a tight bound for the transition rate function---as we were able to derive in this case---is critical to this observed performance.
When using a looser bound (in this case, the bound on the Hessian of the negative log posterior), multiple steps are taken before a change of direction of exploration can happen.  
In practice, we found this lead to performance below that of MH and I-Jump.

HMC provides even better performance, improving over MALA and the I-Jump sampler on these examples.
The irreversible variants, SOL-HMC and hybrid MALA, further improve upon HMC, achieving the highest $\widehat{ESS}$ \emph{per sample} generated.
However, the algorithms are slower per iteration than our proposed I-MALA.
As a result, as seen in Table~\ref{table:Bayes_logistic}, I-MALA has by far the best performance across all datasets.
The I-MALA algorithm combines the benefits of HMC's traversing and MALA's diffusion, plus our previously demonstrated benefits of irreversibility.

\subsubsection{Stochastic Volatility Model}
\label{sec:stoch_vol}
In this section, we follow \cite{RiemannianMALA,JunLiu,Stoch_Vol} to study a stochastic volatility model.
The daily returns $y_t$ are modeled as $y_t = \epsilon_t \beta \exp(x_t/2)$, where $\epsilon_t\sim\mathcal{N}(0,1)$ and the latent volatilities $x_t$ follow an order-$1$ autoregressive process $x_{t+1} = \phi x_t + \eta_{t+1}$ with $x_1\sim\mathcal{N}(0,\sigma^2/(1-\phi^2))$.
The joint probability is given by
\begin{align}
p(\vy,\vx,\beta,\phi,\sigma) = &\prod_{t=1}^T p(y_t|x_t,\beta) p(x_1) \pi(\beta) \nonumber\\ \nonumber
& \prod_{t=2}^T p(x_t|x_{t-1},\phi,\sigma) \pi(\phi) \pi(\sigma) ,
\end{align}
\sloppy with priors on the parameters chosen to be $\pi(\beta)\propto1/\beta$, $\sigma^2\sim{\rm Inv}$-$\chi^2(10,0.05)$, and
$(\phi+1)/2\sim{\rm Beta}(20,1.5)$.
We transform the constrained parameters $\phi$ and $\sigma$ to the real line as $\sigma=\exp(\gamma)$ and $\phi=\tanh(\alpha)$, taking into account the Jacobian of the transformation.
We iteratively sample over latent volatilities $\vx$ and parameters $(\beta, \phi, \sigma)$ within a Gibbs sampling procedure (see \cite{RiemannianMALA} for further details).

\begin{table*}
  \centering
  \begin{tabular}{ c | c | c }
Sampler $\setminus$ Variables & Latent Variables ($d=2000$) & Parameters ($d=3$)\\
\hline
MH & $1.33$, \;$3.24$ & $3.86$, \;$5.17$\\
\hline
I-Jump & $1.51$, \;$3.38$ & $4.15$, \;$6.12$\\
\hline
MALA & $2.41$, \;$3.67$ & $11.72$, \;$9.00$\\
\hline
HMC & $7.59$, \;$6.21$ & $0.96$, \;$1.18$\\
\hline
SOL-HMC & $6.80$, \;$6.92$ & $2.06$, \;$2.71$\\
\hline
Hybrid MALA & $5.65$, \;$5.99$ & $7.40$, \;$9.83$\\
\hline
I-MALA & $\bf{7.71}$, \;$\bf{7.64}$ & $\bf{18.27}$, \;$\bf{13.58}$\\
\end{tabular}
  \caption{Comparison of $\widehat{ESS}$ per second of runtime ({\it Left}: using $\widehat{ESS}^{BW}$ of \eqref{eq:ess} and {\it Right}: using $\widehat{ESS}^{MBM}$ of \eqref{eq:ess_batch}) for various algorithms sampling latent variables and parameters of a stochastic volatility model described in Sec.~\ref{sec:stoch_vol}.
  }\label{table:stoch_vol}
\end{table*}

\paragraph{Hyperparameters} For MH, I-Jump and MALA, the stationary acceptance rates are similar to ones verified in Sec.~\ref{sec:BayesLogisticRegression} (between $20\%$ and $40\%$, $30\%$ and $50\%$, and $40\%$ and $60\%$, respectively).
For HMC, the leapfrog steps decorrelate quickly when sampling the latent variables, but are heavily correlated when sampling the model parameters.
Hence we take $6$ leapfrog steps for the former and $10$ steps for the latter (more leapfrog steps can help HMC decorrelate, but increases runtime).
Step sizes are then tuned to have acceptance rates for HMC between $80\%$ and $90\%$.

For I-MALA, we expand the space as in the HMC algorithm and take $\mD(\vz) = \left(
\begin{array}{cc}
         a^{-1} \mI_{d\times d} & 0 \\
         0 & a \mI_{d\times d}
\end{array}
\right)$
and $\mQ(\vz) = a\left(
\begin{array}{cc}
         0 & - \mI_{d\times d} \\
         \mI_{d\times d} & 0
\end{array}
\right)$, where $d=2000$ for the latent variables and $3$ for model parameters.
The motivation for this choice is as follows.
When $a$ is larger, the dynamics of I-MALA are similar to HMC; when $a$ is smaller, I-MALA becomes closer to MALA.
Thus, the form of $\mD(\vz)$, $\mQ(\vz)$ specified above allows us to combine the benefits of HMC and MALA.
Here, we leverage intuition from statistical mechanics:
high-dimensional variables that are only pairwise correlated move approximately according to Newtonian mechanics perfectly described by the Hamiltonian dynamics while the low-dimensional parameters are similar to the summary quantities (or collective variables) and intrinsically follow stochastic Langevin dynamics.
As such, we choose $a$ to give HMC-like behavior for the latent state sequence sampling and MALA-like behavior for the parameter sampling
by setting $a=20$ for latent variables and $a=5$ for model parameters.

\begin{figure*}[!t]
\centering
\includegraphics[height=0.18\textheight]{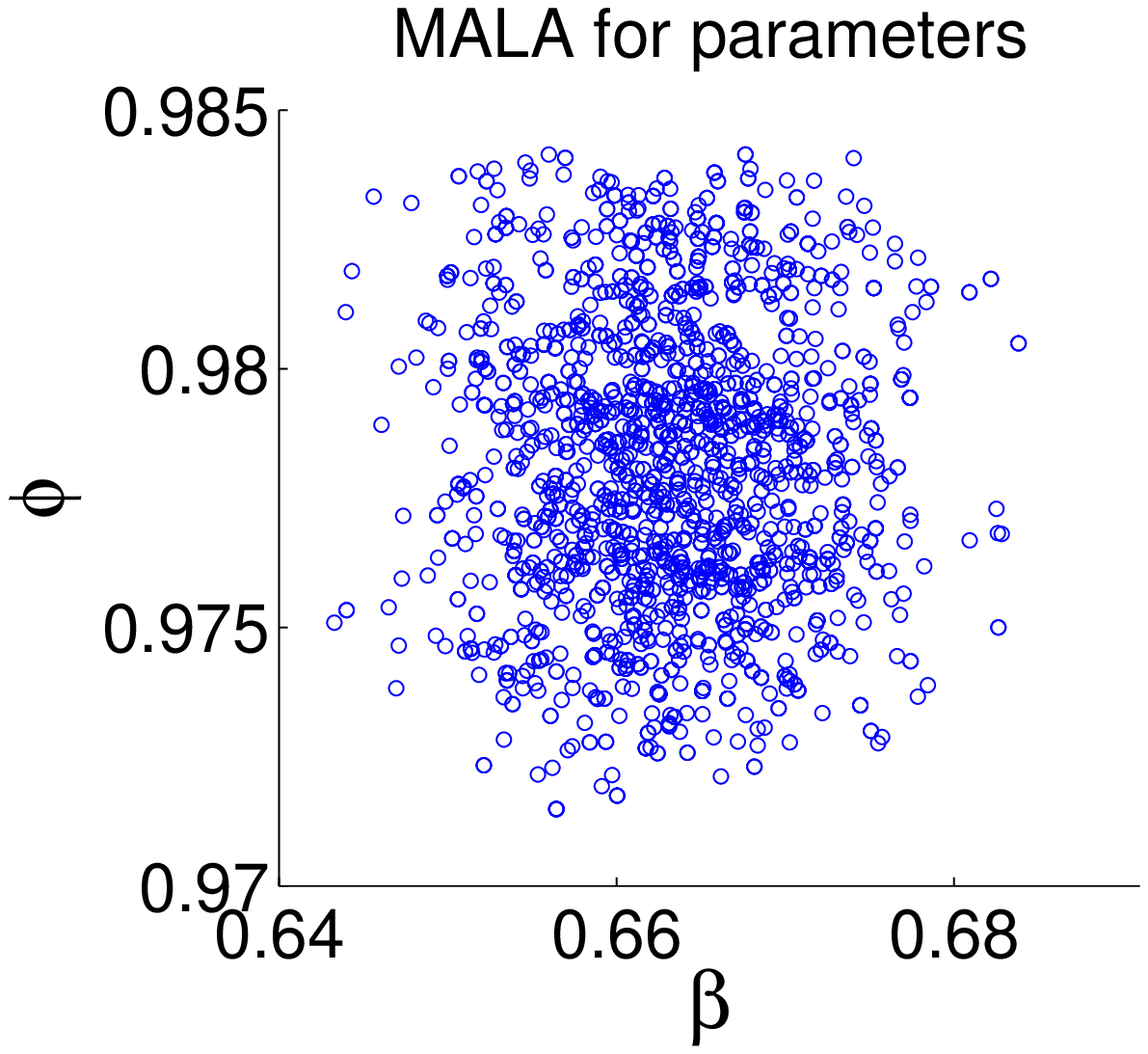}
\includegraphics[height=0.18\textheight]{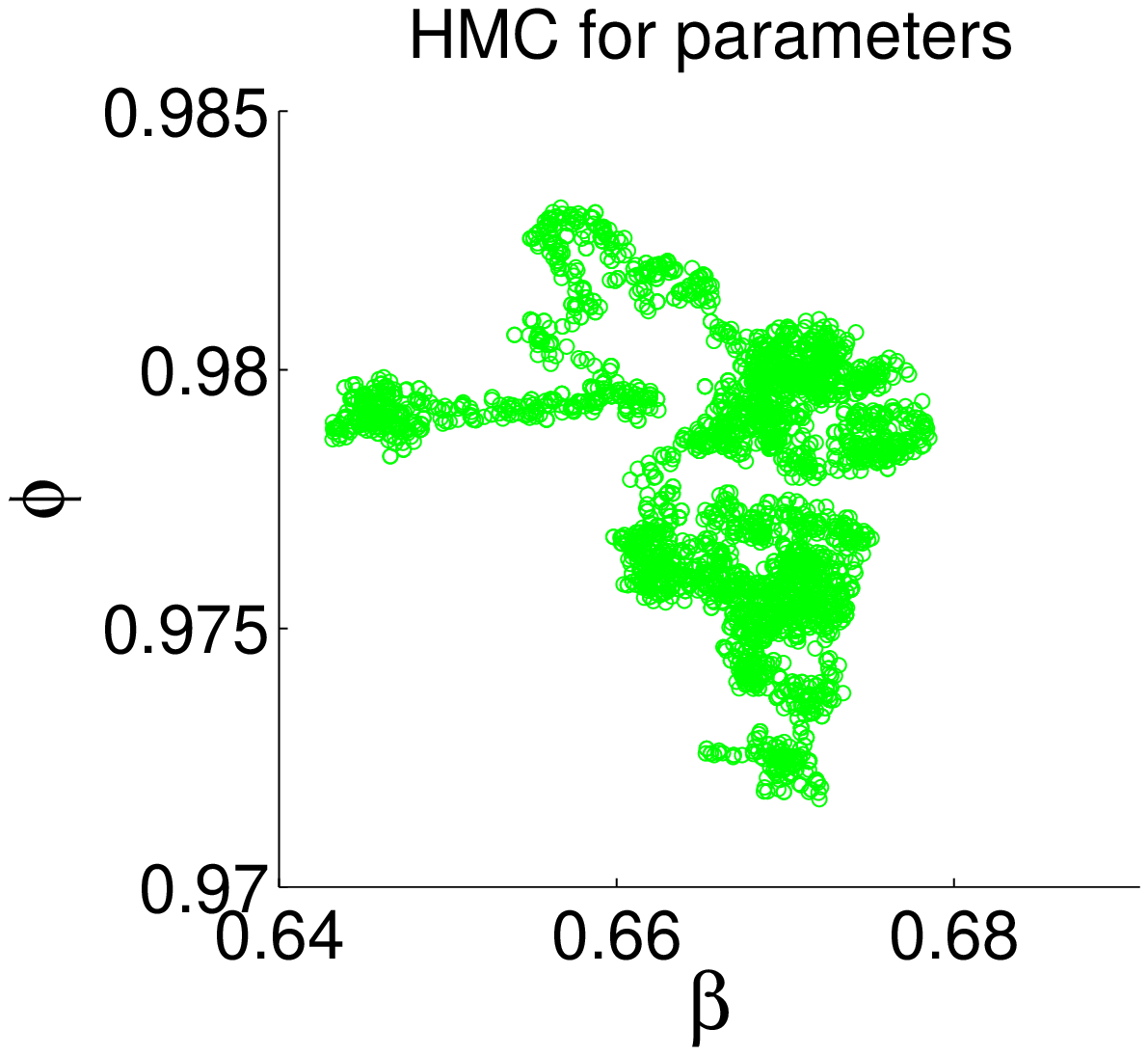}
\includegraphics[height=0.18\textheight]{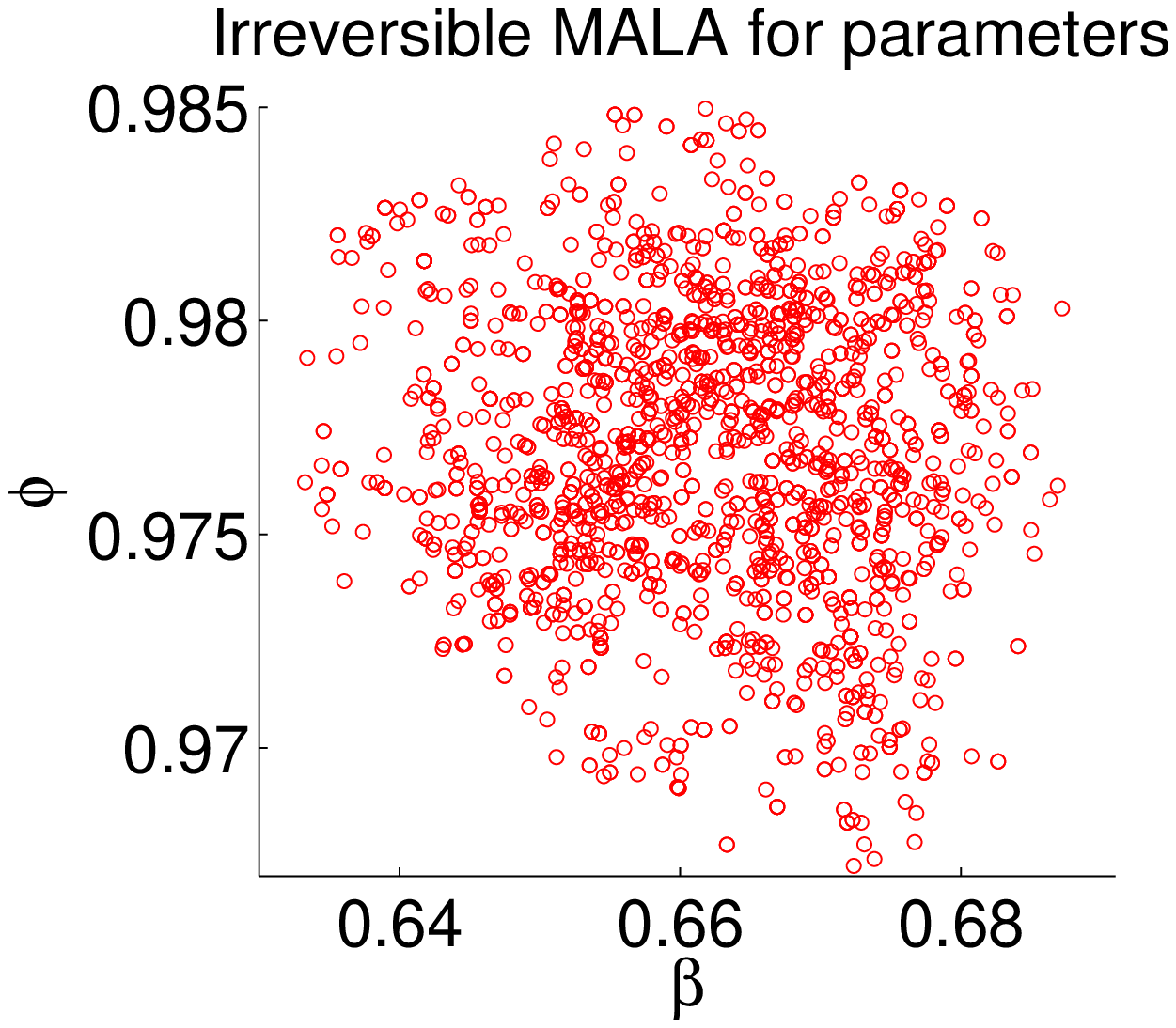}\\
\includegraphics[height=0.18\textheight]{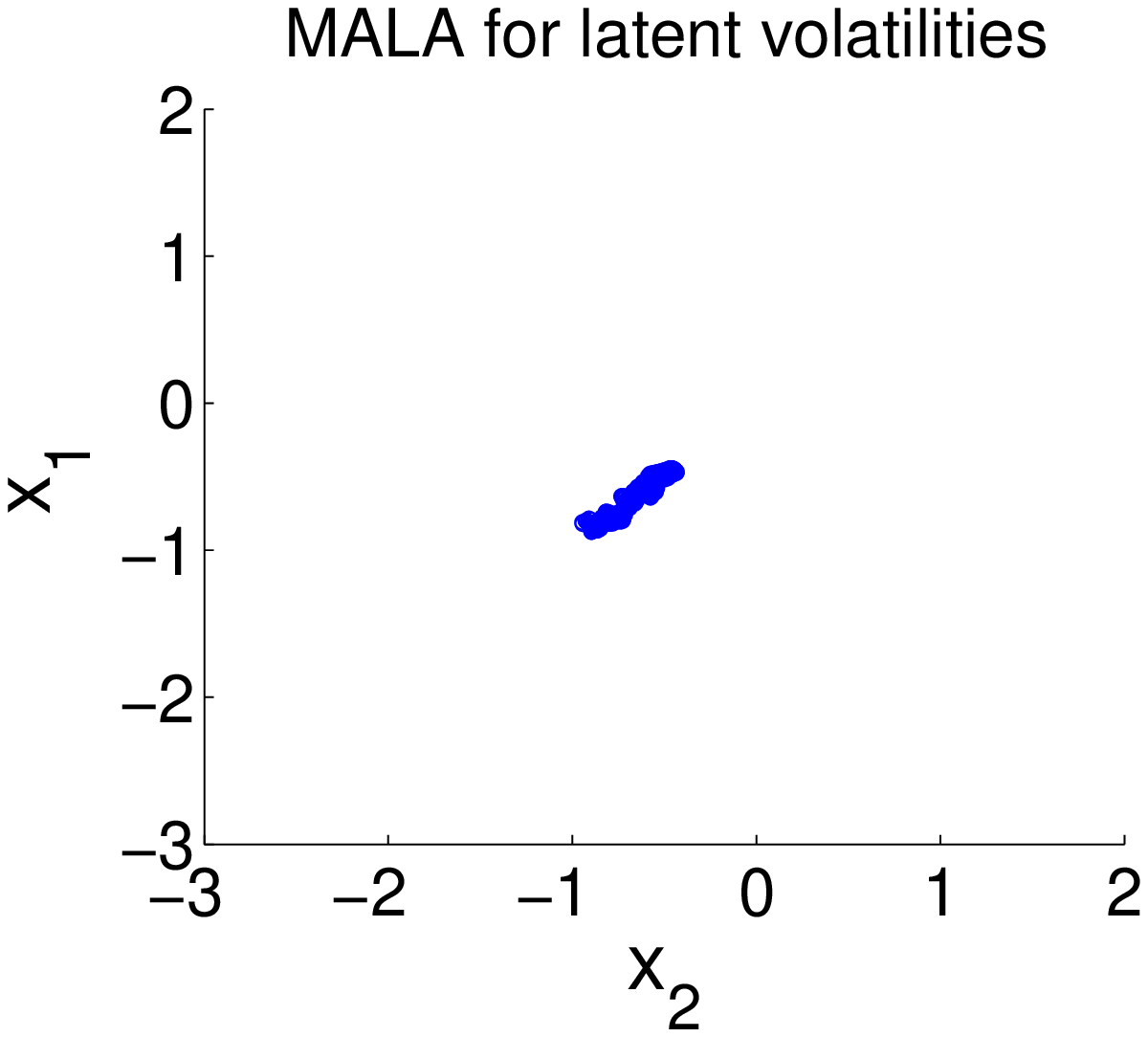}
\includegraphics[height=0.18\textheight]{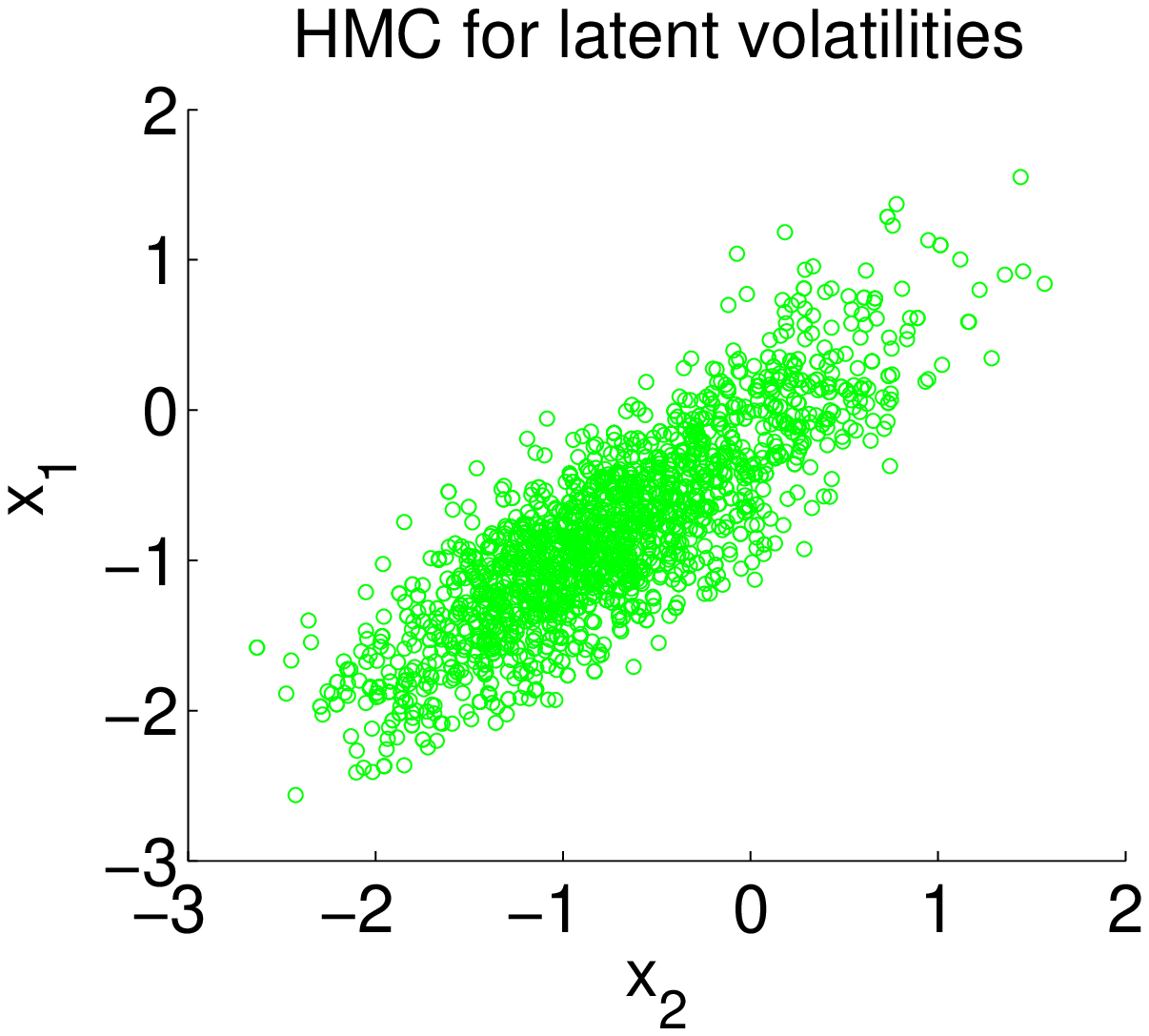}
\includegraphics[height=0.18\textheight]{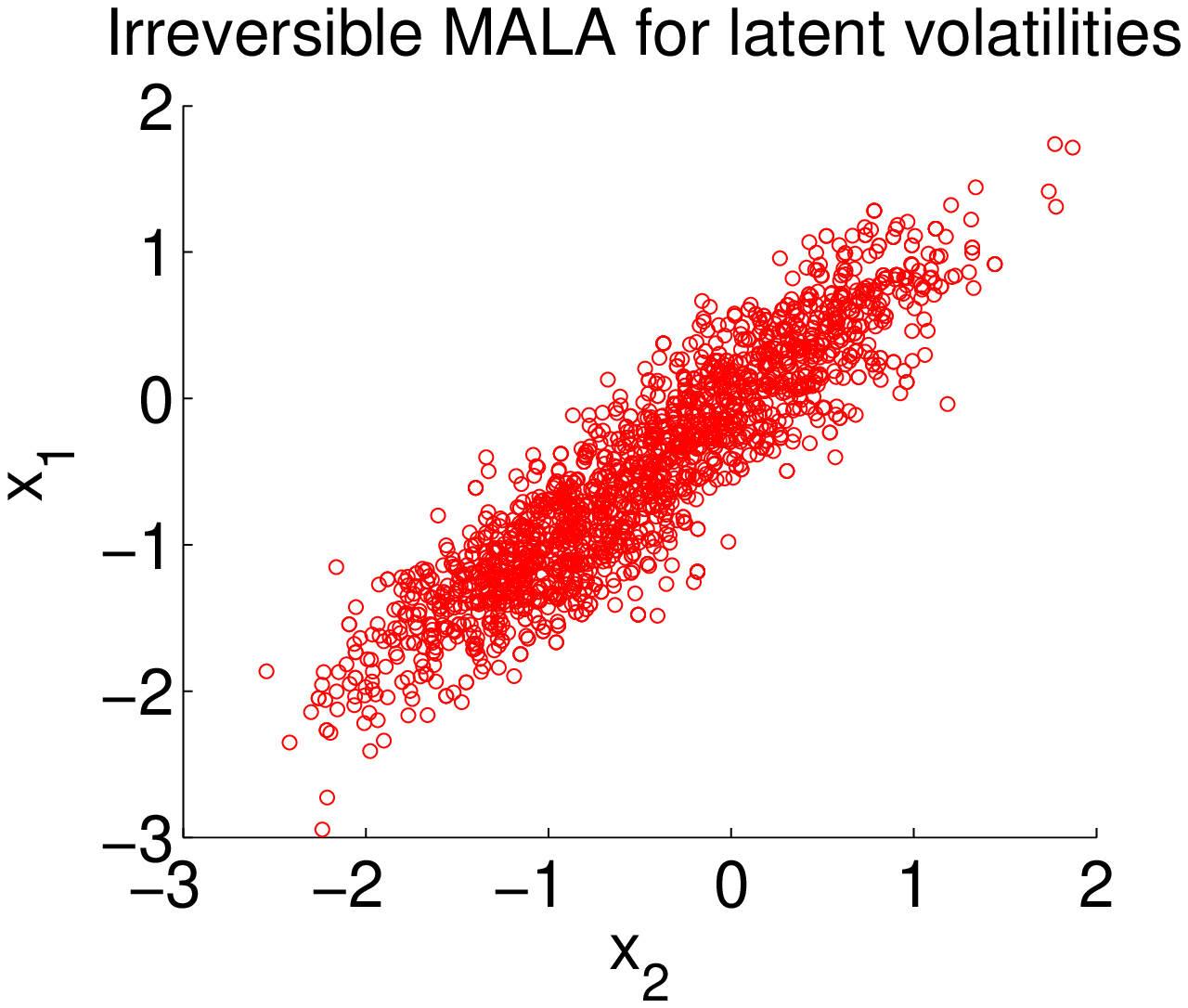}
\caption{
$2,000$ samples of MALA, HMC, and I-MALA for parameters and latent volatilities.
}
\label{fig:Stoch_Vol_trace}
\end{figure*}

\paragraph{Experimental results} The results are summarized in Table~\ref{table:stoch_vol}.
Again the I-Jump sampler improves upon MH.
Furthermore, in this model, MALA provides the best baseline in terms of sampling the parameter space while HMC excels at sampling the latent variables.
This behavior, further explored in Fig.~\ref{fig:Stoch_Vol_trace}, follows the pattern previously described by the statistical mechanics intuition.
Hence, HMC provides faster traversing in the easy-to-explore high-dimensional latent variable space while MALA helps to diffuse faster in the lower-dimensional parameter space that exhibits complex dependencies.

SOL-HMC seems to be able to mitigate the slow mixing in model parameters.
Likewise, hybrid MALA seems to be able to combine some of the benefits from HMC and MALA.
Again, however, the I-MALA algorithm combines the benefits of HMC and MALA and exceeds them to provide the best overall performance.
It appears from the experiments that in high dimensions, the vanilla I-Jump sampler with half-space Gaussian proposal has less improvements over MH than in low dimensional cases.
We further verify this observation by the experimental results in Appendix \ref{appnd_dimension} on a standard normal distribution with doubling dimensions.
However, the benefits of I-MALA over reversible MALA does not seem to have strong correlation with the dimension of the space, thus corroborating with our intuition that a direction of efficient exploration is important for the irreversible samplers. 
This is demonstrated in Table \ref{table:stoch_vol}, where we notice that the improvements of I-MALA over MALA are still pronounced when going from $d=3$ to $d=2000$. 
\section{Conclusion}
\label{sec:conclusion}
In this paper, we primarily focused on developing an irreversible MALA (I-MALA) algorithm (Algorithm~\ref{alg:Irr_MALA}).
The construction of the I-MALA algorithm is comprised of two components: irreversible continuous dynamics as the proposal distribution, and an irreversible jump sampler (I-Jump) to correct for any numerical error in simulating the continuous dynamics.

Building on our preliminary work \cite{completesample}, we first provided a framework for specifying irreversible continuous dynamics defined via a positive semidefinite matrix $\mD$ and a skew-symmetric matrix $\mQ$.
We prove that any choice of these two matrices leaves the target distribution invariant, and any continuous Markov process with the correct stationary distribution has a representation in our formulation with a certain $\mD$ and $\mQ$.

We then designed the I-Jump sampler, which in itself serves as a general purpose MCMC algorithm and excels at various sampling tasks over the MH algorithm.
We achieved this irreversibility by simply switching between different proposal and acceptance distributions upon rejection.
The algorithm is nearly identical structurally and computationally to MH, and thus serves as a simple replacement.

Using these irreversible continuous dynamics as a proposal distribution and the I-Jump sampler to accept or reject the proposal, we arrive at our I-MALA algorithm.
We demonstrated the benefits of I-MALA in sampling Bayesian logistic regression and stochastic volatility models on various real-world and synthetic datasets.

When computing target distributions that represent the posterior distribution of large datasets, scalability of the sampling algorithms becomes important.
To cope with this scenario, stochastic gradient (SG) counterparts to the continuous dynamical samplers have been proposed (see \cite{completesample} for more discussion).
However, these SG-MCMC methods rely on the stepsize to asymptotically approach zero to converge to the correct target distribution.
Otherwise a bias is introduced.
One could use our proposed I-Jump sampler to correct this bias, but as with standard MH, the accept/reject calculations involve touching the whole dataset, exactly what the SG methods seek to avoid.
Instead, we could imagine extending the MH data-subsampling ideas of \cite{Korattikara:2013austerity,Bardenet,TallData} to our I-Jump sampler to decrease the computational burden of the accept-reject step.

Another direction for future work is to examine the effect of incorporating second order information about the target distribution (similar to RMALA) into our I-MALA.
It is well-studied that incorporating the Fisher information metric in $\mD$ and $\mM$ can speed up mixing rates in MALA and HMC \cite{RMALA,RiemannianMALA,Girolami_Geometry}; the benefits can be explained by the fact that using an approximation of the Hessian of the target distribution to precondition the gradient makes the space locally ``flat".
However, in our case there are many ways in which such second order information can be incorporated in $\mD$ and $\mQ$.
Exploring these options and the interplay between $\mD$ and $\mQ$ is a direction of future work.

\section*{Acknowledgments}
This work was supported in part by ONR Grant N00014-15-1-2380, NSF CAREER Award IIS-1350133, and the TerraSwarm Research Center sponsored by MARCO and DARPA.
We thank Samuel Livingstone, Paul Fearnhead, Galin Jones, Hong Qian and Michael I. Jordan for helpful suggestions and discussions.
Y.-A. Ma would like to thank Sebastian J. Vollmer for directing him to the reference \cite{stochastic_fluid_dynamics}.
We also thank the reviewers for their thoughtful comments and suggestions.

\newpage
\appendix

\section{Samplers from Continuous Markov Processes}

\subsection{Proof of Theorem \ref{thm:complete} (existence of $\mQ(\vz)$ in \eqref{Eq:FPE})}
\label{appnd_CompleteProof}

Proof of Theorem~\ref{thm:complete} is comprised of two sets of ideas appearing in different fields.
Here, we write the proof in two steps accordingly.
\begin{enumerate}
\item
Plug the stationary solution $\pi(\vz)$ into \eqref{Eq:ItoFPE} and observe that
\[
\sum_i \dfrac{\partial}{\partial \vz_i} \left\{ \sum_j\dfrac{\partial}{\partial \vz_j}\Big(\mD_{ij}(\vz)\pi(\vz) \Big) - \vf_i\pi(\vz) \right\} = 0.
\]
\item
Constructively prove that if entries of
\[
\left\{ \sum_j\dfrac{\partial}{\partial \vz_j}\Big(\mD_{ij}(\vz)\pi(\vz) \Big) - \vf_i\pi(\vz) \right\}
\]
belong to $L^1(\mathbb{R}^d)$, then there exists a matrix $\mQ(\vz)$ with entries in $W^{1,1}(\pi)$, such that $\sum_{j} \dfrac{\partial}{\partial \vz_j} \Big( \mQ_{ij}(\vz) \pi(\vz) \Big)
= \vf_i\pi(\vz) - \sum_j\dfrac{\partial}{\partial \vz_j}\Big(\mD_{ij}(\vz)\pi(\vz) \Big)$.
\end{enumerate}
Here we denote $L^1(\mathbb{R}^d)$ as the space of Lebesgue integrable functions on $\mathbb{R}^d$, and $W^{1,1}(\pi)$ as the Sobolev space of functions with weak derivatives and function values integrable with respect to the density $\pi$ times the Lebesgue measure.
Step 1 can be found in literatures of continuous Markov processes \cite{qian-jsp,Dembo,qian_jmp,large_deviations,Villani,PavBook}.
Step 2 has been found in earlier works on stochastic models in fluid dynamics and homogenization \cite{stochastic_fluid_dynamics}.

Step 1 is accomplished by noting that the un-normalized density function $\pi(\vz)\propto p^s(\vz)$ is also a stationary solution of \eqref{Eq:ItoFPE}.
Hence,
\begin{equation}
\sum_i \dfrac{\partial}{\partial \vz_i} \left\{ \sum_j\dfrac{\partial}{\partial \vz_j}\Big(\mD_{ij}(\vz)\pi(\vz) \Big) - \vf_i\pi(\vz) \right\} = 0.
\label{eq:appnd_divless}
\end{equation}

Step 2 provides a possible form of $\mQ$ in terms of $\mD$, $\vf$, and $\pi$.
First, compare the right hand sides of \eqref{Eq:ItoFPE} and \eqref{Eq:FPE} and observe that they are equivalent if and only if there exists $\mQ(\vz)$ such that:
\begin{align}
\sum_{j} \dfrac{\partial}{\partial \vz_j} \mQ_{ij}(\vz) \pi(\vz)
= \vf_i\pi(\vz) - \sum_j\dfrac{\partial}{\partial \vz_j}\Big(\mD_{ij}(\vz)\pi(\vz) \Big). \label{eq:appnd_eqQ}
\end{align}
Since entries of $\left\{ \sum_j\dfrac{\partial}{\partial \vz_j}\Big(\mD_{ij}(\vz)\pi(\vz) \Big) - \vf_i\pi(\vz) \right\}$ belong to $L^1(\mathbb{R}^d)$, Fourier transform of it exists:
\begin{align}
\hat{{\mF}}_i({\vk})
= \int_{\mathbb{R}^{d}} &\rd \vz
\left({\vf}_i(\vz) \pi(\vz)
- \sum\limits_{j}
\dfrac{\partial}{\partial \vz_j} \Big({\mD}_{ij}(\vz) \pi(\vz)\Big)\right)
\nonumber\\
& e^{-2\pi {\rm i}\ {\vk}^T \vz}. \nonumber
\end{align}
Therefore, we can take $\mQ(\vz)$ such that
\begin{align}
{\mQ}_{ij}(\vz) \pi(\vz)
= {\displaystyle \int_{\mathbb{R}^{d}} \dfrac{{\vk}_j\hat{{\mF}}_i({\vk})-{\vk}_i\hat{{\mF}}_j({\vk})}{(2\pi {\rm i})\cdot \sum\limits_l {\vk}_l^2}
e^{2\pi {\rm i} \sum\limits_l {\vk}_l \vx_l} \rd {\vk} }, \label{Soln}
\end{align}
where entries of $\mQ(\vz)$ belong to $W^{1,1}(\pi)$.

\begin{remark}
It is worth noting that the condition of
\[
\left\{ \sum_j\dfrac{\partial}{\partial \vz_j}\Big(\mD_{ij}(\vz)\pi(\vz) \Big) - \vf_i\pi(\vz) \right\} \in L^1(\mathbb{R}^d)
\]
can be rewritten for the vector field related to the SDE as: $\left\{ \vf(\vz) + \mD(\vz)\nabla H(\vz) - \Gamma^{\mD}(\vz) \right\} \in L^1(\pi)$, where $\Gamma^{\mD}_i(\vz) = \sum_{j=1}^d \dfrac{\partial}{\partial \vz_j} {\mD}_{ij}(\vz)$.
We see that the condition is relatively mild for the purpose of constructing samplers.
\end{remark}

\subsection{Reversible and Irreversible Continuous Dynamics for Sampling}
\label{appendix:rev}
As has been previously noted \cite{qian-jsp,Dembo,qian_jmp,large_deviations,Villani,PavBook}, the stochastic dynamics of Eqs.~\eqref{Eq:FPE} and the corresponding \eqref{Eq:SDE} can be decomposed into a reversible Markov process, and an irreversible process.
Formally, this can be elucidated by the infinitesimal generator $\mathcal{G}[\cdot]$ of the stochastic process \eqref{Eq:SDE} (see \cite{PavBook,stochastic_fluid_dynamics} for more background).
For ease of derivation, we work with the Hilbert space $L^2(\pi)$ of square integrable functions with respect to $\pi(\vz)$, equipped with inner product $<\phi, \varphi>_\pi = \int_{\mathbb{R}^{d}} \bar{\phi}(\vz) \varphi(\vz) \pi(\vz) \rd \vz$.
For $\phi(\vz)\in W^{2,2}(\pi)$ (i.e., $\phi$ and its second order weak derivatives belong to $L^2(\pi)$),
\begin{align}
\mathcal{G} [\phi(\vz)]
=& \nabla^T \Big( \big(\mD(\vz) - \mQ(\vz)\big) \nabla \phi(\vz) \Big) \nonumber\\
&- \nabla H(\vz)^T \big(\mD(\vz) - \mQ(\vz)\big) \nabla \phi(\vz).
\end{align}
Then adjoint of $\mathcal{G}$ in $L^2(\pi)$ is:
\begin{align}
\mathcal{G}^* [\phi(\vz)]
=& \nabla^T \Big( \big(\mD(\vz) + \mQ(\vz)\big) \nabla \phi(\vz) \Big) \nonumber\\
&- \nabla H(\vz)^T \big(\mD(\vz) + \mQ(\vz)\big) \nabla \phi(\vz).
\end{align}
Therefore, $\mathcal{G}$ decomposes into a self-adjoint part $\mathcal{G}^S [\phi(\vz)] = \nabla^T \Big( \mD(\vz) \nabla \phi(\vz) \Big) - \nabla H(\vz)^T \mD(\vz) \nabla \phi(\vz)$ and an anti-self-adjoint part $\mathcal{G}^A [\phi(\vz)] = \nabla^T \Big( \mQ(\vz) \nabla \phi(\vz) \Big) - \nabla H(\vz)^T \mQ(\vz) \nabla \phi(\vz)$.
The self-adjoint operator $\mathcal{G}^S$ corresponds to the reversible Markov process while the anti-self-adjoint operator $\mathcal{G}^A$ corresponds to the irreversible process.

It can be seen that the reversible process is determined solely by the diffusion matrix $\mD$, where evolution of its probability density function is:
\begin{align}
\centering
\dfrac{\partial}{\partial t} p(\vz;t)
=& \nabla^T \cdot
	\bigg( \mD(\vz) \left[
p(\vz;t) \nabla H(\vz) + \nabla p(\vz;t) \right]\bigg) \nonumber\\
=& \sum_{i,j=1}^d \dfrac{\partial^2}{\partial \vz_i \partial \vz_j} \bigg(\mD_{ij}(\vz) p(\vz;t)\bigg) \label{Eq:Ls} \\ \nonumber
&+ \sum_{i=1}^d \dfrac{\partial}{\partial \vz_i} \left( \left[\sum_j\mD_{ij}(\vz) \dfrac{\partial H(\vz)}{\partial \vz_j} - \Gamma^{\mD}_i(\vz)\right] p(\vz;t) \right).
\end{align}
Here, $\Gamma^{\mD}_i(\vz) = \sum_{j=1}^d \dfrac{\partial}{\partial \vz_j} {\mD}_{ij}(\vz)$.
According to It\^o's convention, \eqref{Eq:Ls} corresponds to reversible Brownian motion in a potential force field on a Riemannian manifold specified by the diffusion matrix $\mD(\vz)$: $\rd \vz = \big[ - \mD(\vz) \nabla H(\vz) + \Gamma^{\mD}(\vz)\big] \rd t + \sqrt{2 \mD(\vz)} \rd \mW$. This is referred to as \emph{Riemannian Langevin dynamics} \cite{RiemannianMALA,RMALA}. When $\mD(\vz)$ is positive definite, the reversible Markov dynamics have nice statistical regularity and will drive the system to converge to the stationary distribution.

The irreversible process is determined solely by $\mQ$, with its probability density function evolving according to:
\begin{align}
\centering
\dfrac{\partial}{\partial t} p(\vz;t)
=& \nabla^T \cdot
	\bigg( \mQ(\vz) \left[
p(\vz;t) \nabla H(\vz) + \nabla p(\vz;t) \right]\bigg) \nonumber\\
=& \nabla^T \cdot
	\Bigg( \Big[\mQ(\vz) \nabla H(\vz) - \Gamma^{\mQ}(\vz) \Big] {p(\vz;t)} \Bigg). \label{Eq:La}
\end{align}
Here, $\Gamma^{\mQ}_i(\vz) = \sum_{j=1}^d \dfrac{\partial}{\partial \vz_j} {\mQ}_{ij}(\vz)$. The last line of \eqref{Eq:La} is a Liouville equation, which describes the density evolution of ${p(\vz;t)}$ according to conserved, deterministic dynamics: $\rd \vz /\rd t = -\mQ(\vz) \nabla H(\vz) + \Gamma^{\mQ}(\vz)$, with $\pi(\vz)$ its invariant measure.

Combining the dynamics of \eqref{Eq:Ls} and \eqref{Eq:La} leads to a general SDE, \eqref{Eq:SDE}, with stationary distribution $\pi(\vz)$.
Previous methods \cite{MALA,RMALA,SGLD,SGRLD,NealHMC} have primarily focused on solely reversible or irreversible processes, respectively, as we make explicit in Sec.~\ref{sec:previous_continuous}.
\section{Irreversible Jump Processes for MCMC}
\subsection{Equivalence of \eqref{Eq:MJP} and \eqref{Eq:Jump_S}}
\label{appnd_NewJump}


We introduce a symmetric bivariate function
$S(\vx,\vz) = S(\vz,\vx) = \dfrac12\Big(W(\vz|\vx)\pi(\vx) + W(\vx|\vz)\pi(\vz)\Big)$,
and an anti-symmetric bivariate function
$A(\vx,\vz) = -A(\vz,\vx) = \dfrac12\Big(W(\vz|\vx)\pi(\vx) - W(\vx|\vz)\pi(\vz)\Big)$
for \eqref{Eq:MJP}.
A different form of the jump process \eqref{Eq:MJP} can be written according to $S(\vx,\vz)$ and $A(\vx,\vz)$ as
%
\begin{align}
\dfrac{\partial \ptrans{\vy}{\vz}{t}}{\partial t}
= \int_{\mathbb{R}^d} \rd \vx
\bigg[ & S(\vx ,\vz) \dfrac{\ptrans{\vy}{\vx}{t}}{\pi(\vx)} - S(\vx,\vz) \dfrac{\ptrans{\vy}{\vz}{t}}{\pi(\vz)} \nonumber\\
& + A(\vx,\vz) \dfrac{\ptrans{\vy}{\vx}{t}}{\pi(\vx)}\bigg].
\nonumber
\end{align}

Plugging $\ptrans{\vy}{\vz}{t} = \pi(\vz)$ into the above equation, we find that as long as $\int_{\mathbb{R}^d} A(\vx,\vz) \rd \vx = 0$, $\pi(\vz)$ is a stationary solution to the equation.
Since $\dfrac{S(\vx,\vz) + A(\vx,\vz)}{\pi(\vx)}$ denotes a transition probability density, ${S(\vx,\vz) + A(\vx,\vz)}>0$ for any $\vx$ and $\vz$.
The restriction that $S(\vx,\vz) \pi^{-1}(\vx)$ and $ A(\vx,\vz) \pi^{-1}(\vx)$ are bounded is imposed for the practical purpose of proposing samples in \eqref{Eq:Jump_Sampler}.
We thereby notice that the requirement that $\pi(\vz)$ is a stationary distribution of the jump process is translated into simpler constraints.

\subsection{Verifying condition \ref{Eq:Jump_Constraints} on $A(\vy,\vy^p,\vz,\vz^p)$ in
Section \ref{sec:irr_jump}}
\label{Appnd_SkewSymm}
The anti-symmetric function $A(\vy,\vy^p,\vz,\vz^p)$ (expressed in \eqref{eq:A_alter}) of \eqref{Eq:IrrFlow} can be written as:
\begin{align}
A(\vy, & \vy^p,\vz,\vz^p) \nonumber\\
= \dfrac{1}{2 \Delta t}
\Big( & \pi(\vy)\pi(\vy^p) \ptrans{\vy,\vy^p}{\vz,\vz^p}{\Delta t} \nonumber\\
& - \pi(\vz)\pi(\vz^p) \ptrans{\vz,\vz^p}{\vy,\vy^p}{\Delta t} \Big) \nonumber\\
= \dfrac{1}{2} \delta(\vz^p & -\vy^p)
\Big(\mathfrak{F}(\vy,\vy^p,\vz,\vz^p) - \mathfrak{F}(\vz,\vz^p,\vy,\vy^p)\Big) \nonumber\\
- \dfrac{1}{2} \delta( & \vz^p+\vy^p)\delta(\vz-\vy) \nonumber\\
\cdot \int_{\mathbb{R}^d} & \Big(\mathfrak{F}(\vy,\vy^p,\vx,-\vz^p) - \mathfrak{F}(\vz,\vz^p,\vx,-\vy^p)\Big)\rd \vx .
\nonumber
\end{align}
Below we prove that, as required,
\[
\int_{\mathbb{R}^{d+d^p}} A(\vy,\vy^p,\vz,\vz^p) \ \rd \vy \ \rd \vy^p = 0.
\]
\begin{proof}
\begin{align}
& \int_{\mathbb{R}^{d+d^p}} A(\vx,\vx^p,\vz,\vz^p) \ \rd \vx \ \rd \vx^p \nonumber\\
= & \dfrac{1}{2} \int_{\mathbb{R}^{d}} \Big(\mathfrak{F}(\vy,\vz^p,\vz,\vz^p)
- \mathfrak{F}(\vz,\vz^p,\vy,\vz^p)\Big) \ \rd \vy
\nonumber\\
& - \dfrac{1}{2} \int_{\mathbb{R}^{d}} \Big(\mathfrak{F}(\vz,-\vz^p,\vx,-\vz^p) - \mathfrak{F}(\vz,\vz^p,\vx,\vz^p)\Big)\rd \vx.
\nonumber
\end{align}
One can check that in \eqref{Eq:f&g} and \eqref{Eq:f&g_1D},
\[
\widetilde{f}(\vz,\cdot \ |\vy, -\vy^p) = \widetilde{g}(\vz,\cdot \ |\vy,\vy^p).
\]
Hence,
\[
\mathfrak{F}(\vy,-\vy^p,\vz,-\vz^p)=\mathfrak{F}(\vz,\vz^p,\vy,\vy^p).
\]
Therefore
\begin{align}
& \int_{\mathbb{R}^{d+d^p}} A(\vx,\vx^p,\vz,\vz^p) \ \rd \vx \ \rd \vx^p \\ \nonumber
= & \dfrac{1}{2} \int_{\mathbb{R}^{d}} \Big(\mathfrak{F}(\vz,-\vz^p,\vy,-\vz^p)
- \mathfrak{F}(\vz,\vz^p,\vy,\vz^p)\Big) \ \rd \vy
\\ \nonumber
& - \dfrac{1}{2} \int_{\mathbb{R}^{d}} \Big(\mathfrak{F}(\vz,-\vz^p,\vx,-\vz^p) - \mathfrak{F}(\vz,\vz^p,\vx,\vz^p)\Big)\rd \vx
= 0.
\nonumber
\end{align}

\end{proof}

\section{Proof of Theorem \ref{thm:adjoint} (relation between forward process and adjoint process)}
\label{append_adjoint}

We first prove that for the infinitesimal generators, the backward transition probability density following the adjoint process and the forward transition probability density are related as: $\pi(\vy) P\big(\vz | \vy; \rd t \big) = \pi(\vz) P^\dag\big(\vy | \vz; \rd t \big)$.
Taking path integrals with respect to the infinitesimal generators leads to the conclusion.

As is standard,
we use two arbitrary smooth test functions $\psi(\vy)$ and $\phi(\vz)$.
Then we use the definition of the infinitesimal generator of the process $P$ and $P^\dag$:
$\mathcal{G} [\phi(\vy)] \rd t = \int P\big(\vz | \vy; \rd t \big) \phi(\vz) \rd \vz - \phi(\vy)$ and obtain
\begin{align}
& \int\int \rd \vy \rd \vz \ {P\big(\vz | \vy; \rd t\big)} {\pi(\vy)} \ \psi(\vy) \phi(\vz) \nonumber\\
= & <\psi, (I+ \rd t \ \mathcal{G})[\phi]>_\pi,
\nonumber
\end{align}
and
\begin{align}
& \int\int \rd \vy \rd \vz \ {P^\dag\big(\vy | \vz; \rd t\big)} {\pi(\vz)} \ \psi(\vy) \phi(\vz) \nonumber\\
= & <(I+ \rd t \ \mathcal{G}^*)[\psi], \phi>_\pi.
\nonumber
\end{align}
Since $\mathcal{G}$ and $\mathcal{G}^*$ are adjoint in $L^2(\pi)$: $<\psi, \mathcal{G}[\phi]>_\pi = <\mathcal{G}^*[\psi], \phi>_\pi$,
\begin{align}
\pi(\vy) P\big(\vz | \vy; \rd t \big) = \pi(\vz) P^\dag\big(\vy | \vz; \rd t \big).
\nonumber
\end{align}

Then we take path integrals over the forward path and the backward one.
Using the Markov properties,
\begin{align}
&P(\vz_N,t_N|\vz_0,t_0)
\nonumber\\
= &\int\cdots\int\prod_{i=1}^{N-1}\rd \vz_i\prod_{i=0}^{N-1} P(\vz_{i+1},t_{i+1}|\vz_i,t_i);
\nonumber
\end{align}
and
\begin{align}
& P^\dag(\vz_0,t_N|\vz_N,t_0) \nonumber\\
= &\int\cdots\int\prod_{i=1}^{N-1}\rd \vz_i\prod_{i=0}^{N-1} P^\dag(\vz_{i},t_{i+1}|\vz_{i+1},t_i)
\nonumber\\
= &\int\cdots\int\prod_{i=1}^{N-1}\rd \vz_i\prod_{i=0}^{N-1} \dfrac{\pi(\vz_i)}{\pi(\vz_{i+1})} P(\vz_{i+1},t_{i+1}|\vz_{i},t_i) \nonumber\\
= &\int\cdots\int\prod_{i=1}^{N-1}\rd \vz_i\prod_{i=0}^{N-1} \dfrac{\pi(\vz_i)}{\pi(\vz_{i+1})} P(\vz_{i+1},t_{i+1}|\vz_{i},t_i) \nonumber\\
= &\dfrac{\pi(\vz_0)}{\pi(\vz_{N})} P(\vz_N,t_N|\vz_0,t_0).
\nonumber
\end{align}
Taking the time interval between $t_i$ and $t_{i+1}$ to be infinitesimal, we obtain that $\dfrac{P(\vz^{(T)},T|\vz^{(t)},t)}{P^\dag(\vz^{(t)},T|\vz^{(T)},t)} = \dfrac{\pi(\vz^{(T)})}{\pi(\vz^{(t)})}$.
The same conclusion can be reached by proving that the semigroups $e^{t\mathcal{G}}$ and $e^{t\mathcal{G}^*}$ generated by $\mathcal{G}$ and $\mathcal{G}^*$ are also adjoint with each other \cite{Markov_Dual,Semigroups,poncet}.

\section{Experiments}

\subsection{Parameter settings for the irreversible jump sampler}
\label{appnd_Experiment}

In the 1D experiments, we use $\beta = 1.2$ for the normal distribution case (where the length of the region of definition is $10$) and $\beta = 0.8$ for the log-normal distribution (where the length of the region of definition is $5$).
The acceptance rate is around $50\%$ in these cases.
Due to the irreversibility of the sampler, a high acceptance rate can be maintained while reducing the autocorrelation time.

In the visual comparison of samplers of Fig.~\ref{fig:TracePlot}, we use $\beta=0.15$ (where the lengths of the region of definition is $2 \times 2$).
In the 2D bimodal experiments, we use $\beta=0.4$ (where the lengths of the region of definition is $6 \times 3$).
In the 2D multimodal experiments, we take $\beta=1.5$ (where the lengths of the region of definition is $14 \times 14$).
For the 2D correlated distribution, we take $\beta=0.25$ (where the lengths of the region of definition is $5 \times 2$).

\subsection{Effect of Dimensionality on I-Jump versus MH}
\label{appnd_dimension}
In this appendix, we explore how the relative improvement of I-Jump over MH scales with dimensionality.
We consider a standard normal distribution and double the dimension from one comparison to the next.
For the I-Jump sampler, we use the vanilla half-space Gaussian proposal.
It can be observed from Table \ref{table:Gauss} that the potential benefits of irreversibility in the I-Jump sampler slowly diminish with increasing dimensionality.

\begin{table}
  \centering
  \begin{tabular}{ c | c | c | c }
Dimensions & MH & I-Jump & I-Jump:MH \\
\hline
10 & $16683.95$ & $20717.96$ & $1.24$ \\
\hline
20 & $7101.00$ & $9648.46$ & $1.36$ \\
\hline
40 & $3464.73$ & $3772.41$ & $1.09$ \\
\hline
80 & $1555.40$ & $1787.41$ & $1.15$ \\
\hline
160 & $537.87$ & $561.94$ & $1.04$ \\
\hline
320 & $72.46$ & $76.97$ & $1.06$ \\
\hline
640 & $28.18$ & $28.99$ & $1.03$
\end{tabular}
  \caption{Comparison of $\widehat{ESS}$ per second of runtime for I-Jump versus MH with different dimensions of Gaussian target distributions.
  }\label{table:Gauss}
\end{table}

\end{document}